\documentclass[journal]{IEEEtran}

\usepackage[T1]{fontenc}
\usepackage{ae,aecompl}
\usepackage{pslatex}
\usepackage[samesize]{cancel}
\usepackage{graphicx}
\usepackage{setspace}
\usepackage{amsmath}
\usepackage{amsfonts}
\usepackage{subcaption}
\usepackage{amssymb}
\usepackage{booktabs} 
\usepackage{xcolor}
\usepackage{booktabs}
\usepackage{float}
\usepackage{comment}
\usepackage{mathptmx} 
\usepackage{lineno}
\usepackage{url}
\usepackage{calc}
\usepackage{tabto}
\usepackage{lettrine}
\usepackage{multirow}
\usepackage{upgreek}
\usepackage{float}
\usepackage{placeins}

\begin{document}




\title{Anomaly distinguishability in an asteroid analogue using quasi-monostatic experimental radar measurements}

    \author{Yusuf ~Oluwatoki ~Yusuf,
        Astrid ~Dufaure,
        Liisa-Ida ~Sorsa,
        Christelle ~Eyraud,
        Jean-Michel ~Geffrin,
        Alain ~Hérique,
        and~Sampsa ~Pursiainen
\thanks{ Y. O. Yusuf, L-I. Sorsa, and S. Pursiainen are with the Computing Sciences, Tampere University (TAU), PO Box 692, 33101 Tampere, Finland.}
\thanks{Y. O. Yusuf, A. Dufaure, C. Eyraud, and J-M. Geffrin are with Aix Marseille Univ, CNRS, Centrale Marseille, Institut Fresnel, Marseille, France.}
\thanks{A. Hérique is with Univ. Grenoble Alpes, CNRS, CNES, IPAG, F-38000 Grenoble, France.}
\thanks{This study was carried out through the Research Council of Finland projects 336151 and 359198; and was supported by the Research Council of Finland Centre of Excellence in Inverse Modelling and Imaging, 2018-2025, 353089, as well as Flagship of Advanced Mathematics for Sensing, Imaging and Modelling, 359185. Y.O. Yusuf has also been supported by the Magnus Ehrnrooth Foundation through the graduate student scholarship.} 
\thanks{Manuscript submitted on January 29, 2024. Corresponding author:
Yusuf Oluwatoki Yusuf (yusuf.yusuf@tuni.fi).}}

\maketitle
\begin{abstract}
This study conducts a quantitative distinguishability analysis using quasi-monostatic experimental radar data to find a topographic and backpropagated tomographic reconstruction for an analogue of asteroid Itokawa (25143). In particular, we consider a combination of travel-time and wavefield backpropagation tomography using the time-frequency representation (TFR) and principal component analysis (PCA) approaches as filtering techniques. Furthermore, we hypothesise that the travel time of the main peaks in the signal can be projected as a topographic imaging of the analogue asteroid while also presenting a tomographic reconstruction based on the main peaks in the signal. We compare the performance of several different filtering approaches covering several noise levels and two hypothetical interior structures: homogeneous and detailed. Our results suggest that wavefield information is vital for obtaining an appropriate reconstruction quality regardless of the noise level and that different filters affect the distinguishability under different assumptions of the noise. The results also suggest that the main peaks of the measured signal can be used to topographically distinguish the signatures in the measurements, hence the interior structure of the different analogue asteroids. Similarly, a tomographic reconstruction with the main peaks of the measured signal can be used to distinguish the interior structure of the different analogue asteroids. 
\end{abstract}




\IEEEpeerreviewmaketitle

\section{Introduction}
\label{sec:intro}

The goal of imaging the interior structure of small solar system bodies (SSSBs) such as asteroids and comets is motivated by the need to understand the evolution, composition, and most importantly the threat posed by these objects \cite{michel2015asteroids}. This motivation has led to a multidimensional view of studying SSSBs in fields such as Planetary and Imaging sciences via the methods of mathematical modelling and signal processing. In particular, tomographic radar imaging aims to understand unique sometimes intersecting elements relating to the knowledge of SSSBs' interior \cite{herique2018direct}. These elements can vary from physical properties such as porosity, or mass density to mechanical ones such as velocity, or trajectory and electrical properties e.g., dielectric constant (permittivity) or resistivity.  SSSBs' complex and largely unknown interior structures will be investigated by future planetary space missions, most prominently by the European Space Agency's (ESA's) HERA space mission which will explore its target Dimorphos using its tomographic radar instrument Juventas Radar \cite{michel2022esa}.

Radar probing of SSSBs has been approached using various forward and inverse techniques, since the emergence of such techniques, see e.g. \cite{knaell1995radar}. The CONSERT experiment used radio-wave transmission to probe the deep interior structure of comet 67P/Churyumov–Gerasimenko thus inferring its dielectric properties as part of ESA's Rosetta mission \cite{Kofman2007,Kofman2015}. CONSERT was based on travel-time tomography to reconstruct the propagation speed of electromagnetic waves as they travel through and interact with the target from one end to the other \cite{Kofman2020, Herique2016, Herique2019}. The direct propagation-path tomography was selected on Rosetta since forward propagation simulations through the comet nucleus models expect low scattered and refracted wave power \cite{kofman1998}. The method is robust with respect to noise and modelling errors \cite{barriot1999two}. However, its limitation is its low resolution with regards to obtaining a sharp contrast in the material properties \cite{sava2015}. Other techniques based on the reflection of electromagnetic waves from the object are able to overcome this challenge, thus, providing a high resolution of the dielectric properties distribution in the target via reflection-based methods e.g., wavefield tomography \cite{sava2018tomography,eyraud2018imaging,Sorsa2019,deng2021ei} and Synthetic Aperture Radar (SAR) tomography \cite{gassot2021ultra,herique2019radar}. There are possibilities to advance this investigation in ESA's upcoming Hera mission through the Juventas CubeSat \cite{michel2022esa,herique2019juventas}. The target of the HERA mission, Dimorphos has recently been visited by the National Aeronautics and Space Administration's (NASA's) DART (Double Asteroid Redirection Test) mission \cite{rivkin2021double,hirabayashi2022double,daly2023successful}.


In this study, we aim to infer the deep interior structure of a laboratory-scale analogue through a quantitative distinguishability analysis including time-frequency analysis, backpropagation and filtered backpropagation, thus identifying backscattered signatures of different model compartments. In particular, we consider the distinguishability of anomalous details which is a central question in the analysis of mission data, namely, it is not necessarily evident if the data presents a homogeneous or a non-homogeneous interior. Our approach coupled with previously established forward and inverse models \cite{pursiainen2016orbiter} is aimed to help understand and analyse data of future space missions such as the Hera mission. 

Motivated by the robustness of travel-time tomography \cite{sobolev1999robust}, we investigate time-frequency filtering as a potential way to, isolate the main peaks of the measured signal while maintaining the major reflections of the wavefield data, which we hypothesise to be important with respect to the reconstruction quality. Time-frequency analysis of signals has been extensively studied in the past decades, particularly as it allows characterising signals simultaneously with respect to both time and frequency \cite{daubechies1990wavelet,cohen1995time}. The study of time-frequency analysis originates from Fourier analysis \cite{grochenig2001foundations} and currently constitutes one of the core signal processing methods in the various radar applications of geophysics, civil engineering, and biomedical engineering. Its techniques are primarily categorised as linear transforms \cite{hlawatsch1992linear} such as short-time-Fourier-Transform (STFT), Gabor expansion, wavelet transform, and quadratic transforms \cite{qian1999joint} such as the spectrogram (short-time-Fourier-Transform STFT-based) and Wigner-Vile Distribution (WVD) \cite{grochenig2001foundations,boashash2015time}. To investigate the applicability of the linear and quadratic time-frequency transforms on the distinguishability of laboratory measurement data,  we focus on the wavelet transform and the spectrogram as one of the distinguishability approaches vis-à-vis the topographic reconstruction of the laboratory scale analogue. Furthermore, we consider the backpropagation and filtered backpropagation approaches which have been extensively discussed as tomographic methods for inferring the deep interior structure of objects \cite{esmersoy1989backprojection,l2012filtered,koljonen2019mathematical,dufaure2023imaging,sorsa2023imaging}.

Our results suggest that the wavefield information is vital for obtaining an appropriate reconstruction quality and that filters based on wavelet processing and principal component analysis can improve the reconstruction quality if the signal-to-noise ratio (SNR) of the measurement is above 10 dB. The results also suggest that the main peaks of the measured signal can be used to distinguish the signatures in the measurements topographically, hence the interior structure of the different asteroid analogues. Similarly, a tomographic reconstruction with the main peaks of the measured signal can be used to distinguish the interior structure of the different analogues.





\section{Materials and methods}
\label{sec:methods}

\subsection{Experimental data acquisition}
\label{analogue}
\begin{figure}[!ht]
    \centering  
    \begin{scriptsize}
    \begin{minipage}{2.55cm} \centering \hskip -0.5cm
    \includegraphics[width=2.55cm]{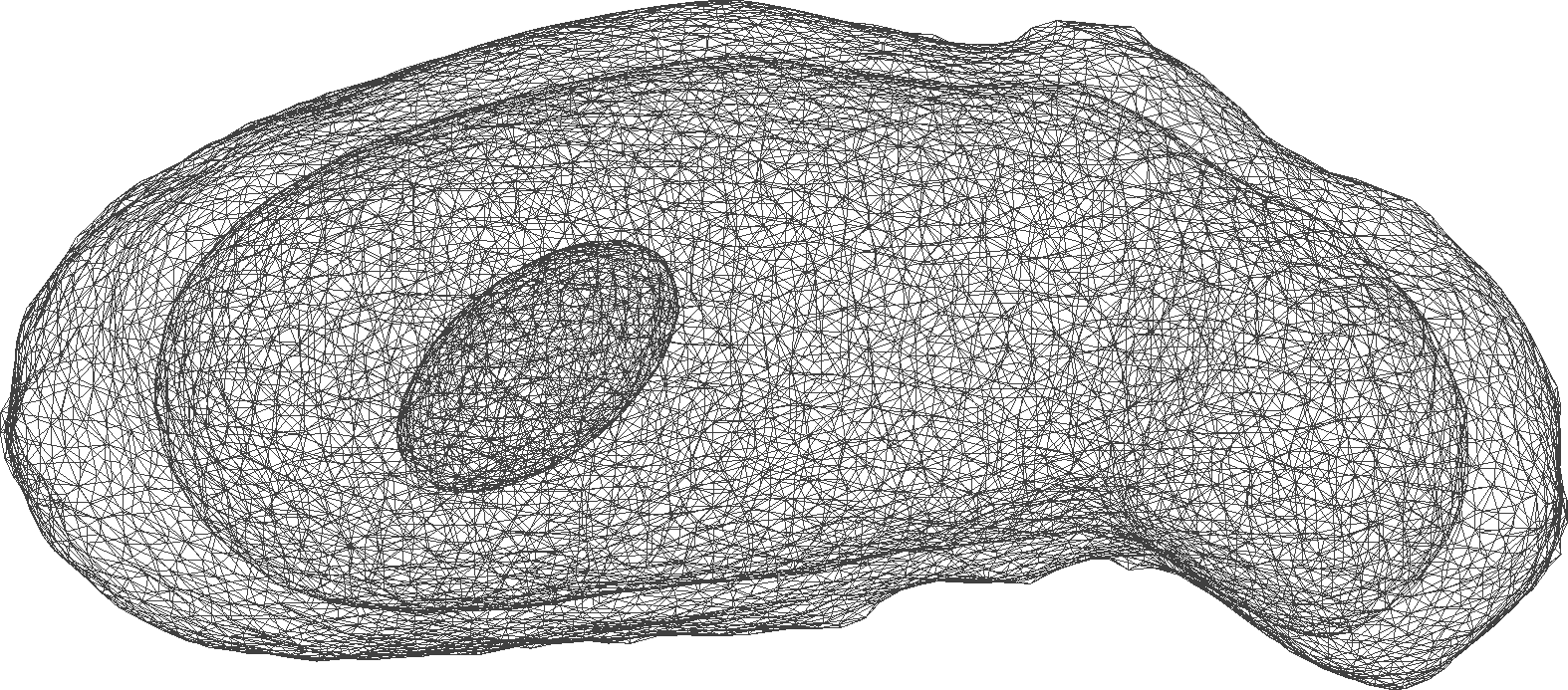} 
\end{minipage} \begin{minipage}{2.55cm} \centering
    \includegraphics[width=2.55cm]{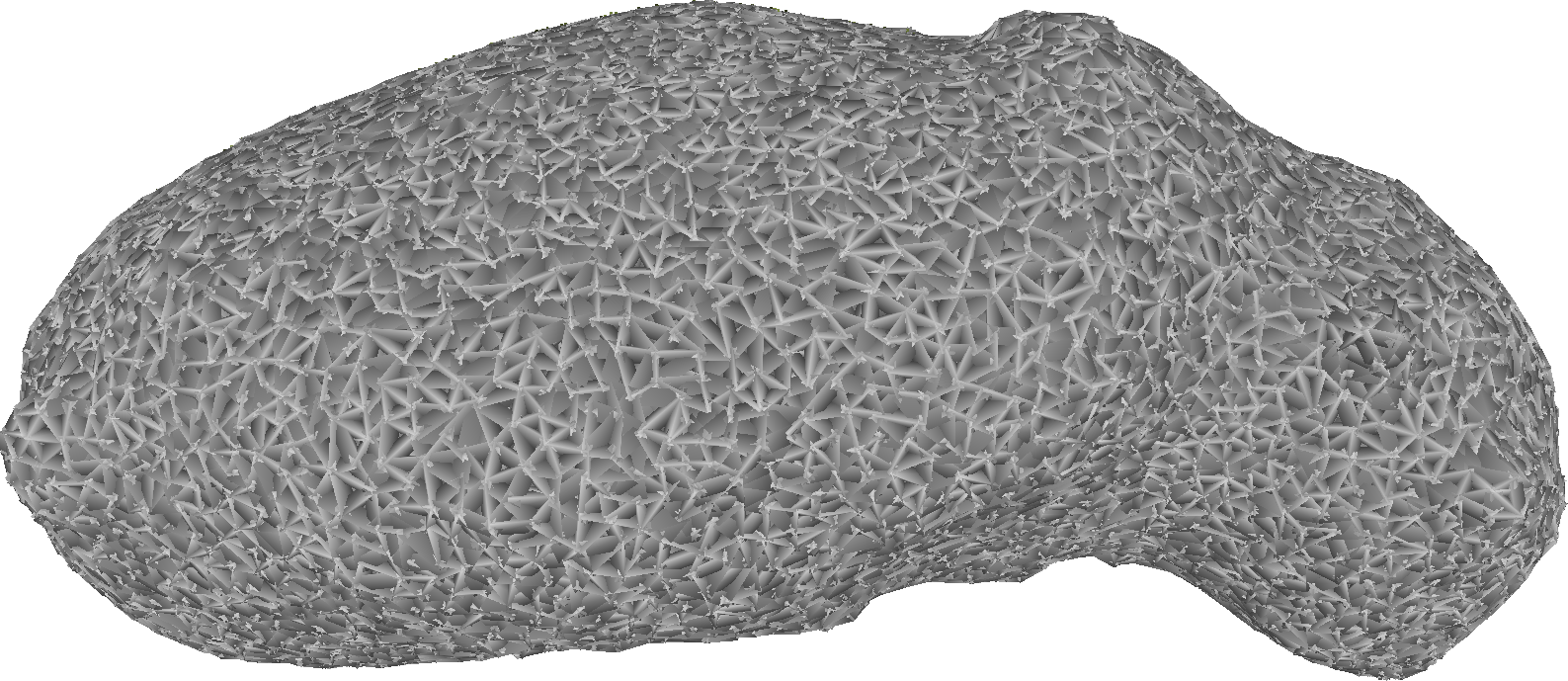} 
\end{minipage}
\end{scriptsize}
    \caption{3D printed Analogue of the asteroid Itokawa. {\bf{Left:}} Detailed model (DM) showing the surface layer, interior, and the deep interior void. {\bf{Right:}} Homogeneous model (HM) showing a uniform distribution with no compartments.} 
    \label{fig1}
\end{figure}

Two 3D printed analogues with the external shape of asteroid Itokawa (25143) \cite{kawaguchi2008hayabusa} are used as the tomographic target in this study. The structure, composition and variation of the analogue are consistent with the previous studies on the investigation of the interior structure of this target \cite{sorsa2021analogue,Eyraud2020analog,sorsa2021analysis}. The detailed model (DM, see Fig. \ref{fig1}, Left) consists of a surface layer $\varepsilon_r = 2.56 + j0.02$), an interior compartment
($\varepsilon_r = 3.40 + j0.04$), and a deep interior void ($\varepsilon_r = 1.00$), while the homogeneous model (HM, see Fig. \ref{fig1}, Right) has a constant relative permittivity of ($\varepsilon_r = 3.40 + j0.04$). The values of the relative permittivity $\varepsilon_r$ have been measured in previous studies describing how mixture models such as the classical Maxwell Garnett (MG) can be used to estimate the permittivity of 3D printed analogues \cite{sorsa2021analogue,Eyraud2020analog}. The experimental data was measured in an anechoic chamber using a quasi-monostatic configuration with a total of 2732 measurement positions with a 12-degree elevation angle separation between the transmitter and receiver antenna. The measurement frequency ranges from $f_{\min} = 2$ GHz to $f_{\max} = 18$ GHz with a sampling step of 0.05 GHz covering a total of 321 frequency points. Each analogue has a diameter of 20.5 cm which is 12.3 times the wavelength of the highest measured frequency in vacuum, and  22.7 times the wavelength in the medium ($\varepsilon_r = 3.40 + j0.04$). The measurements were performed from a 1.794 m distance which is 8.7 times the analogue diameter.

\begin{figure}[!ht]
    \centering  
    \begin{scriptsize}
    \begin{minipage}{2.55cm} \centering
    \includegraphics[width=2.55cm]{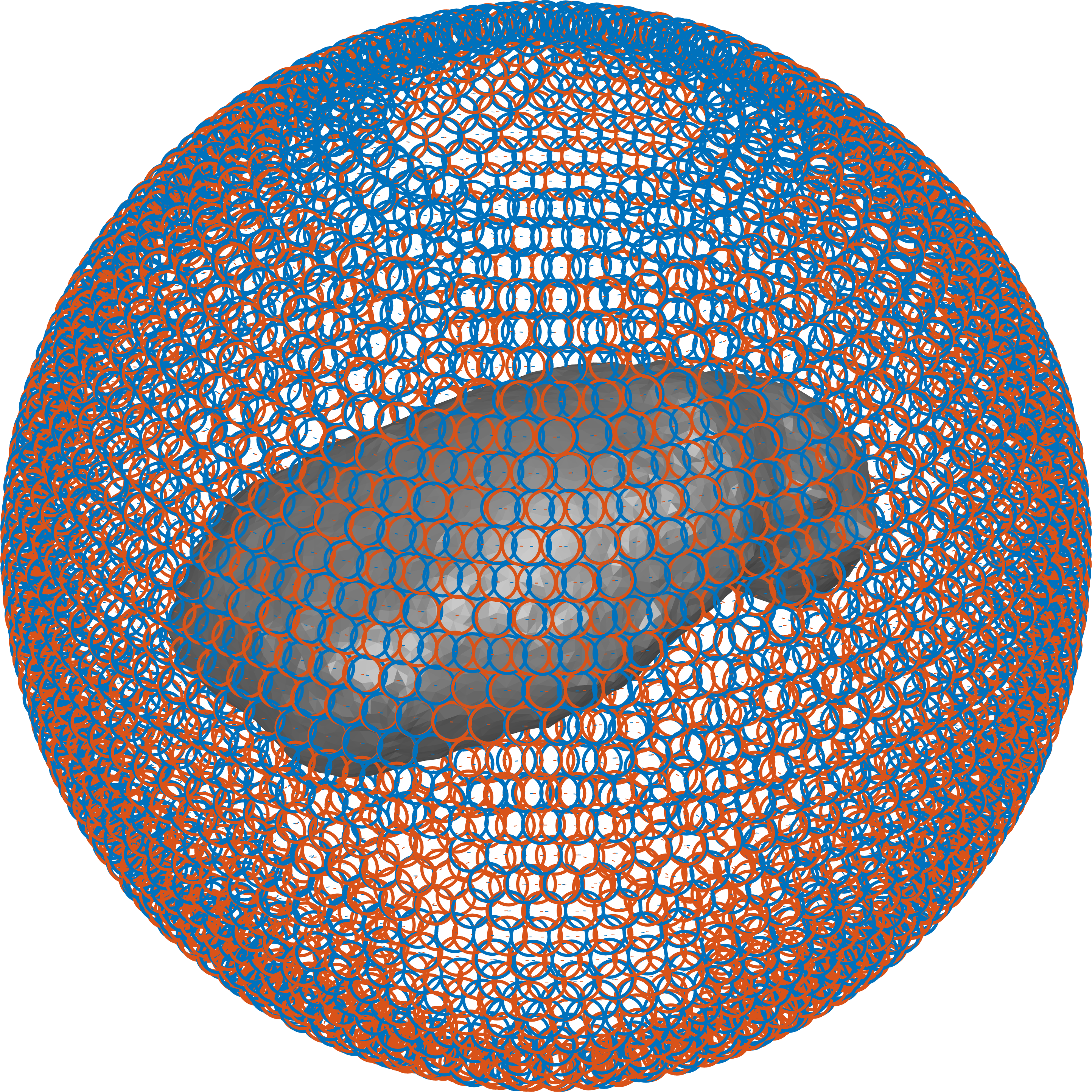} 
\end{minipage} \\ \vskip 0.5cm
\begin{minipage}{4.55cm} \centering
    \includegraphics[width=4.55cm]{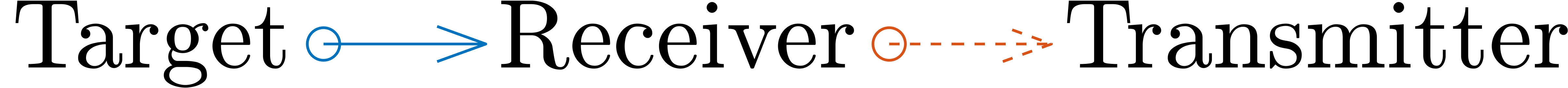} 
\end{minipage}
\end{scriptsize}
    \caption{The quasi-monostatic signal configuration of 2732 measurement positions with a 12-degree elevation angle separation between the transmitter and receiver. } 
    \label{fig2}
\end{figure}
\subsection{Linear forward model}
\label{linear_forward_model}

As the unknown of the tomography, we consider a discretized induced current distribution inside the asteroid, i.e., vector ${\bf x}$, which is connected to the measurement data via a linear mapping of the form ${\bf y} = \mathcal{G}{\bf x} $. The data vector ${\bf y}$ is composed of wavefield measurements performed in the set of measurement positions described above. The  Green's function $\mathcal{G}$ under the far-field condition, due to the distance of the target to the receiver and the wavelength-to-diameter ratio can be approximated as a vacuum space propagation. However, such an approximation introduces an error term which comprises the measurement and modelling error $\tilde{\bf y} $. We formulate the forward model as: 
\begin{equation}
\label{forward_model}
{\bf y} = {\bf G} {\bf x} + \tilde{\bf y}, 
\end{equation}
where ${\bf G}$ depicts a matrix of the Green's function in a vacuum space under the far-field condition, ${\bf G} {\bf x}$ is the scattered field and $\tilde{\bf y}$ an unknown residual vector. The residual effect can be reduced by presenting (\ref{forward_model}) in the following filtered form
\begin{equation}
\label{filtered_forward_model}
{\bf F}_1 {\bf y} = {\bf G} {\bf F}_2 {\bf x} + {\bf F}_1 \tilde{\bf y} 
\end{equation}
where ${\bf F}_1$ is a data filter suppressing higher-order scattering effects in the data so that $\|{\bf F}_1 \tilde{\bf y} \| << \| \tilde{\bf y} \|$,  and ${\bf F}_2$ is a weighting matrix, which aims at correcting the modelling bias related to first-order scattering, i.e., the difference between free space propagation and true signal propagation in the target.

\begin{figure}[!ht]
    \centering  
    \begin{scriptsize}
    \begin{minipage}{4.55cm} \centering
    \includegraphics[width=4.55cm]{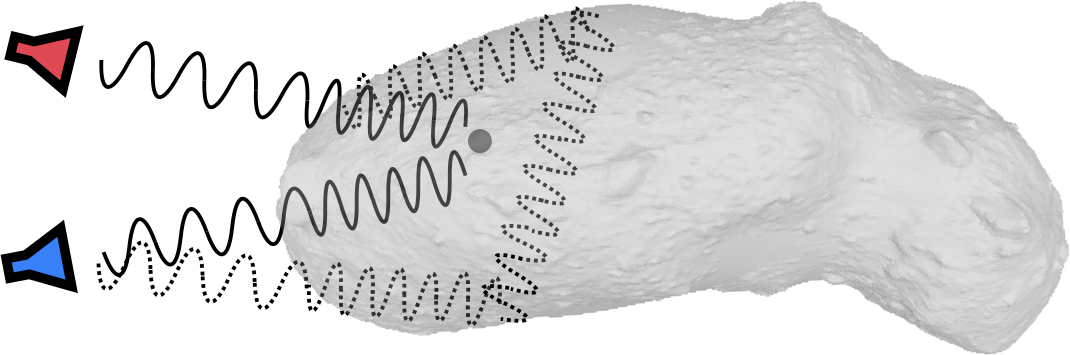} 
\end{minipage} 
\end{scriptsize}
    \caption{Schematic representation of a first-order scattering from the target to the receiver in thick lines versus the higher-order scattering in the dashed lines.  } 
    \label{fig2b}
\end{figure}
In travel-time tomography, the first-order scattering effects are often associated with the first received signal wavefront. Distinguishing the first-order scattering effects from the measurement data can be considered with these algorithms for obtaining an appropriate reconstruction quality. This part of the signal which contains the majority of the signal energy shown in Figs. \ref{fig2b}, is often dominated by surface reflections. To approximate first-order scattering while suppressing the surface scattering, we isolate the signal peak with maximum energy using time-frequency filtering techniques, which are briefly reviewed below.
\subsection{Time-frequency representation (TFR)}
\label{transforms}
The theory of time-frequency and wavelet transform have been extensively discussed in literature such as \cite{cohen1995time,daubechies1990wavelet,daubechies1992ten,grochenig2001foundations,jaffard2001wavelets}. A time-frequency transform provides a two-dimensional representation of a signal characterised by its spatio-temporal resolution. Wavelet transform constitutes a special case in which the split between time and frequency domains is obtained with respect to a wavelet function basis whose entries are well-localised in both spatial and temporal domains. We focus on the spectrogram (STFT-based) which follows from the short-time-Fourier-Transform, and the Wavelet transforms as examples of quadratic and linear time-frequency transforms on the distinguishability of laboratory measurement data. 

\subsubsection{Spectrogram}
For a continuous signal in time $s(t) \in \mathbb{C}$ and its Fourier transform $s(\omega) = \int_{-\infty}^{\infty} s(t)\exp({-j{\omega}t}) dt$, the power spectrum is defined as 
\begin{equation}
    \label{power spectrum}
    P(\omega) = \left| s(\omega)\right|^2 = \int_{-\infty}^{\infty} s(t)s^*(t-\tau)\exp({-j{\omega}t}) dt
\end{equation}

The short-time-Fourier-transform (STFT) replaces the time-lagged signal $s^*(t-\tau)$ by a window function $\gamma^*(t-mT)$ and the phase by $\exp({-jn{\Omega}t})$. Here, the complex conjugate of the signal $s^*$ is
used in the assumption that the signal $s(t)$ is complex-valued, $\tau$ is the time-delay, $T$  and $\Omega$ are time and frequency steps, respectively. Taking the Fourier transform of the signal on a block basis, defined by the window length $mT$ where $m$ is a fraction of the signal length $T$. The STFT is hence defined as 
\begin{equation}
    \label{STFT}
    STFT[mT, n\Omega] = \int_{-\infty}^{\infty} s(t)\gamma^*(t-mT)\exp({-jn{\Omega}t}) dt.
\end{equation}
A way to show the time dependency of the signal is to take the square of the STFT in equation \eqref{STFT}. This gives the energy localisation of the spectral component for each time frame considered.
\begin{equation}
    \label{spectrogram}
    P(t, \omega) = \left| \int_{-\infty}^{\infty} s(t)\gamma^*(t-\tau)\exp({-j{\omega}t}) dt \right|^2.
\end{equation}
Equation \eqref{spectrogram} is the spectrogram of the signal $s(t)$. 

The time-frequency techniques are such that each technique is characterised by its inherent strengths and weaknesses. The most widely used technique is the spectrogram due to its simplicity, and can be written in the Cohen class form \cite{cohen1995time,qian1999joint,stankovic1994method}, however, it may not be suitable for applications that require high resolution in time and frequency \cite{cohen1995time}. The spectrogram, for example, also suffers from the window effect where the width of the window function determines the resolution of the signal in time and frequency.

\subsubsection{Continuous wavelet transform}
The continuous wavelet transform of the signal $s(t)$ is given by:
\begin{equation}
    \label{wavelet}
    \mathrm{WT}(a,b)  = \frac{1}{\sqrt{a}}\int s(t) \psi^*\left(\frac{t-b}{a}\right)dt,
\end{equation}
where $a$ and $b$ are the scale and translation parameters, respectively, and $\psi^{a,b} = \frac{1}{\sqrt{a}} \psi^*\left(\frac{t-b}{a}\right)$ denotes the wavelet basis. Here, $\psi^*$ is the complex conjugate of the mother wavelet $\psi$ as in equation \ref{wavelet}.

The weighted energy of the continuous wavelet transform coefficients of the signal (scalogram) $\mathrm{PWT}(a,b)$ is hence defined as 
\begin{equation}
\label{scalogram}
    \mathrm{PWT}(a,b) = \frac{1}{2\pi Ca^2}|WT(a,b)|^2,
\end{equation}

where the constant $C$ is the admissibility condition that ensures the signal energy is preserved in the scalogram \cite{debnath2003wavelets,daubechies1992ten}.


While the window function in the spectrogram and the wavelet basis function in the wavelet transform both provide a time-frequency description of the signal $s(t)$, the difference lies in their shape and width which is the same over the signal in the case of the window function but adaptive with regards to the frequency in the case of the wavelet. Hence, the wavelet transform gives a multiresolution of the signal (zooming in on high frequency) \cite{daubechies1992ten}.

\subsection{Principal Component Analysis (PCA)}
\label{pca}
PCA is a powerful tool with diverse applicability ranging from dimensionality reduction, data transformation, regression, and data reconstruction. This study uses PCA specifically for dimensionality reduction and data reconstruction.

Given the measurement data $Y = \{y_{(i)}\}$ with dimensions $n\times p$, where the $n$ rows represent the different measurement observations and the $p$ columns represent the frequencies. The principal component (PC) of $Y$ is the projection 
\begin{equation}
\label{pca2}
    T = YW = \{t_{k(i)}\} = \{y_{(i)}.w_{(k)}\},
\end{equation} where $i = 1, \cdots, n$ and $k = 1, \cdots, p$. The elements of the $p\times p$ weight matrix $W$ are the PC loadings where $w_{(k)} = (w_{(1)},\cdots,w_{(p)})_{(k)}$ while the elements of $T_{n\times p}$ are the PC scores.

For dimensionality reduction, the first $\ell<p$ principal components can be selected to give a truncated transformation of the form
\begin{equation}
\label{dim_red}
    T_l = YW_\ell = \{t_{k(i)}\} = \{y_{(i)}.w_{(k)}\},
\end{equation}  where $i = 1, \cdots, n$ and $k = 1, \cdots, \ell$, resulting into a $n\times \ell$ matrix. This matrix maximises the variance in the measurement data while minimising the data reconstruction error \cite{jolliffe2002principal,jolliffe2016principal}. 

The reconstructed $n\times p$ data $\hat{Y}$ based on the first $\ell<p$ principal components is given by 
\begin{equation}
\label{data_rec}
    \hat{Y} = T_{\ell}W^T_\ell = \{\hat{y}_{(i)}\} = \{t_{k(i)}.w^T_{(k)}\}.
\end{equation} 

To reduce the dimension of the reconstructed data, the column-wise variance $\sigma^2 = \mathrm{Var(y_k)}$ is ranked to give the variables with the largest contribution to the data variance. Thus reducing the weight vector $w_{(1)} = (w_{(1)},\cdots,w_{(m)})_{(1)}$ to the dimension of the largest $m<p$ variables contributing to the variance of $Y$.

The reconstructed $n\times m$ data $\hat{Y}_m$ based on the first $\ell<p$ principal components and the largest $m<p$ variables contributing to the variance of $Y$ is given by 
\begin{equation}
\label{data_rec2}
    \hat{Y}_m = T_{\ell}W^T_{(m)\ell} = \{\hat{y}_{(i)}\} = \{t_{k(i)}.w^T_{(k)}\}.
\end{equation} where $k = 1, \cdots, \ell$ and $W$ is reduced to a $m\times \ell$ matrix.

Since the PCA can segment the data into principal components in increasing order of variance, and the time-frequency methods can present the time and frequency information in the data; the combination of the PCA and the time-frequency methods can be used to analyse the signature of principal parts of the data. We combined these approaches by taking the PCA on the data using the set of measurement positions as the variable. This is implemented by transposing the data (rows represent the frequencies, and the columns represent the different measurement observations) since the analysis would be done according to the set of frequencies. The TFR of the 1st--3rd PC scores (representation of the data in the principal component space) is presented in Section \ref{sec:results}, Fig. \ref{result1}.

\subsection{Topographic projection}
Our approach to the topographic projection of the depth profile on the analogue surface is motivated by travel time tomography as widely used in vertical seismic profiling \cite{bording1987applications,tarantola1984seismic}. We assume that the travel time of the maximum energy derived from the time-frequency methods contains not only surface reflection but also information on the deep interior structure. This approach can be linked to time migration in seismic inversion where it is assumed that there are only mild lateral variations in the velocity profile at the target subsurface \cite{schuster2017seismic}. The distance ${\bf d}_{r}$ from the receiver to discretised analogue points is given by 
\begin{equation}
    \label{distance}
    {\bf d}_{r} = ||\Omega - \Gamma_{r}||, 
\end{equation}
where $\Omega$ is the points in the analogue and $\Gamma_{r}$ is the receiver position. To capture the information of the deep interior structure, we normalised the distances by the maximum distance from the analogue to the receiver as 

\begin{equation}
    \label{depth}
    {\bf u}_r = \exp\left({\frac{-\gamma\times\max{({\bf d}_{r}^2)}}{{\bf d}_{r}^2 + \delta}}\right), 
\end{equation}
where $\gamma$ and $\delta$ are the smoothing and condition parameters, respectively.
A smooth depth profile is obtained as the topographic projection defined by 
\begin{equation}
    \label{project}
    {\bf P} = \sum_{r=1}^n {\mathtt{c}_0}\times{\mathtt{t}}\times{\bf u}_r, 
\end{equation}
where ${\mathtt{c}_0}$ is the speed of light in vacuum, $n$ is the number of receiver positions, and ${\mathtt{t}}$ is the travel time of the maximum energy as obtained from the time-frequency methods. To ensure that this approach is not biased by the longest dimension of the analogue, we also considered as unit sphere as the target, hence replicating the methods in equation \eqref{distance}, \eqref{depth}, and \eqref{project}. The topographic projection of the depth profile on the surface of the analogue and a unit sphere is presented in Figs. \ref{result3} and \ref{result4}.
\subsection{Inversion}
\label{inverse}

There is a growing body of literature on the suitability of different inversion schemes for scattering problems \cite{esmersoy1989backprojection,bertero2021introduction}. In this study, we reconstruct the induced current distribution inside the target using the backpropagation technique, see more details in \cite{dufaure2023imaging}, which is simple and fast when compared to other iterative inversion schemes \cite{l2012filtered}. 

Given a scattered field ${\bf y}$, the backpropagation inversion is defined as
\begin{equation}
\label{backpropagation}
   {\bf x}^\ddag  =  {\bf G}^\dagger {\bf y},
\end{equation}
where the operator, ${\bf G}^\dagger$, is the transpose-conjugate matrix of the dyadic Green's function in free space and ${\bf x}^\ddag$ is the reconstructed map of the induced current in the target.

In the filtered backpropagation (FBP), the reconstruction is found in the following filtered form 
\begin{equation}
{\bf x}^\ddag = {\bf F}_2^\ast {\bf G}^\dagger {\bf F}_1{\bf y}, 
\end{equation}
which follows by backpropagating the filtered forward model (\ref{filtered_forward_model}), and can be assumed to improve (\ref{backpropagation}), if the filters ${\bf F}_1$ and ${\bf F}_2$ are chosen so that the residual term of (\ref{filtered_forward_model}) is marginalised. Thus, in FBP, Green's function does not operate directly on the measurement data but on a function with respect to the data.
Here, the wavelet scalogram in equation \eqref{scalogram}, and PCA reconstruction in equation \eqref{data_rec2} are used as the filtering function ${\bf F}_1{\bf y}$.

\subsubsection{Attenuation correction (AC)}
\label{attenuat}
As the forward model correction ${\bf F}_2$, we test a model which takes into account the signal decay in the target (see, e.g., \cite{chang1978method,koljonen2019mathematical}), which can be approximated to be exponential; a signal with an original amplitude $A$ has the amplitude $A_d = \exp(-\alpha d)\times A$ after travelling the distance $d = 2\times0.205$ m (which is two times the maximum dimension of the target) inside the target. Considering the central frequency $f_c = 10$ GHz, the relative permittivity ($\varepsilon_r = \varepsilon^{'}_{r} + j\varepsilon^{''}_{r} = 3.40 + j0.04$), speed of light in vacuum ${\mathtt{c}_0} = 2.987\times 10^8$ m/s,  and the attenuation constant $\alpha = \frac{2\pi f_c}{\mathtt{c}_0} \sqrt{ \frac{\varepsilon'_r}{2} \sqrt{1+\left(\frac{\varepsilon''_r}{\varepsilon'_r}\right)^2}-1}$. We approximate the attenuation constant to be $\alpha = 2.27$, which matches roughly the real signal attenuation inside the analogues \cite{sorsa2021analogue,yusuf2022investigation} and a rate of 3.6 dB/km, when converted to the real scale of the 535 m diameter asteroid 25143 Itokawa.  For $i$-th inversion point inside the target, $d$ is approximated as $d_i = \mathcal{D} \max{j} \mathcal{F_{i,j}}$, where $\mathcal{D}$ is the diameter of the asteroid and $\mathcal{F}_{i,j}$ the maximum mass fraction on the straight line segment between the $i$-th inversion and $j$-th  measurement point in the homogeneous analogue. The points outside the target are associated with the value $d = 0$. As a result, ${\bf F}_2$ becomes a diagonal matrix in which the $i$-th  diagonal entry is of the form   $\exp(-\alpha d_i)$. A similar correction has been found to be an effective strategy in many laboratory applications of tomographic imaging, such as fluorescence tomography \cite{koljonen2019mathematical}.

\subsection{Evaluation criteria}

To quantify the distinguishability of the reconstructed structures ${\bf x}^\ddag$ (interior, and void) with respect to a  theoretical map ${\bf x}$, we considered the root mean squared error (RMSE) of the different parts of the analogue i.e., exterior, interior, void, and all (a combination of the parts) and the relative overlap (RO) of the void structure in the case of the detailed measurement \cite{sorsa2023imaging}. Since the homogeneous analogue does not present a void structure, the RMSE of the void and RO are excluded. The RMSE measures are best when close to zero while the RO ranges from 0 to 1, and best when close to one.

\label{measure}


\section{Results}
\label{sec:results}

The result section is split into three parts, We first examine the filtered data topographically. We consider in particular, the travel time of the maximum energy to enlighten its relevance in the present experimental setting. In the second part, backpropagated tomographic reconstructions of filtered waveform signals are analysed. We consider both laboratory data as presented in  \cite{sorsa2023imaging,dufaure2023imaging}, where the measurement SNR is greater than 20 dB \cite{Eyraud2020analog}. In the third part, we used noisy data corresponding to additive and uncorrelated Gaussian random errors for the noise characterisation and robustness of methods implemented in the tomographic imaging.

\begin{figure}[h]
    \centering  
    \begin{scriptsize}   \begin{minipage}{8.4cm}\centering \textbf{HM} \hskip1.5cm 1st PC \hskip1.5cm \textbf{DM} \\ \vskip0.2cm
    \begin{minipage}{2.15cm} \centering
    \includegraphics[width=2.15cm]{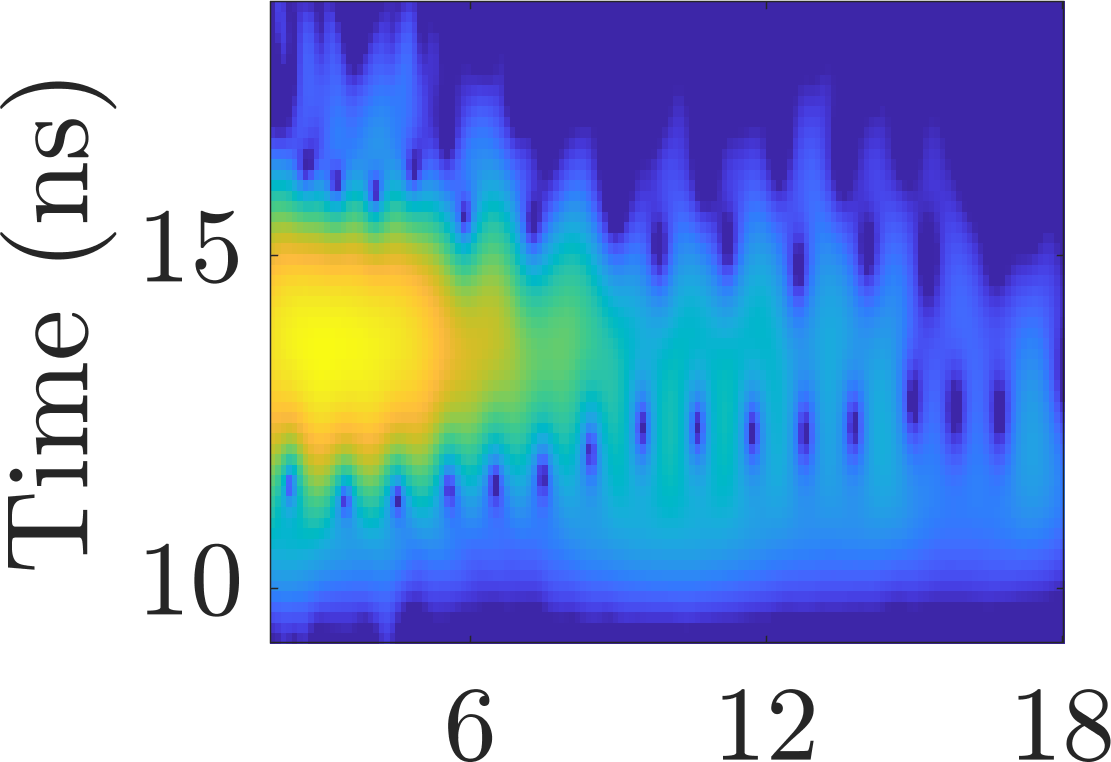} 
\end{minipage}
\begin{minipage}{2.0cm} \centering
    \includegraphics[width=2.0cm]{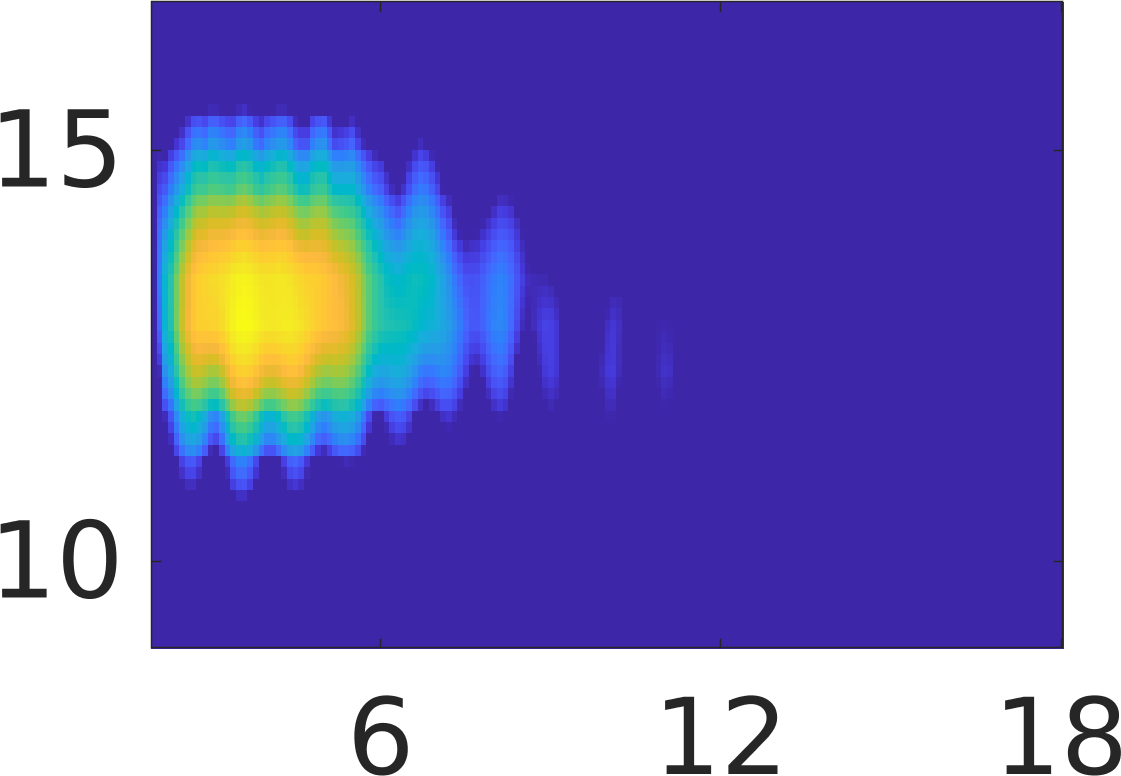} 
\end{minipage}
\begin{minipage}{2.0cm} \centering
    \includegraphics[width=2.0cm]{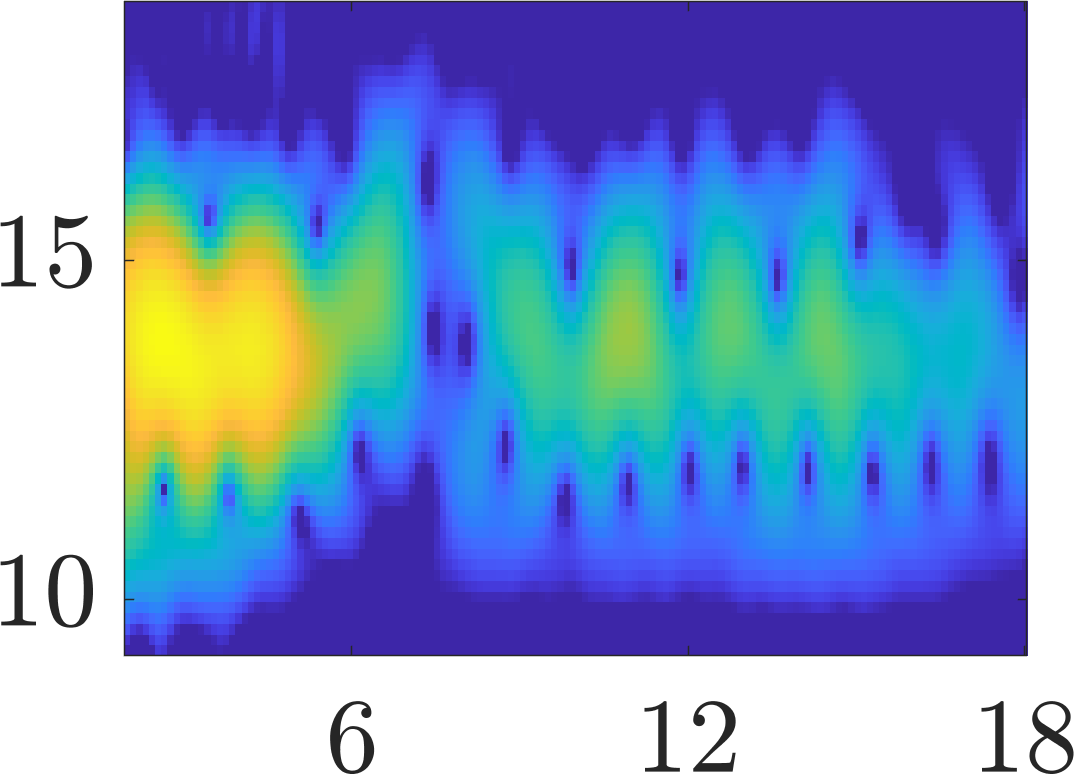} 
\end{minipage}
\begin{minipage}{2.0cm} \centering
    \includegraphics[width=2.0cm]{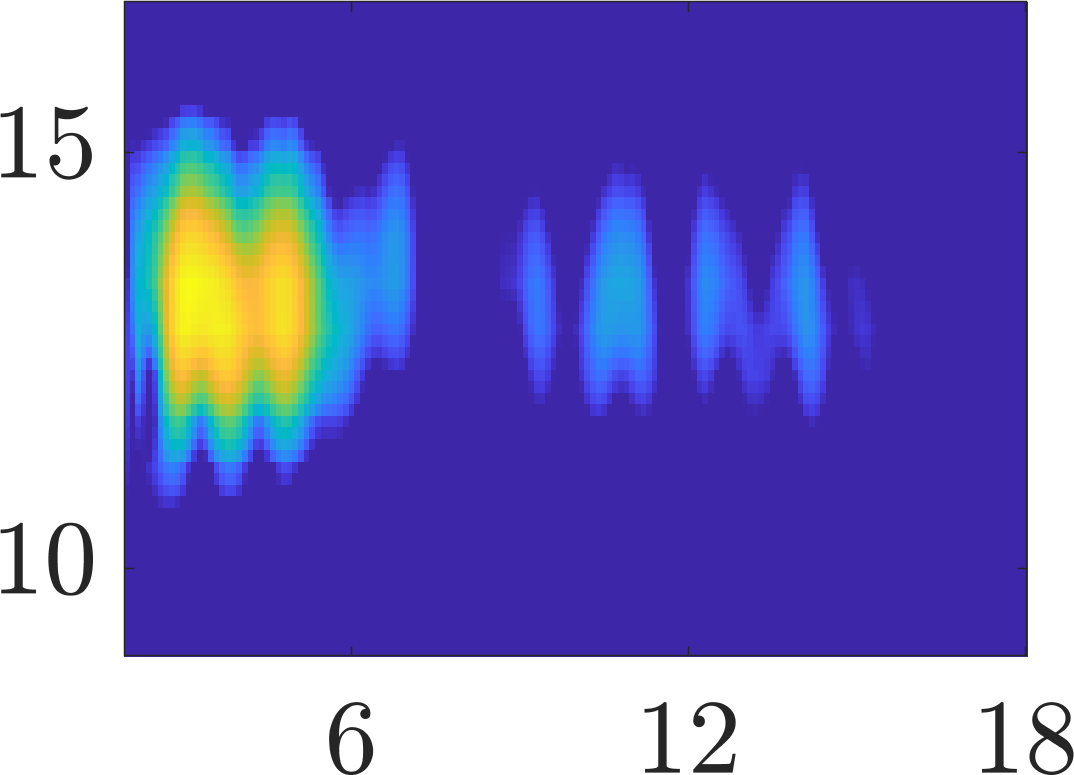} 
\end{minipage}\\ \vskip0.2cm  
2nd PC \\ \vskip0.2cm 
  \begin{minipage}{2.0cm} \centering
    \includegraphics[width=2.0cm]{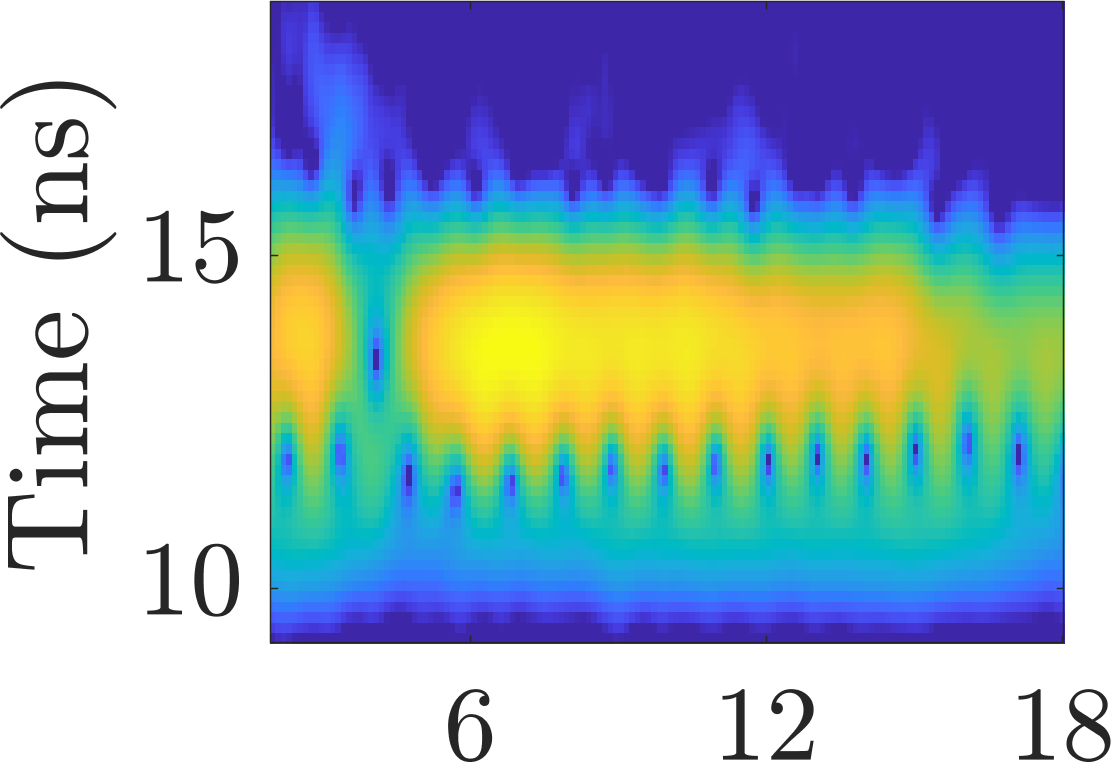} 
\end{minipage}
\begin{minipage}{2.0cm} \centering
    \includegraphics[width=2.0cm]{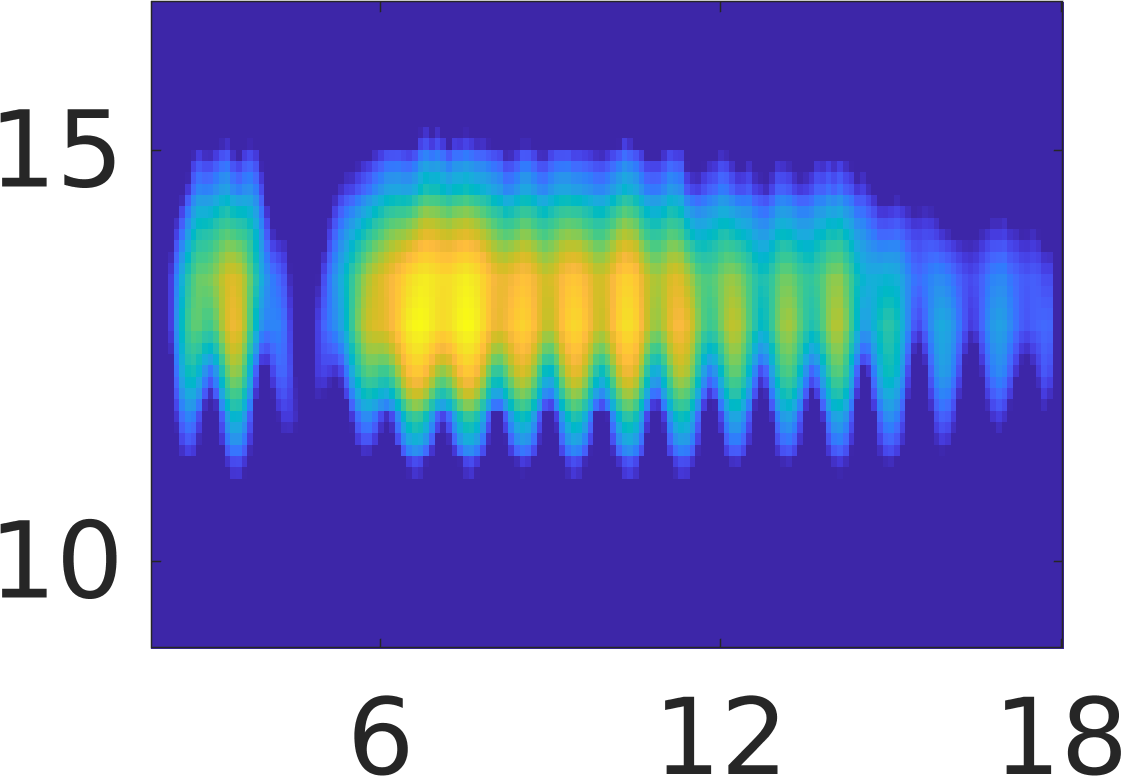} 
\end{minipage}
\begin{minipage}{2.0cm} \centering
    \includegraphics[width=2.0cm]{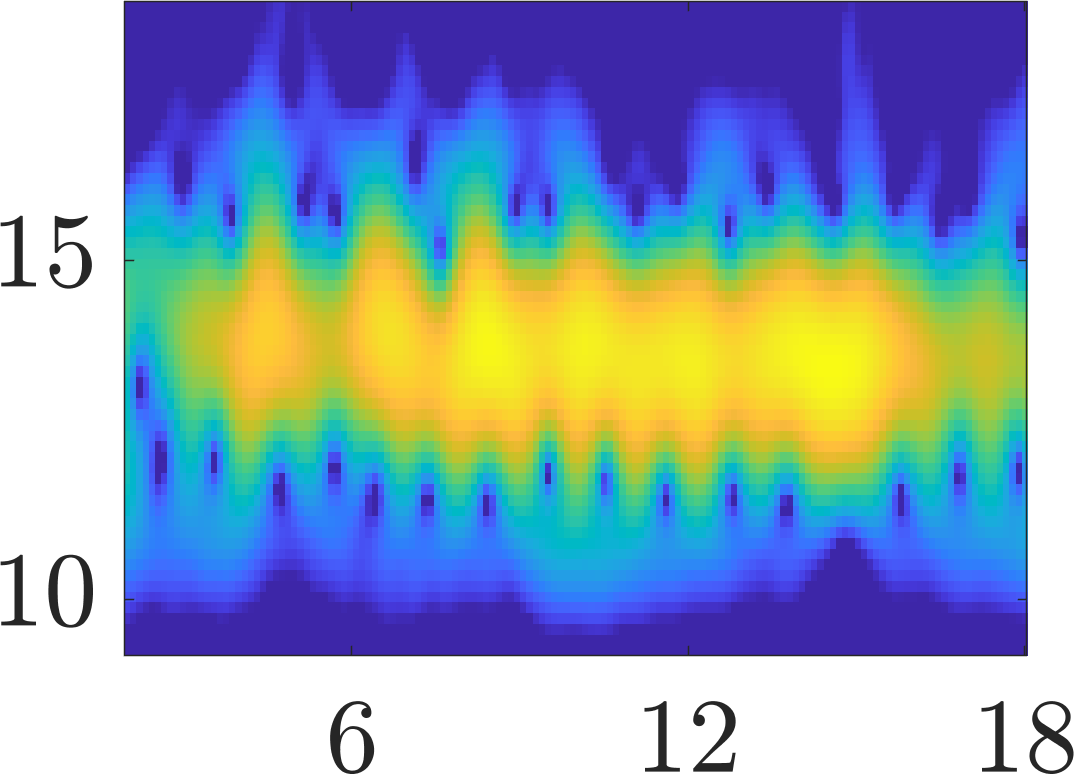} 
\end{minipage}
\begin{minipage}{2.0cm} \centering
    \includegraphics[width=2.0cm]{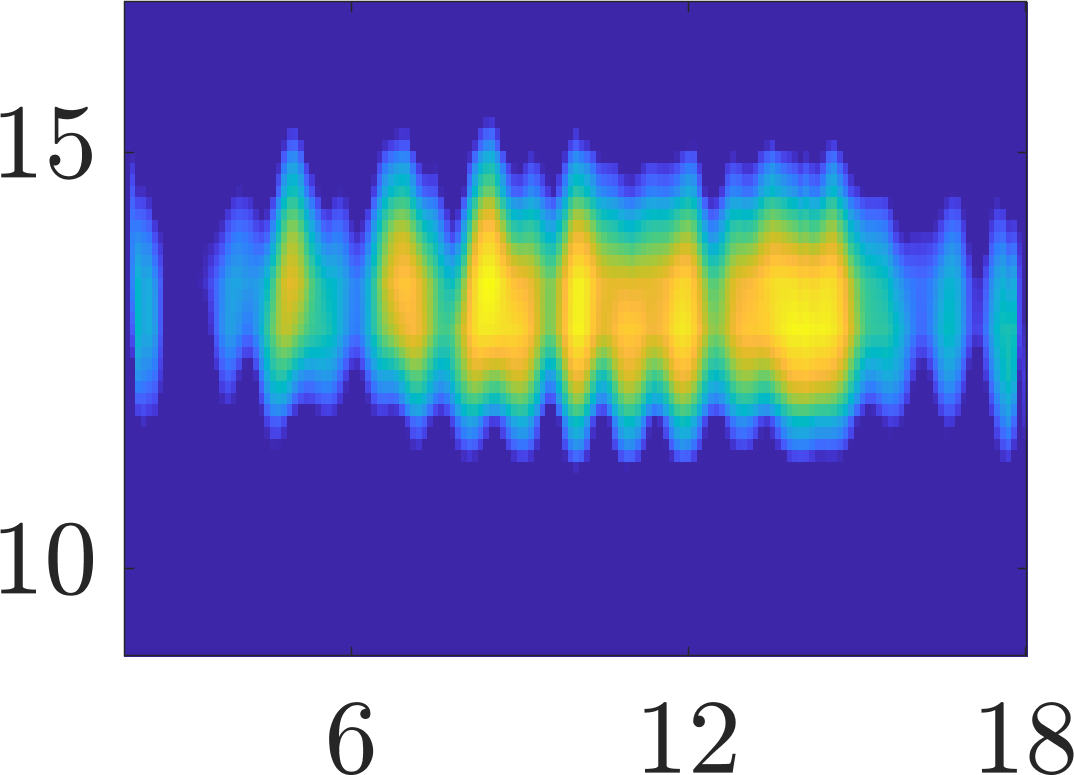} 
\end{minipage}\\ \vskip0.2cm 

3rd PC \\ \vskip0.2cm 
  \begin{minipage}{2.15cm} \centering
    \includegraphics[width=2.15cm]{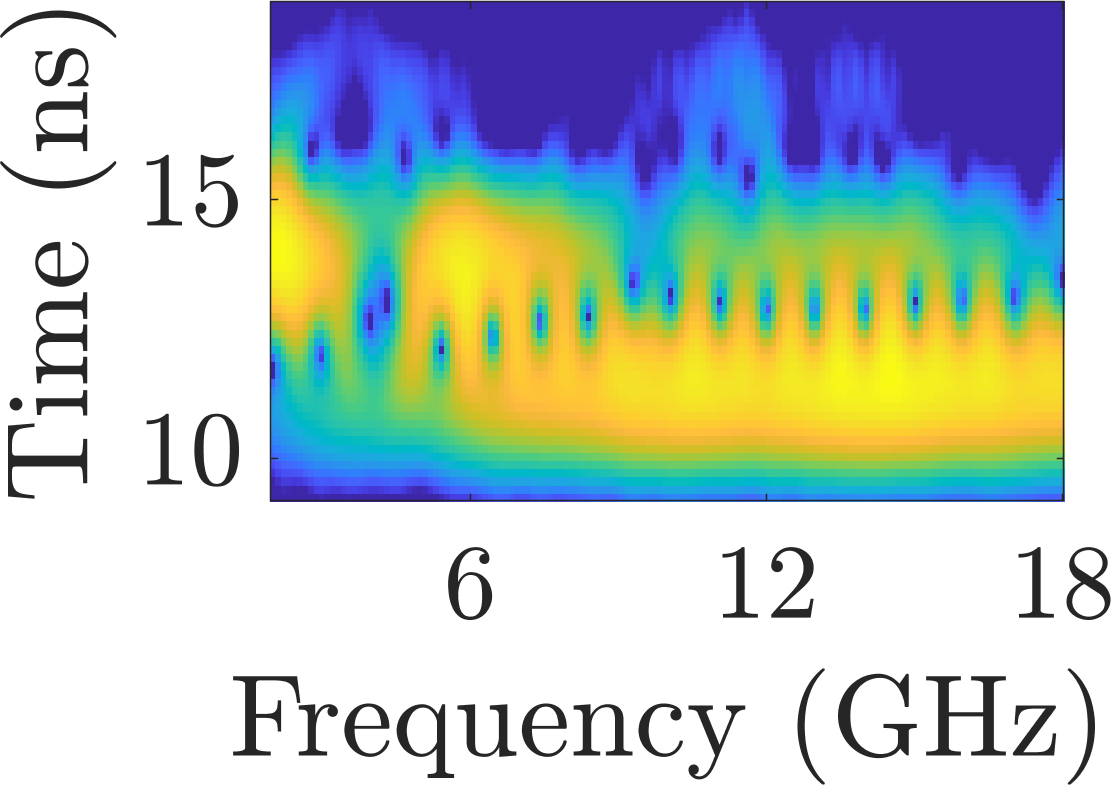} 
\end{minipage}
\begin{minipage}{2.0cm} \centering
    \includegraphics[width=2.0cm]{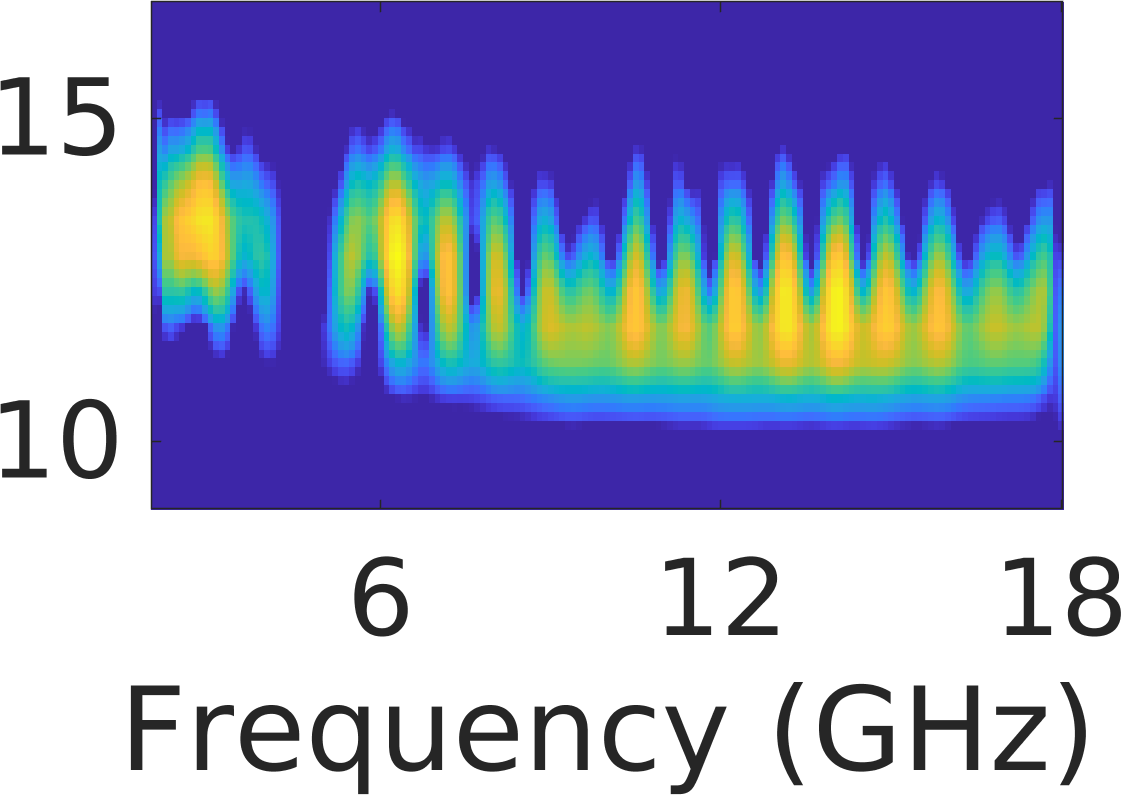} 
\end{minipage}
\begin{minipage}{2.0cm} \centering
    \includegraphics[width=2.0cm]{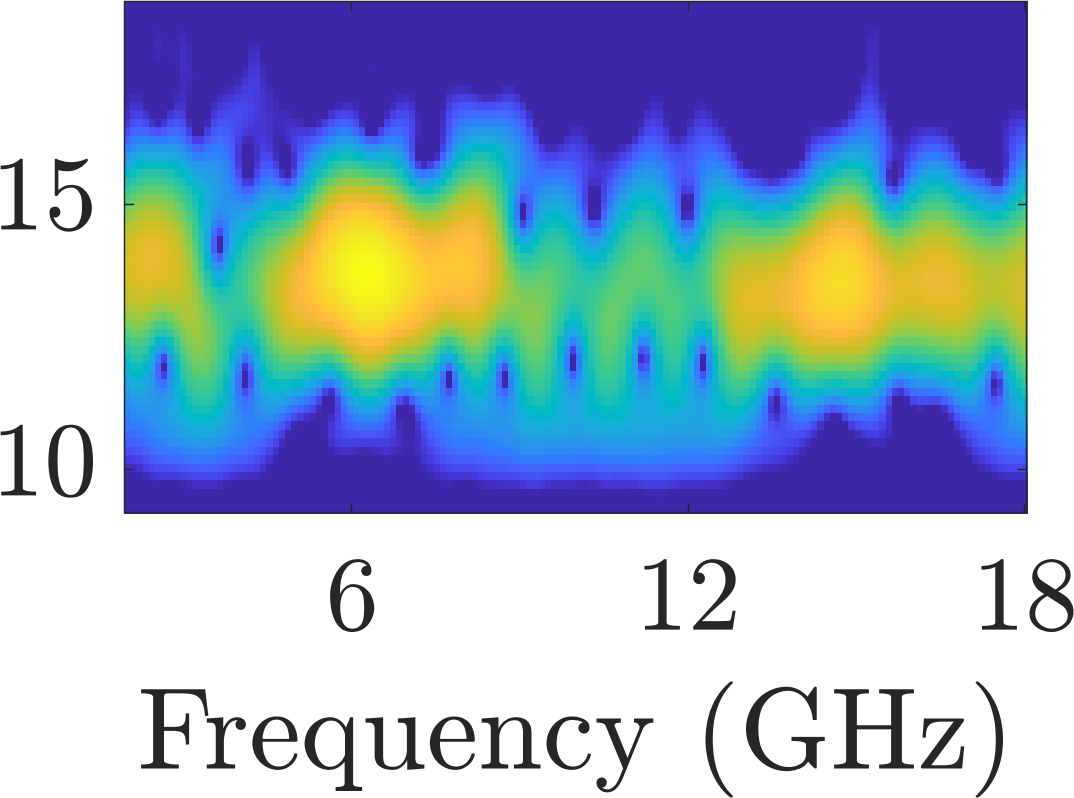} 
\end{minipage}
\begin{minipage}{2.0cm} \centering
    \includegraphics[width=2.0cm]{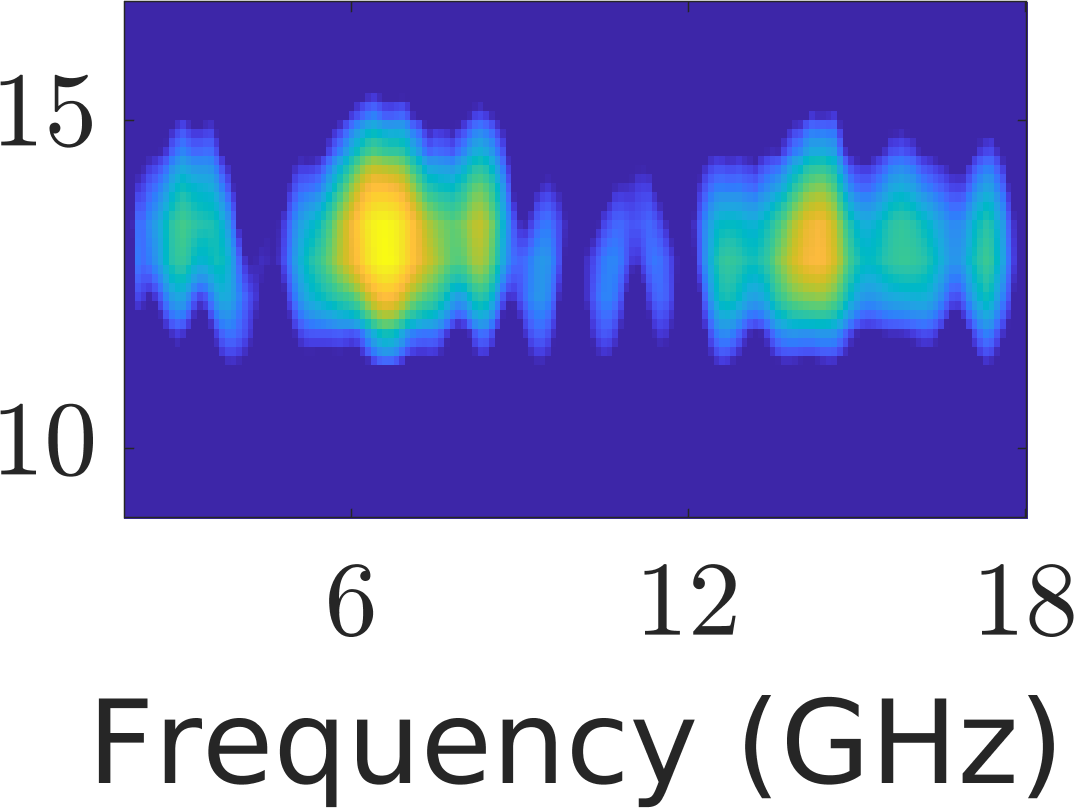} 
\end{minipage}\\ \vskip0.2cm 
  \hskip-0.2cm 
          \begin{minipage}{2.05cm} \centering        
        ({\bf I}): Spectrogram \&  Homogeneous 
             \end{minipage}
                     \begin{minipage}{2.05cm} \centering 
   ({\bf II}): Wavelet \&  Homogeneous 
               \end{minipage}
    \begin{minipage}{2.05cm} \centering 
   ({\bf III}): Spectrogram \&  Detailed  
            \end{minipage}
    \begin{minipage}{2.05cm} \centering 
   ({\bf IV}): Wavelet \&  Detailed 
             \end{minipage} \end{minipage} \hskip-0.1cm 
             \begin{minipage}{0.10cm} \centering 
    \includegraphics[height=4.05cm]{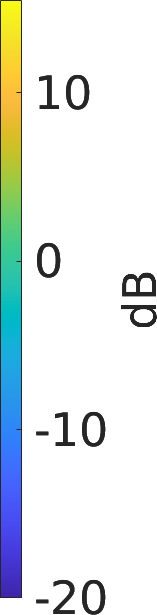}
             \end{minipage} \vskip0.2cm
\end{scriptsize}
\caption{Time-frequency maps for the 1st--3rd principal component obtained using spectrogram and wavelet approach. The spectrogram and wavelet TFRs have bandwidths of $\approx 0.11$ and $\approx 0.1$ GHz, respectively. The datasets of the homogeneous and detailed object are considered in the 1st--2nd and 3rd--4th columns, respectively. }
\label{result1}
\end{figure}

\begin{figure}[!ht]
    \centering 
    \begin{scriptsize} \begin{minipage}{8.5cm}\centering Spectrogram \\ \vskip0.2cm 
    \begin{minipage}{2cm} \centering
    \includegraphics[height=2cm]{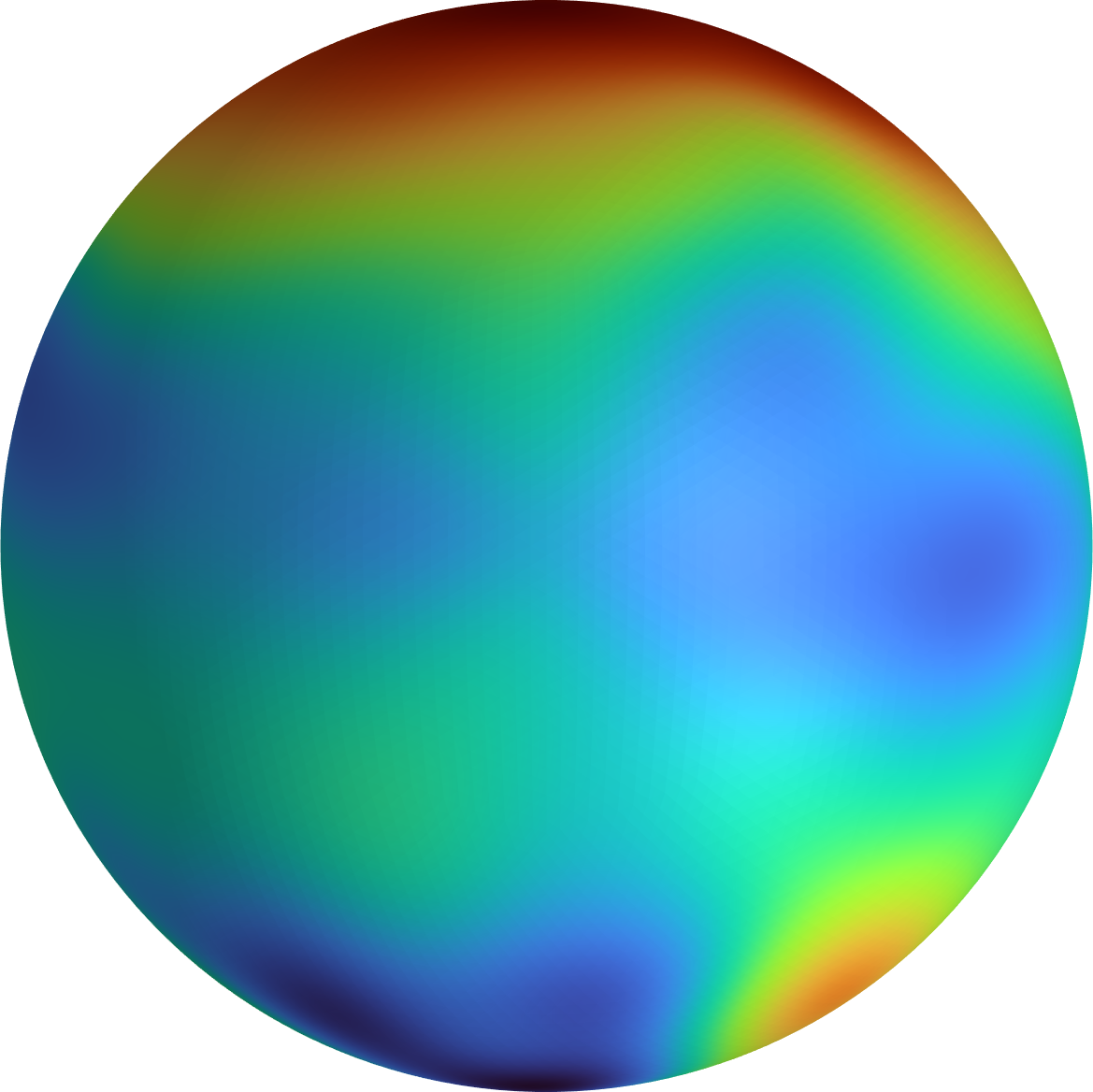} 
\end{minipage}
\begin{minipage}{2cm} \centering
    \includegraphics[height=2cm]{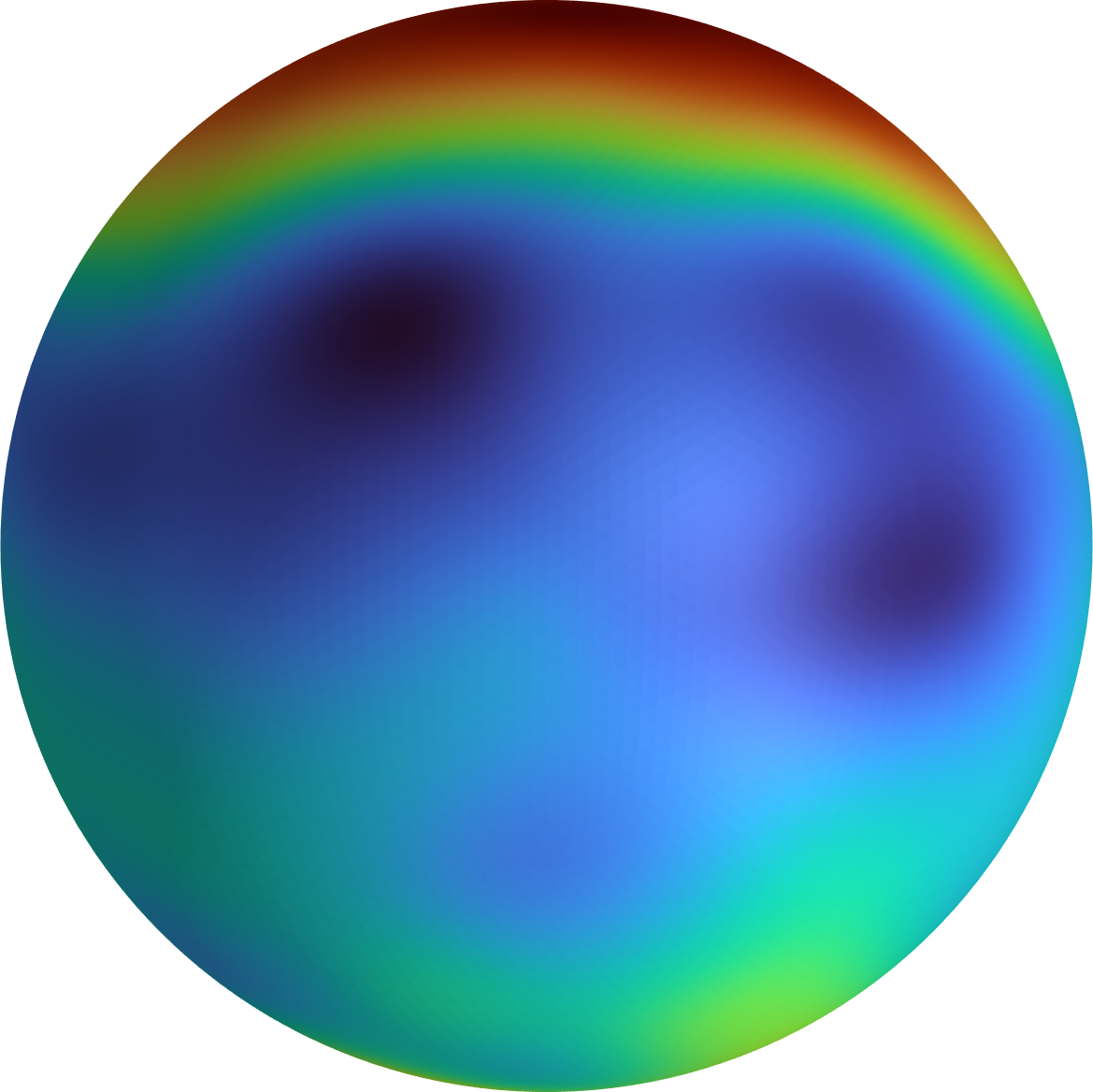} 
\end{minipage}
\begin{minipage}{2cm} \centering
    \includegraphics[height=2cm]{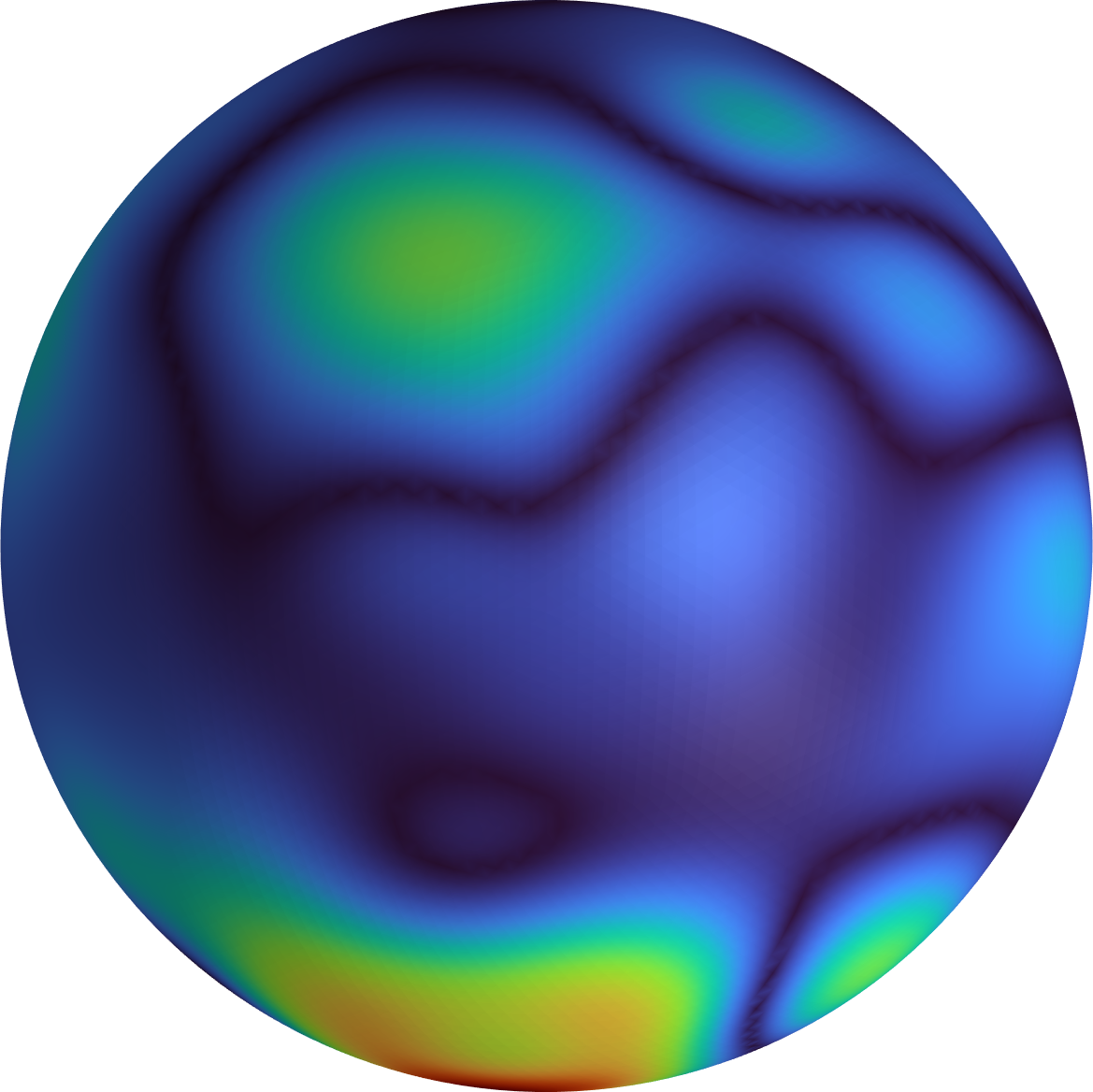} 
\end{minipage}
\begin{minipage}{1.5cm} \centering
  \includegraphics[height=2.5cm]{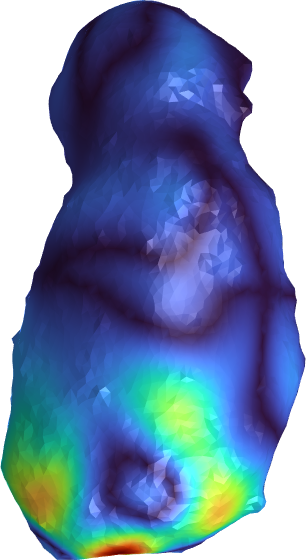} 
\end{minipage}\\  \vskip0.2cm 
%
 Wavelet \\ \vskip0.2cm 
    \begin{minipage}{2cm} \centering
    \includegraphics[height=2cm]{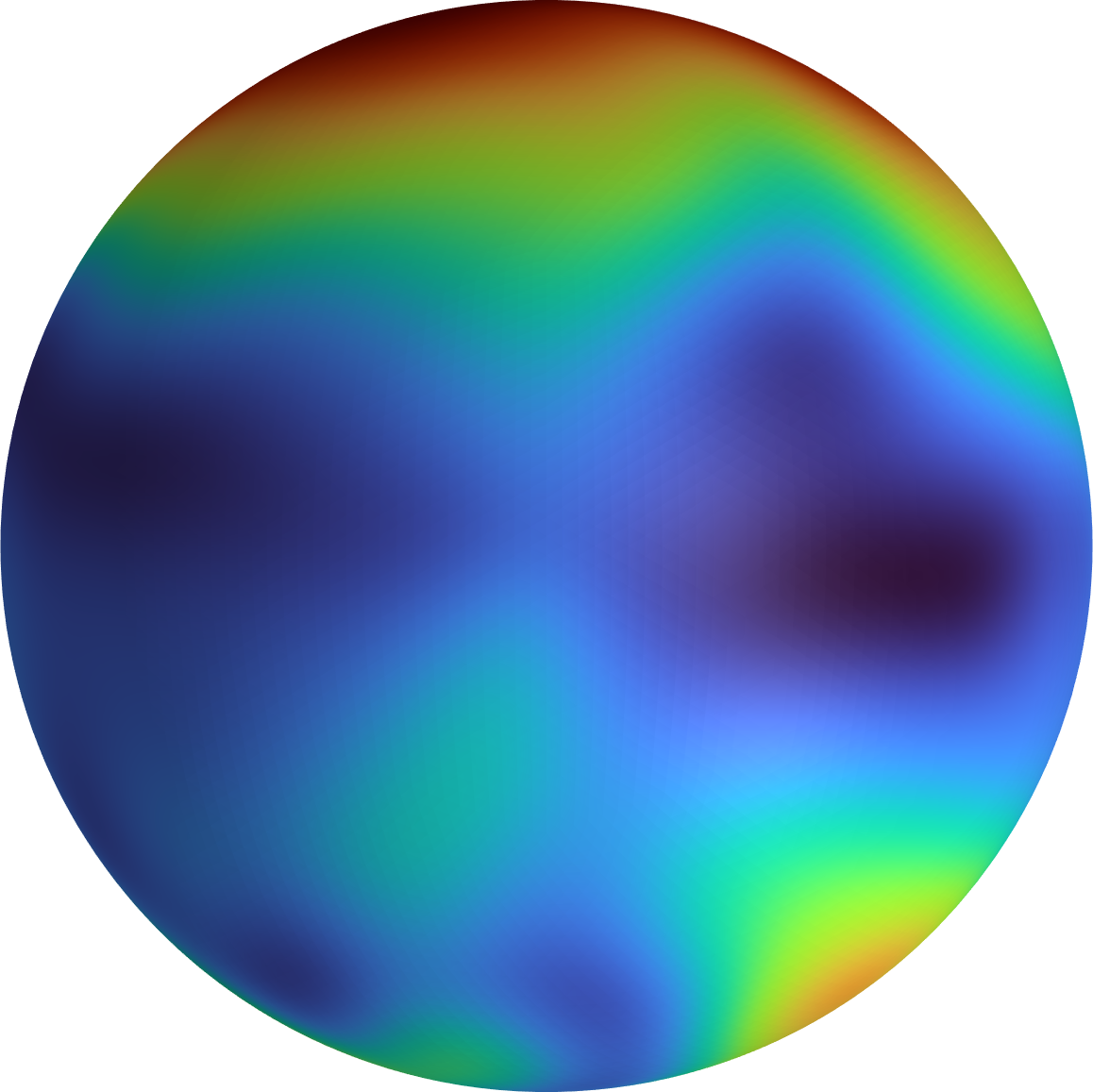} 
\end{minipage}
\begin{minipage}{2cm} \centering
    \includegraphics[height=2cm]{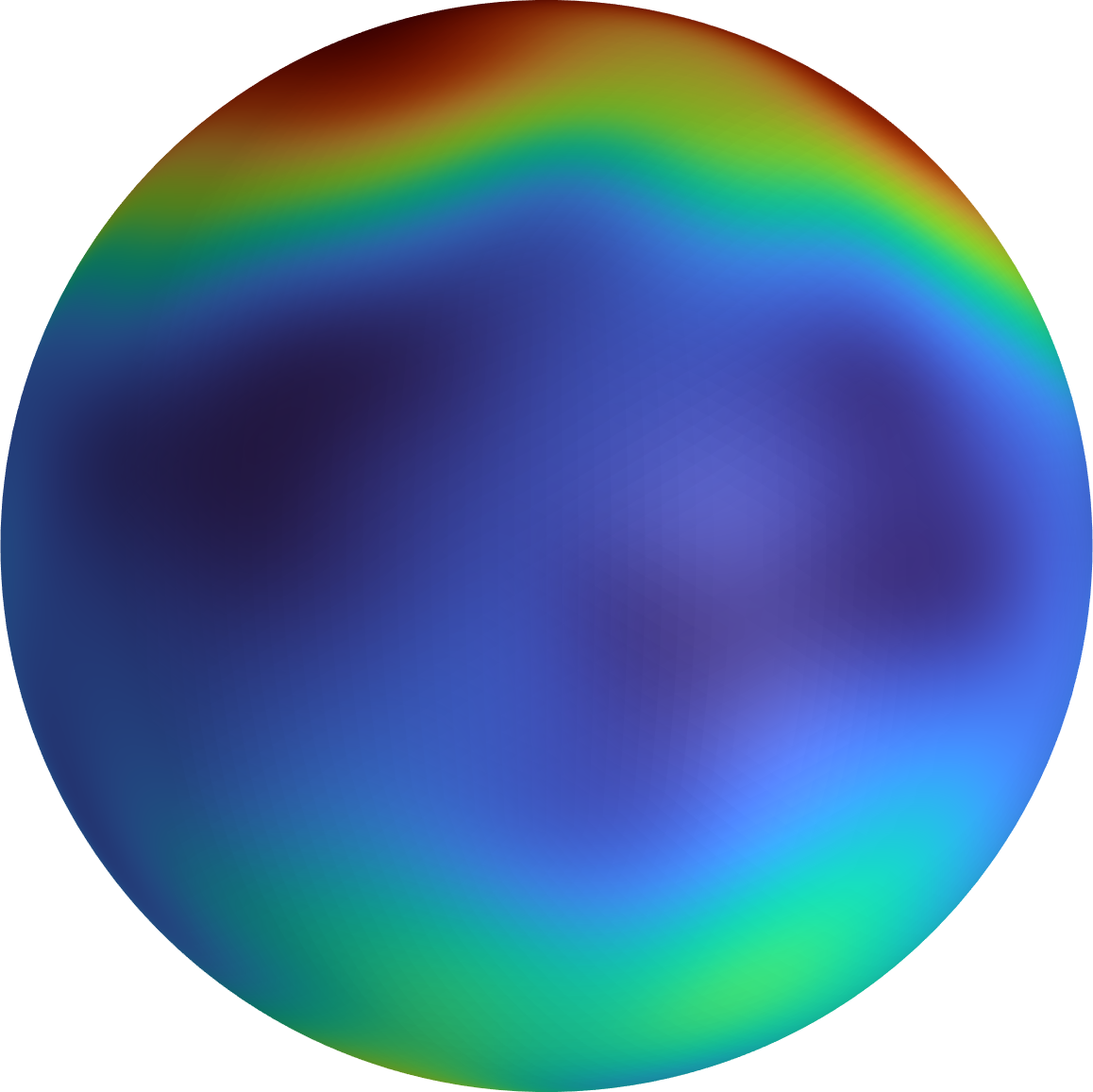} 
\end{minipage}
\begin{minipage}{2cm} \centering
    \includegraphics[height=2cm]{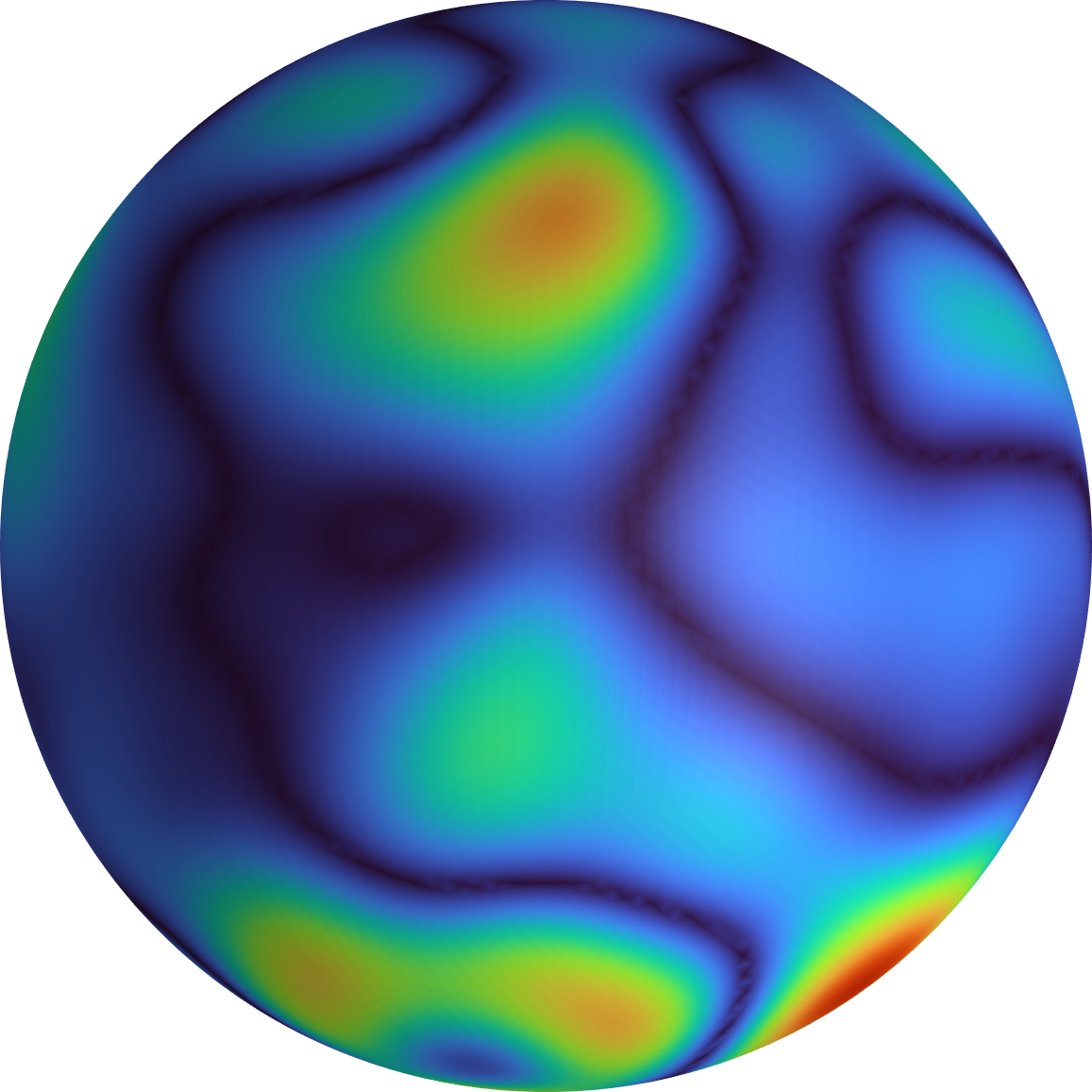} 
\end{minipage}
\begin{minipage}{1.5cm} \centering
  \includegraphics[height=2.5cm]{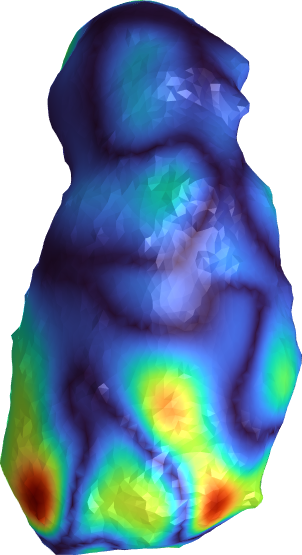} 
\end{minipage}
\\ \vskip0.2cm 
\vskip0.2cm 
          \begin{minipage}{2.05cm} \centering        
        ({\bf I}): Detailed
             \end{minipage}
    \begin{minipage}{2.05cm} \centering 
   ({\bf II}): Homogeneous 
            \end{minipage}
    \begin{minipage}{2.05cm} \centering 
   ({\bf III}): Difference
               \end{minipage}\end{minipage} 
             \begin{minipage}{0.10cm} \centering 
    \includegraphics[height=5.05cm]{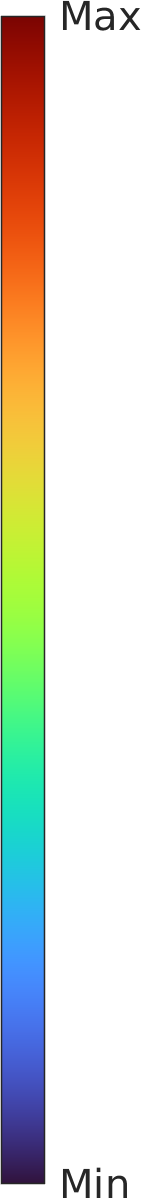}
             \end{minipage}
\end{scriptsize}
\caption{Topographic projection of the maximum energy on the surface of a unit sphere and the difference on the analogue using the spectrogram and wavelet approach.}    
\label{result3}
\end{figure}
 
\begin{figure}[!ht]
    \centering  
    \begin{scriptsize} 
    \begin{minipage}{8.5cm}\centering  Spectrogram \\ \vskip0.2cm 
    \begin{minipage}{2cm} \centering
    \includegraphics[height=2cm]{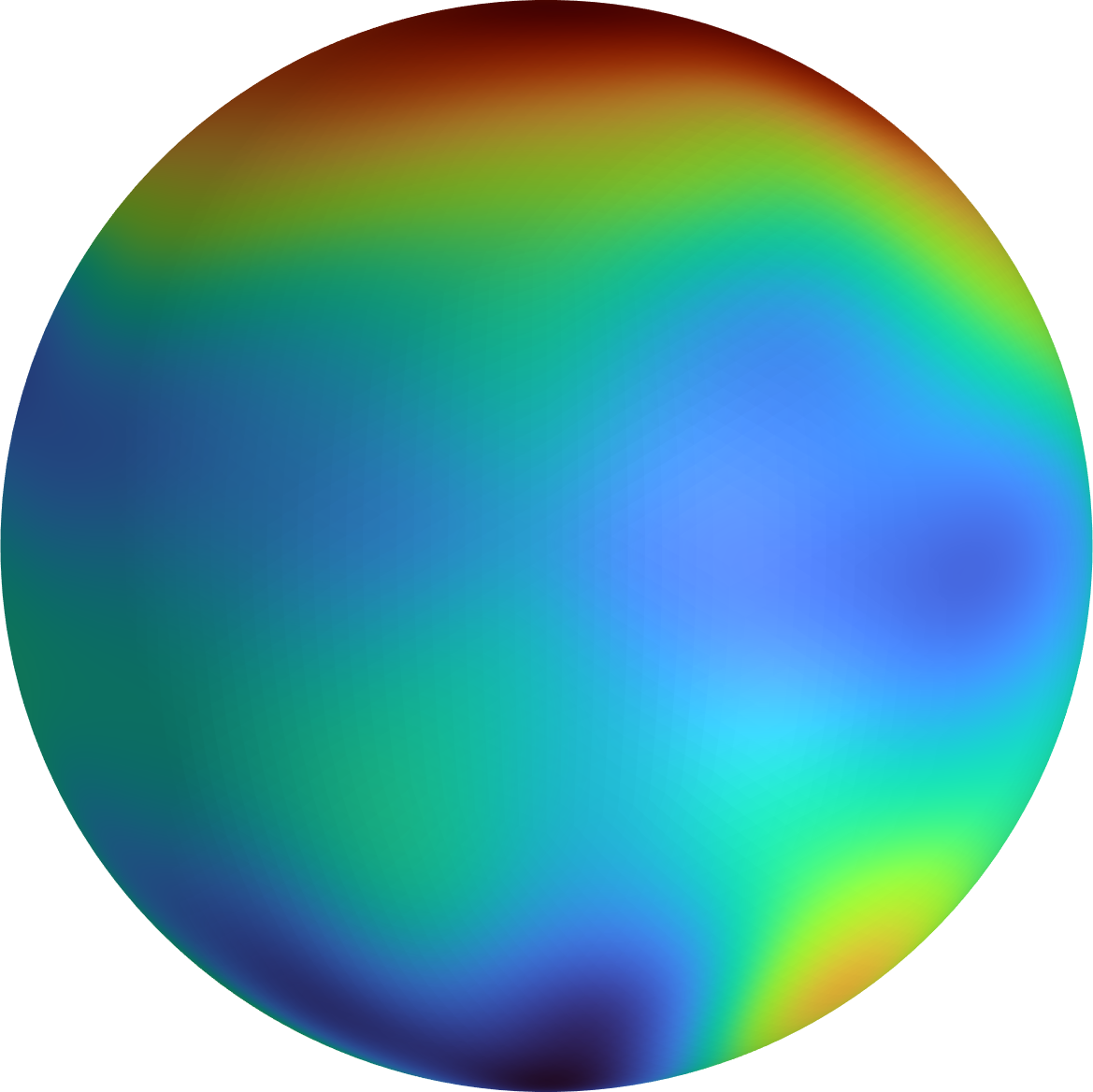} 
\end{minipage}
\begin{minipage}{2cm} \centering
    \includegraphics[height=2cm]{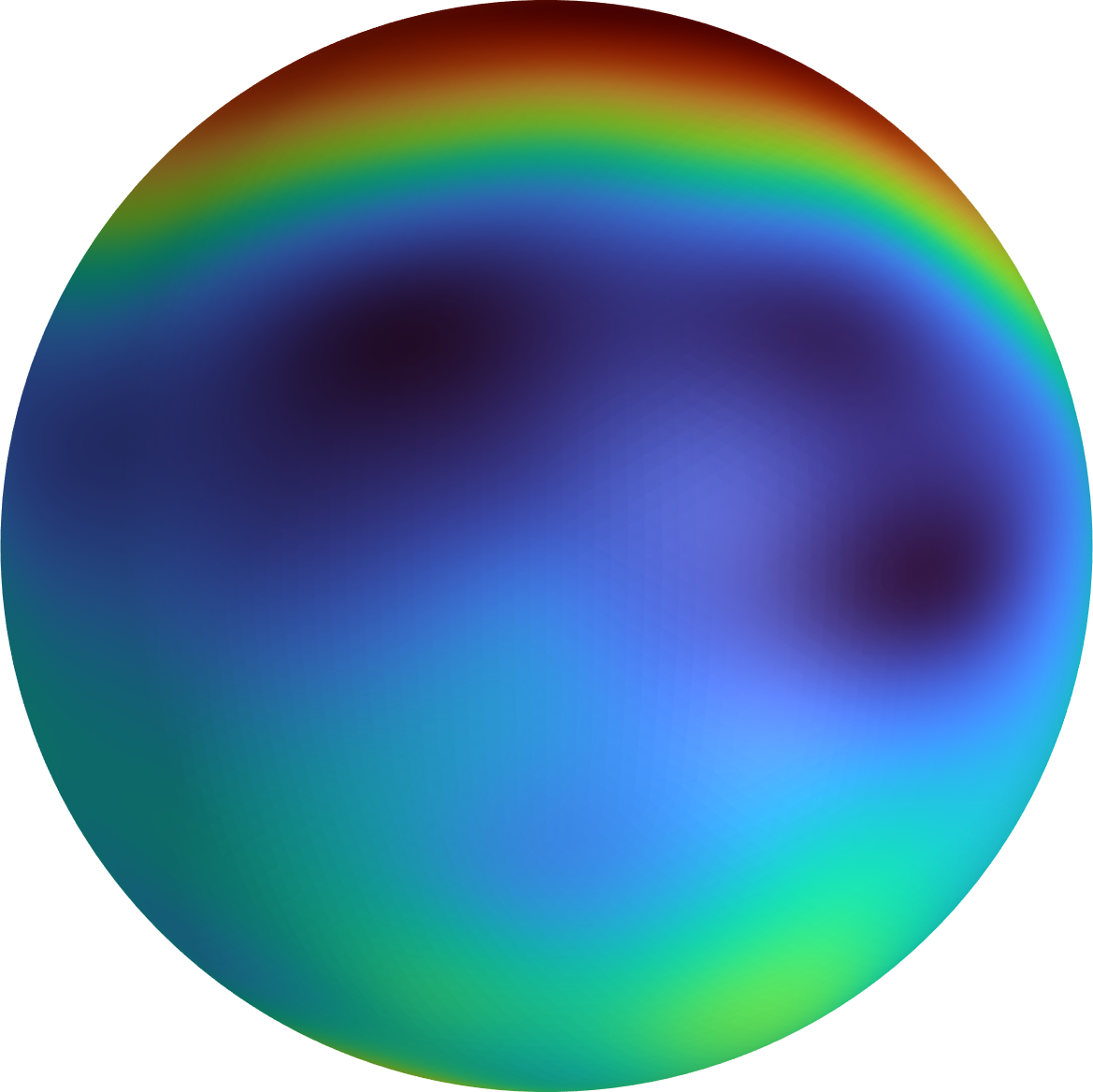} 
\end{minipage}
\begin{minipage}{2cm} \centering
    \includegraphics[height=2cm]{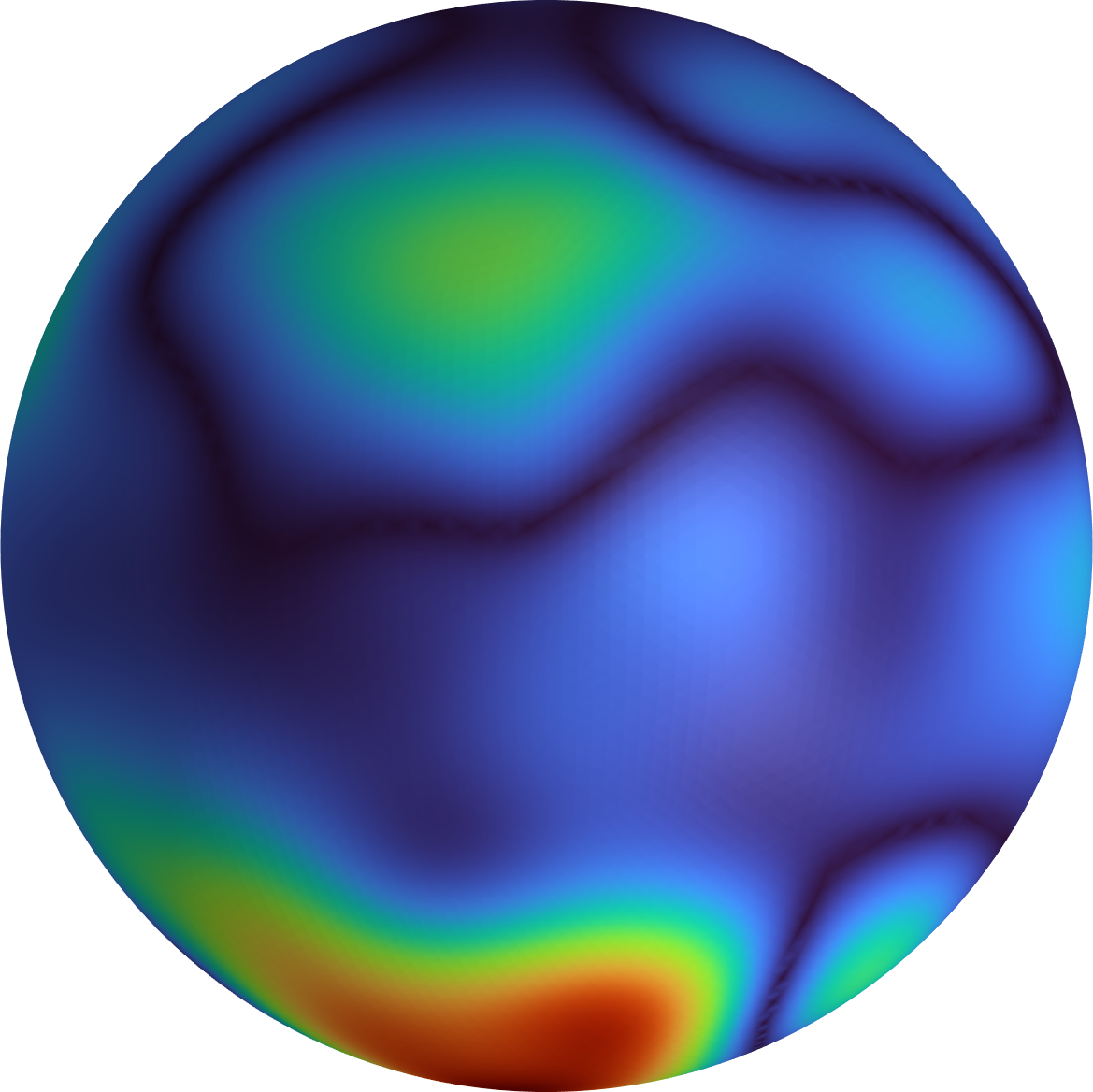} 
\end{minipage}
\begin{minipage}{1.5cm} \centering
  \includegraphics[height=2.5cm]{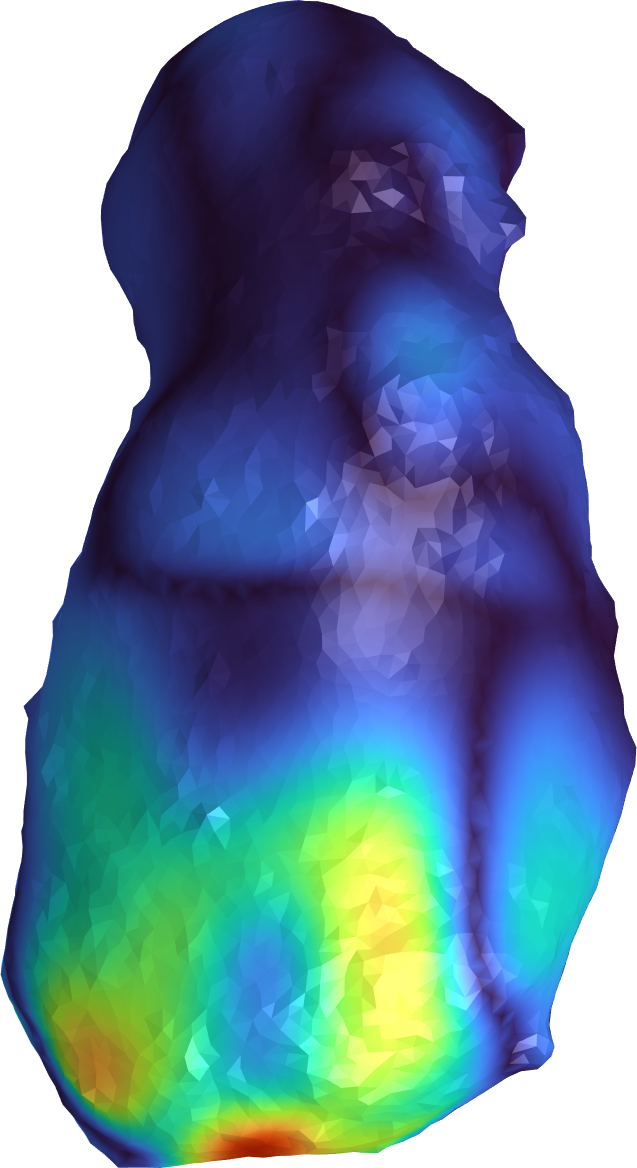} 
\end{minipage}\\  \vskip0.2cm 
 Wavelet \\ \vskip0.2cm 
    \begin{minipage}{2cm} \centering
    \includegraphics[height=2cm]{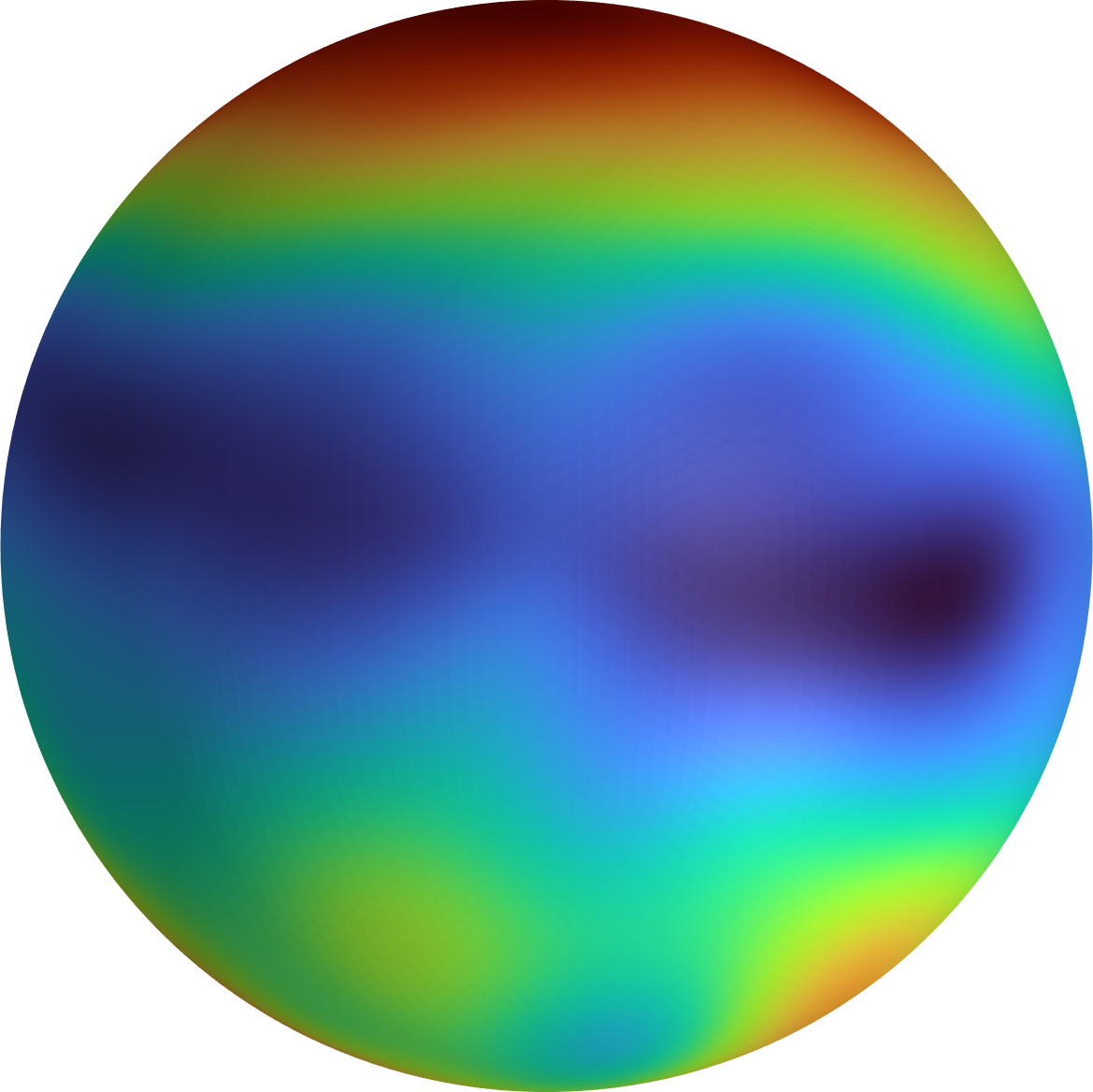} 
\end{minipage}
\begin{minipage}{2cm} \centering
    \includegraphics[height=2cm]{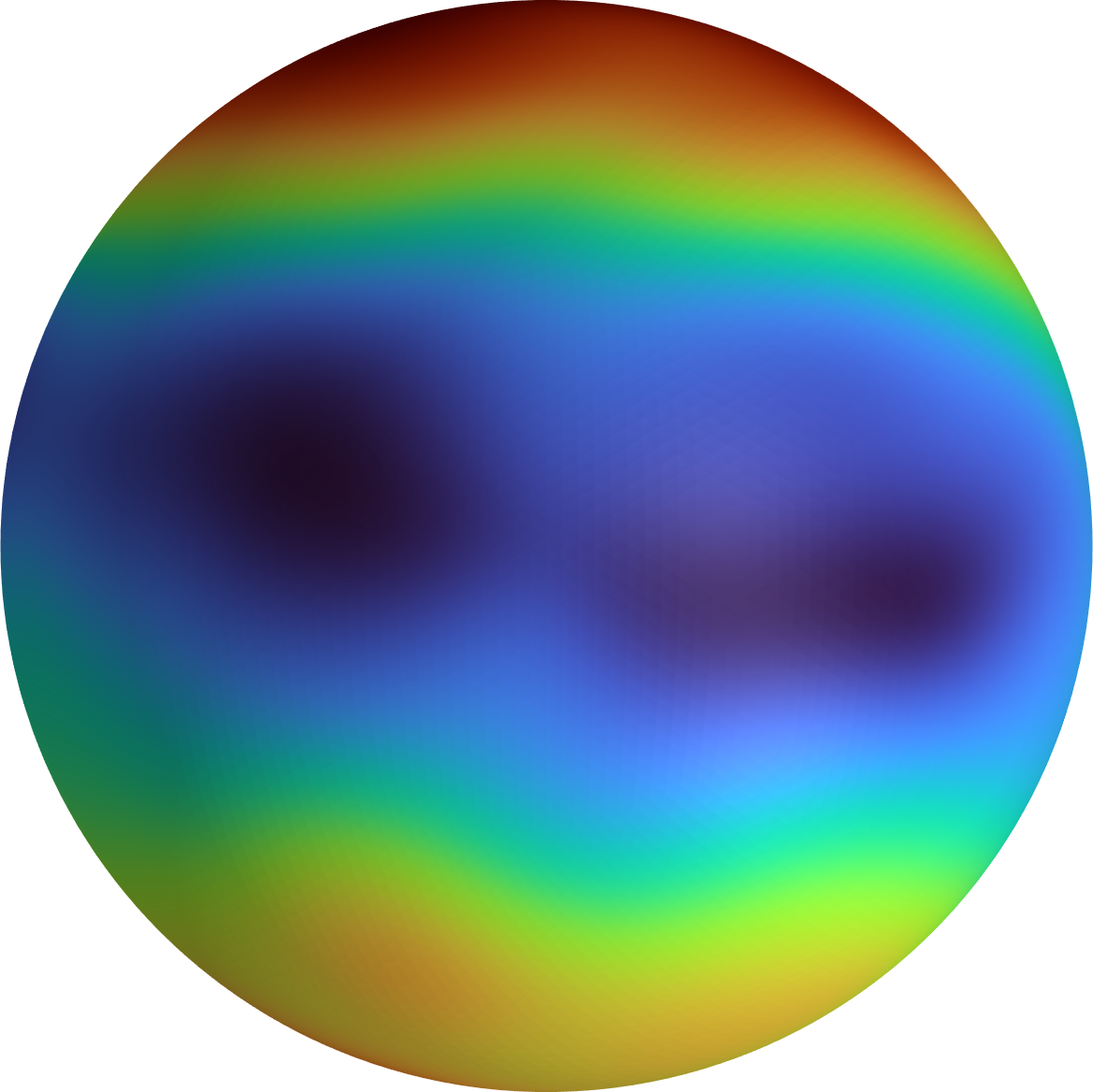} 
\end{minipage}
\begin{minipage}{2cm} \centering
    \includegraphics[height=2cm]{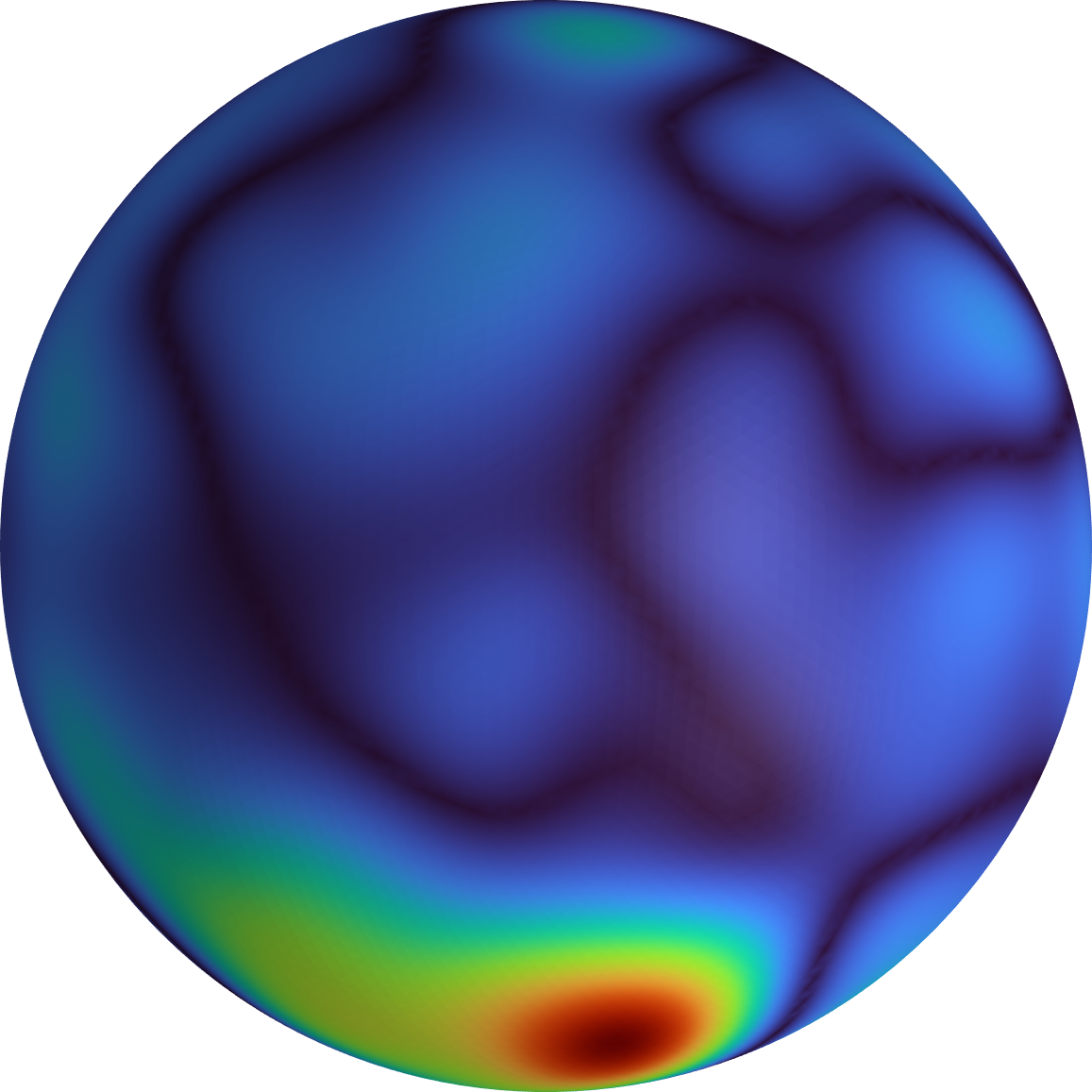} 
\end{minipage}
\begin{minipage}{1.5cm} \centering
  \includegraphics[height=2.5cm]{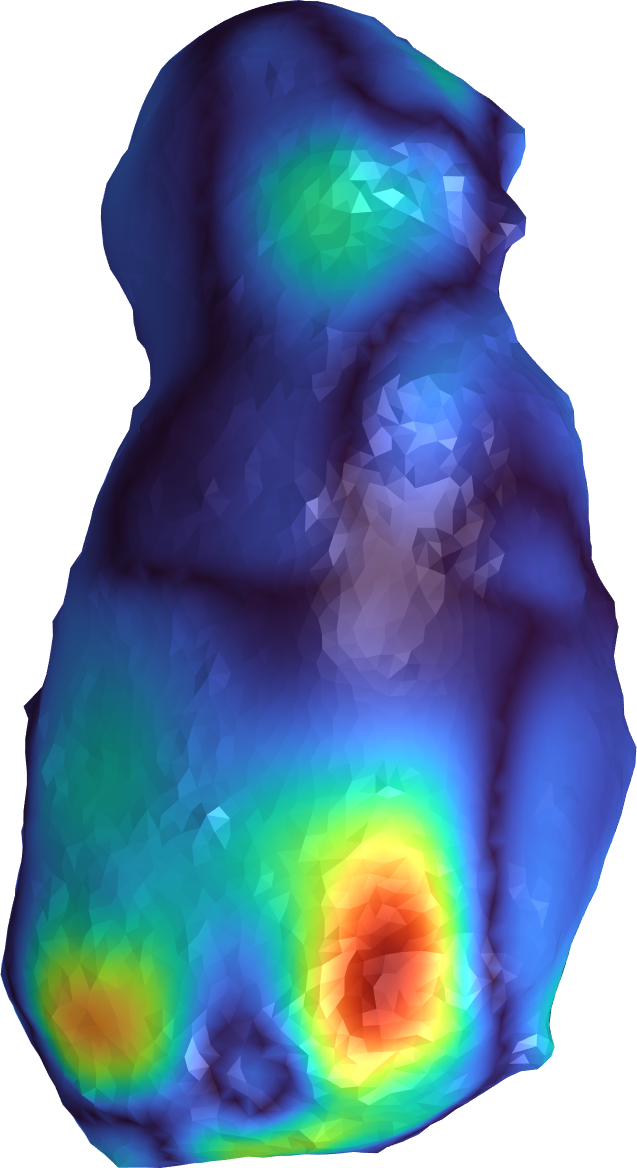} 
\end{minipage}\\ \vskip0.2cm 
\vskip0.2cm 
          \begin{minipage}{2.05cm} \centering        
        ({\bf I}): Detailed
             \end{minipage}
    \begin{minipage}{2.05cm} \centering 
   ({\bf II}): Homogeneous 
            \end{minipage}
    \begin{minipage}{2.05cm} \centering 
   ({\bf III}): Difference
               \end{minipage} \end{minipage} 
             \begin{minipage}{0.10cm} \centering 
    \includegraphics[height=5.05cm]{figs/reconstr/colourbar_sphere.png}
             \end{minipage}
\end{scriptsize}
\caption{Averaged topographic projection of the noisy maximum energy (10 dB peak SNR) on the surface of a unit sphere and the difference on the analogue using the spectrogram and wavelet approach.}
\label{result4}
\end{figure}



   

\begin{figure}[!ht]
    \centering  
    \begin{scriptsize}
    \begin{minipage}{9.55cm} \centering \hskip -0.5cm
    \includegraphics[width=8.55cm]{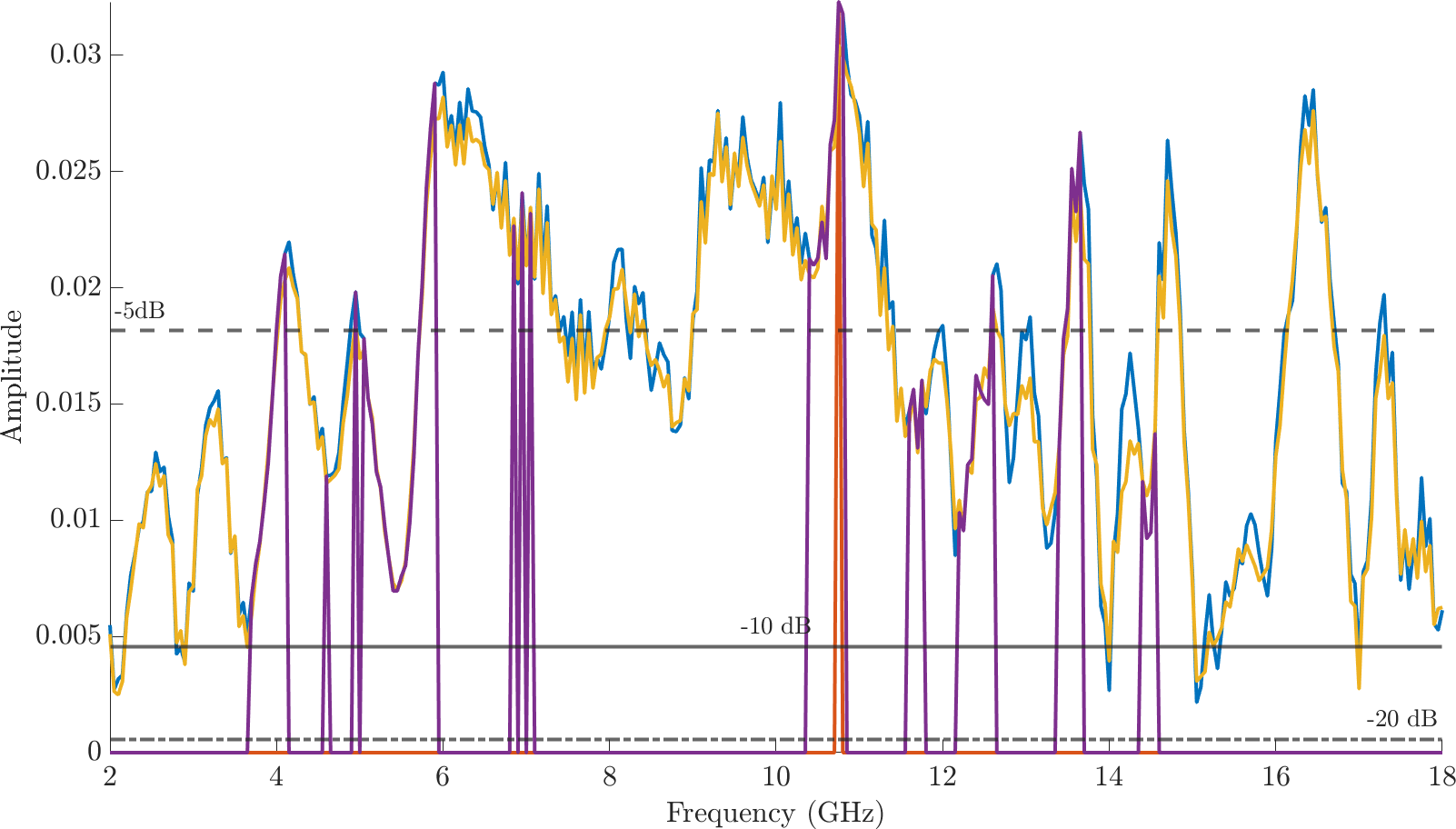} 
\end{minipage}\vskip0.2cm
  \begin{minipage}{5.55cm} \centering
    \includegraphics[width=4.85cm]{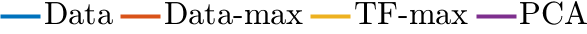} 
\end{minipage}
\end{scriptsize}
\caption{Comparison of an observation point after different processing techniques (data max, TF max, and PCA). The noise levels at different SNRs are shown with the dashed, thick, and dotted lines (20, 10 and 5-dB).}
\label{result8}
\end{figure}

\begin{figure}[!h]
    \centering  
    \begin{scriptsize}   \rotatebox{90}{\hskip-.3cm Base}
    \begin{minipage}{2.55cm} \centering 
    \includegraphics[width=2.55cm,height=2.55cm]{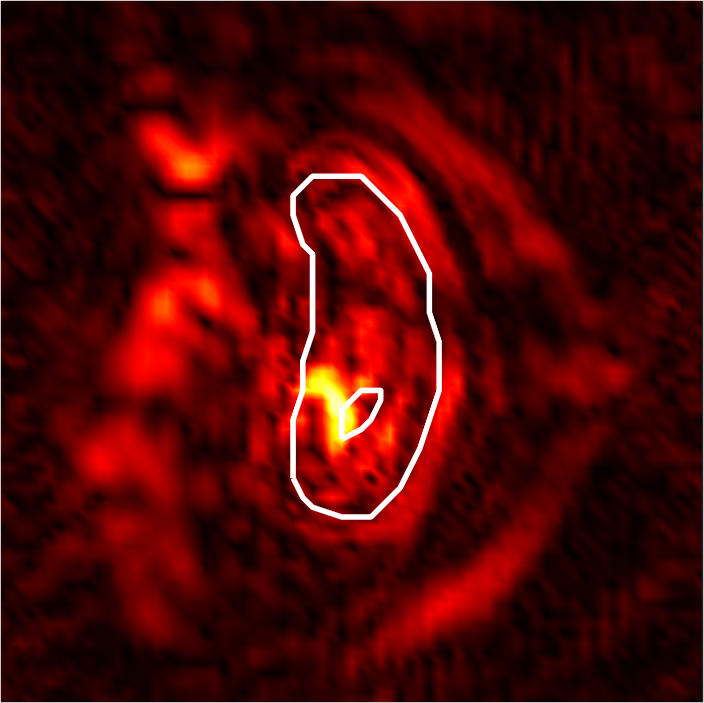} 
\end{minipage}
\begin{minipage}{2.55cm} \centering
    \includegraphics[width=2.55cm,height=2.55cm]{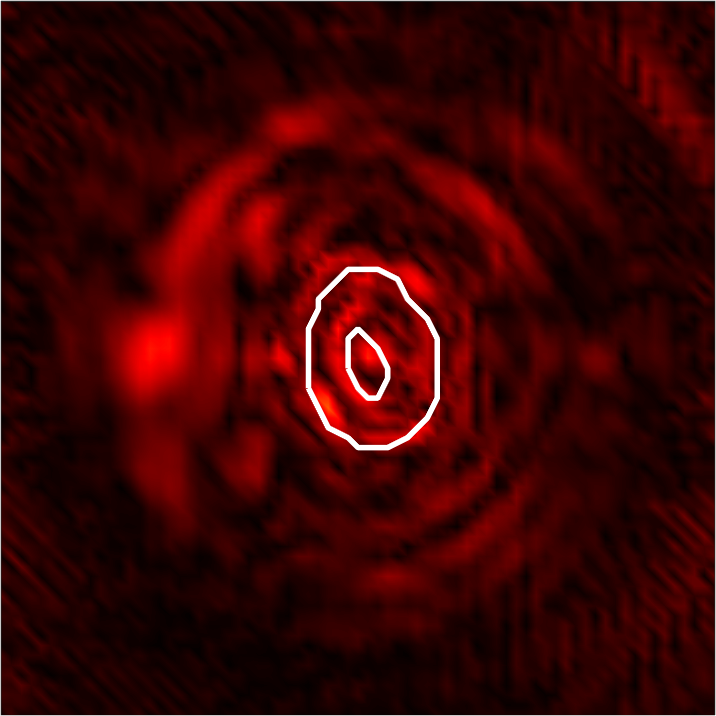} 
\end{minipage}
\begin{minipage}{2.55cm} \centering
    \includegraphics[width=2.55cm,height=2.55cm]{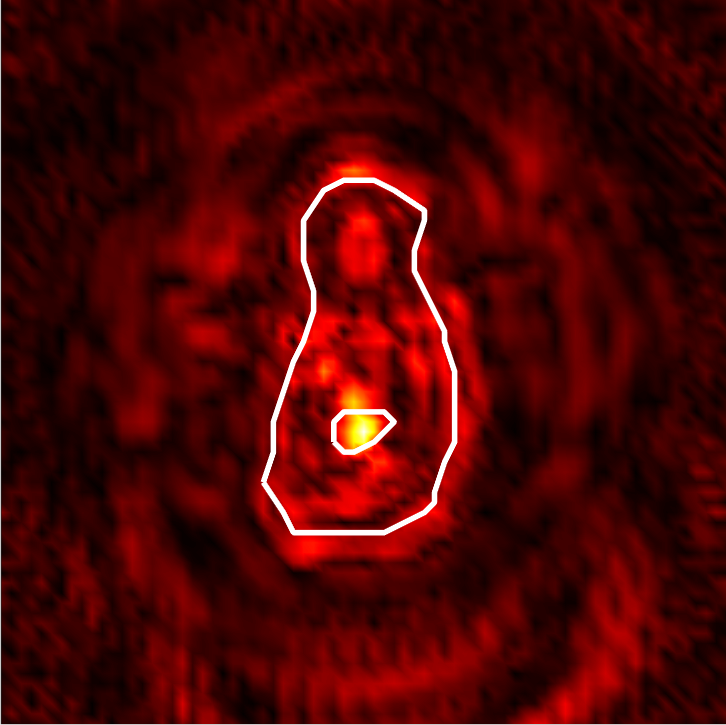} 
\end{minipage}
\\ \vskip0.05cm 
\rotatebox{90}{\hskip-.3cm Data Max} 
    \begin{minipage}{2.55cm} \centering
    \includegraphics[width=2.55cm,height=2.55cm]{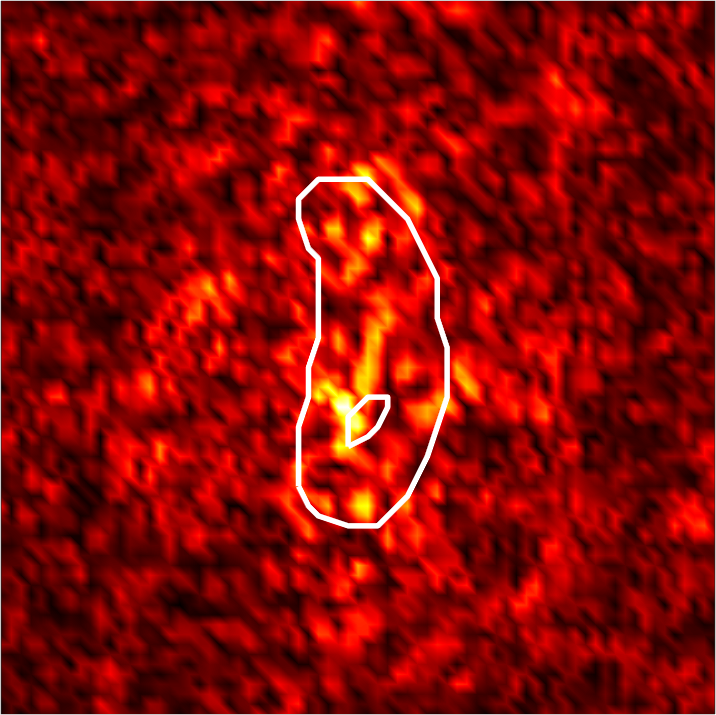} 
\end{minipage}
\begin{minipage}{2.55cm} \centering
    \includegraphics[width=2.55cm,height=2.55cm]{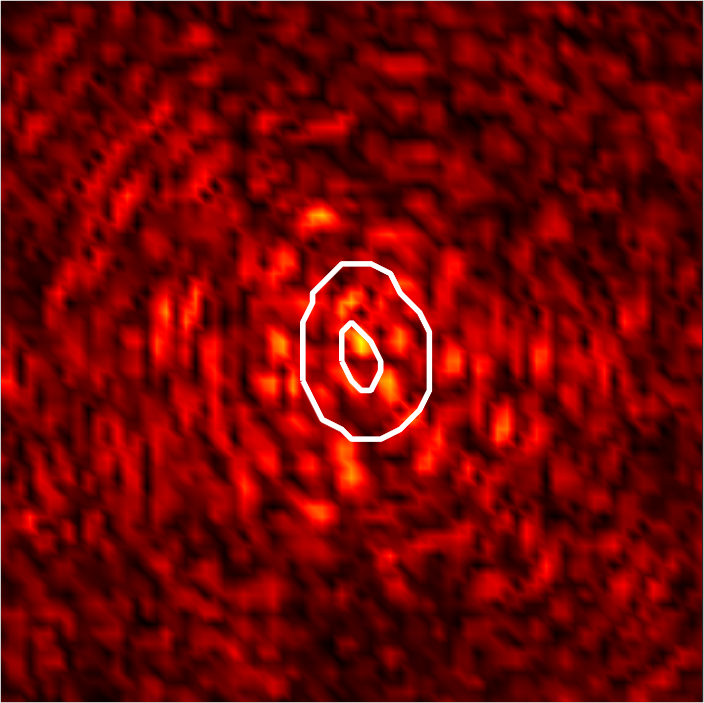} 
\end{minipage}
\begin{minipage}{2.55cm} \centering \vskip-0.03cm
    \includegraphics[width=2.55cm,height=2.55cm]{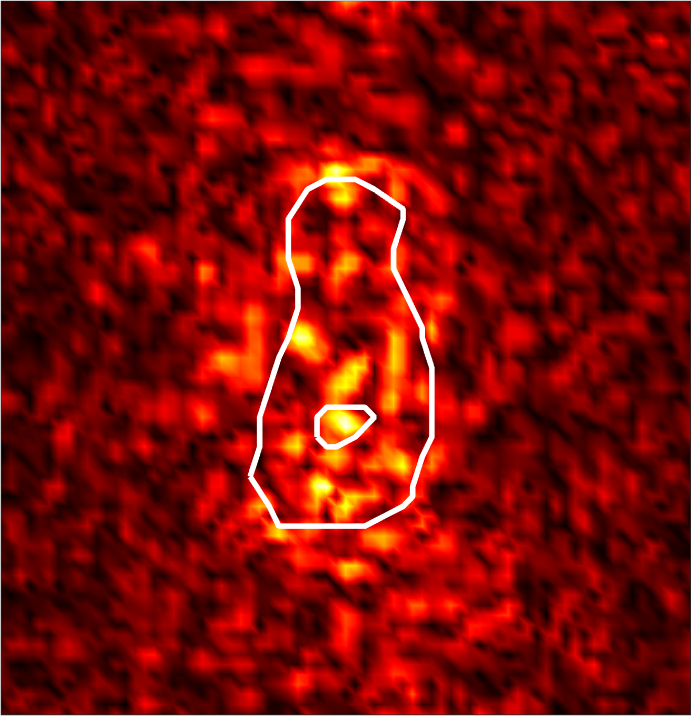} 
\end{minipage}
\\ \vskip0.05cm 
\rotatebox{90}{\hskip-.3cm TF Max}
    \begin{minipage}{2.55cm} \centering
    \includegraphics[width=2.55cm,height=2.55cm]{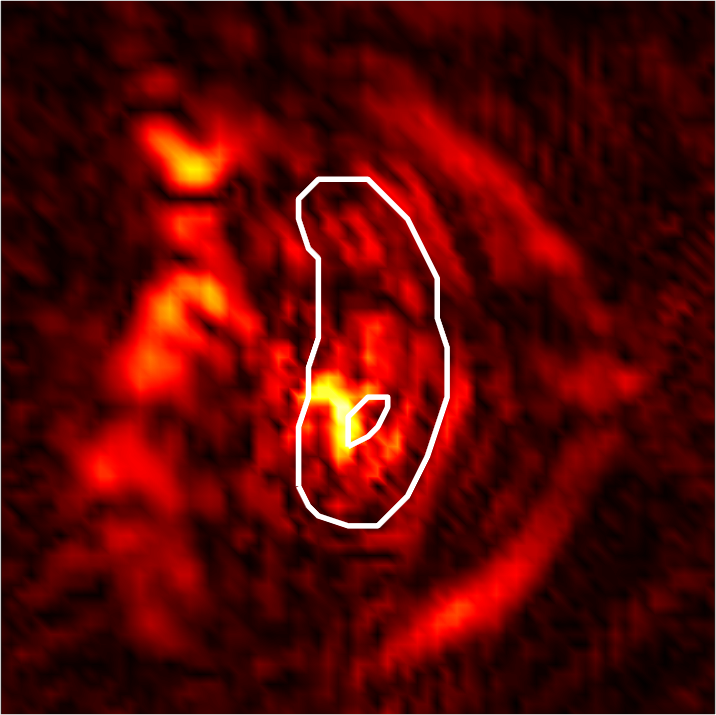} 
\end{minipage}
\begin{minipage}{2.55cm} \centering
    \includegraphics[width=2.55cm,height=2.55cm]{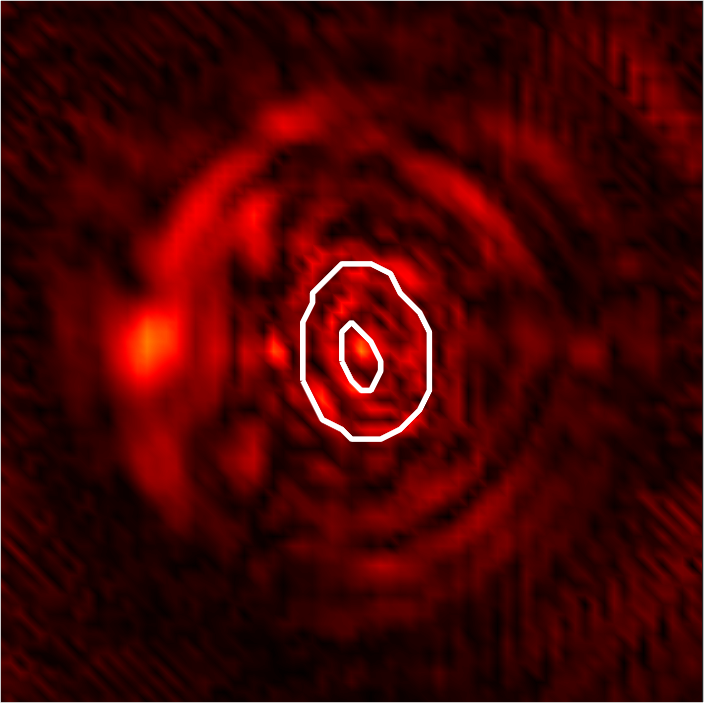} 
\end{minipage}
\begin{minipage}{2.55cm} \centering \vskip-0.04cm
    \includegraphics[width=2.55cm,height=2.55cm]{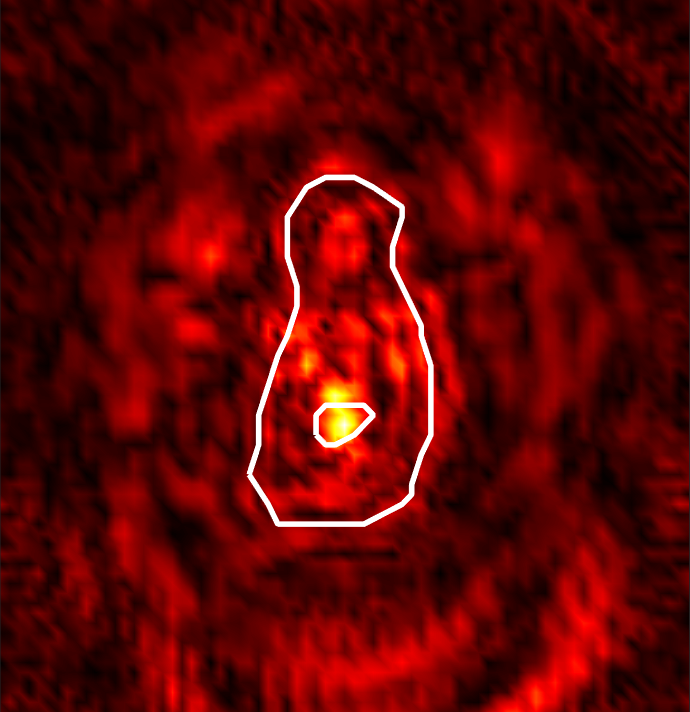} 
\end{minipage}
\\  \vskip0.05cm
\rotatebox{90}{\hskip-.3cm PCA}
    \begin{minipage}{2.55cm} \centering
    \includegraphics[width=2.55cm,height=2.55cm]{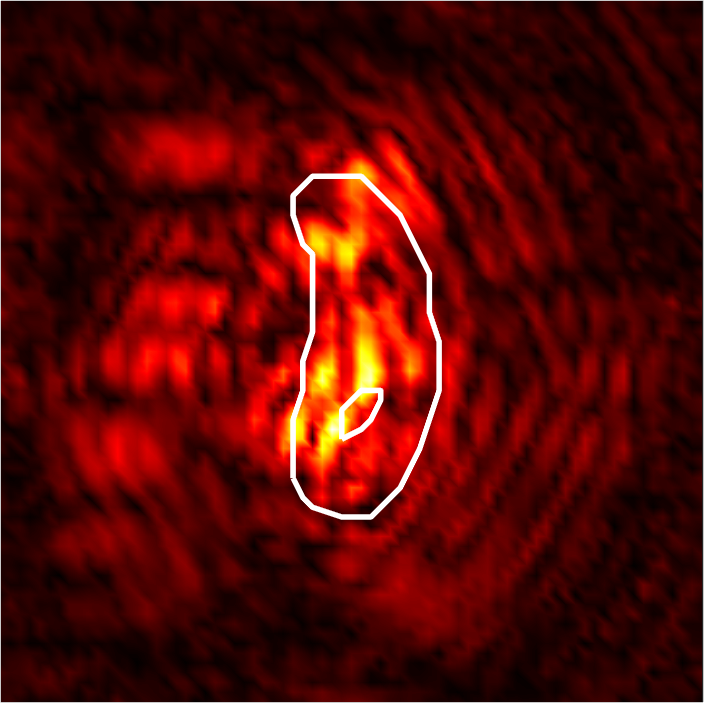} 
\end{minipage}
\begin{minipage}{2.55cm} \centering
    \includegraphics[width=2.55cm,height=2.55cm]{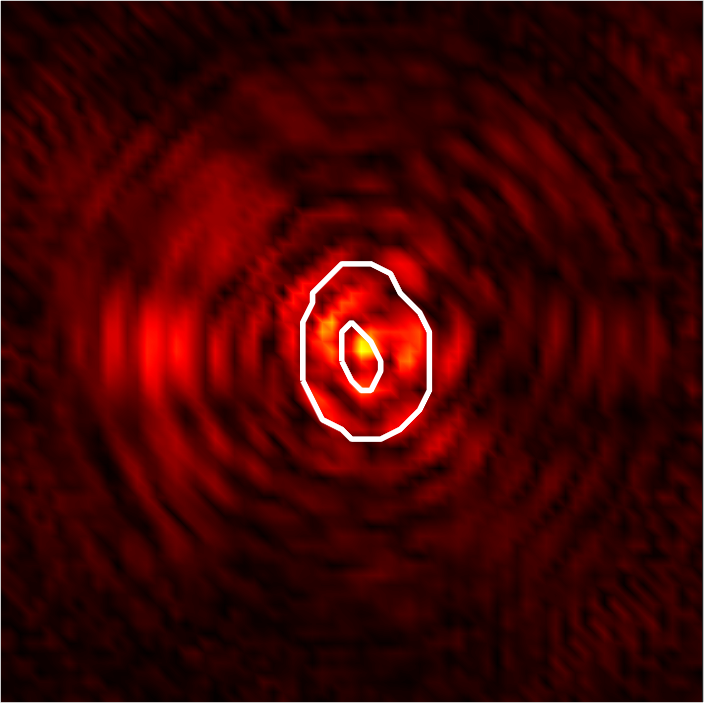} 
\end{minipage}
\begin{minipage}{2.55cm} \centering \vskip-0.03cm
    \includegraphics[width=2.55cm,height=2.55cm]{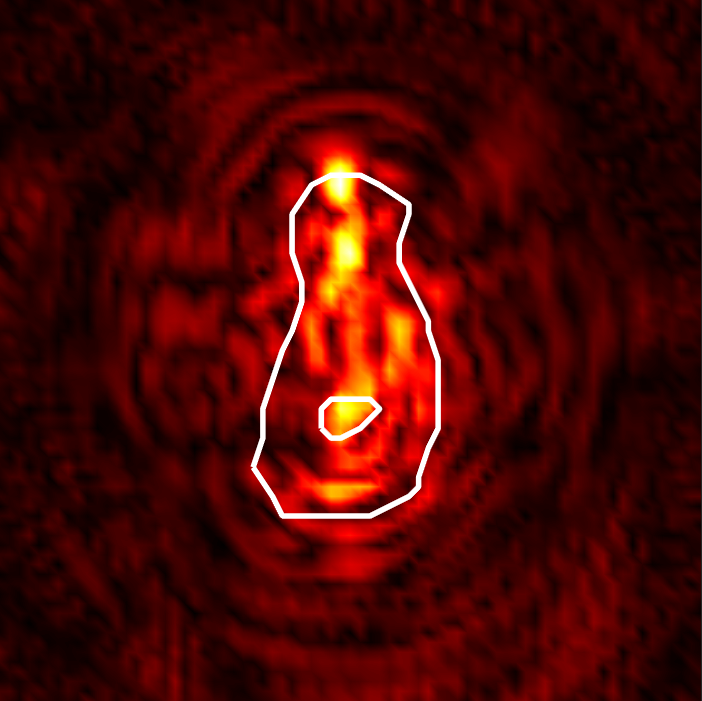} 
\end{minipage}
\\ \vskip0.25cm
 
          \begin{minipage}{2.55cm} \centering        
        ({\bf I}): ZY
             \end{minipage}
    \begin{minipage}{2.55cm} \centering 
   ({\bf II}): ZX 
            \end{minipage}
    \begin{minipage}{2.55cm} \centering 
   ({\bf III}): XY
               \end{minipage}
\end{scriptsize}
\caption{Backpropagated reconstructions of the detailed analogue without  attenuation correction (AC) over a limited frequency band (3.5--12 GHz). Of the applied filters, Base refers to a plain bandpass filter, while data max evaluates, for the given frequency range, the maximum energy of the signal, TF Max evaluates the maximum energy of the continuous wavelet transform, and PCA is based on the $\ell$ largest PCs. Cut in the ZY, ZX, and XY directions (from left to right). }
\label{result5}
\end{figure}
\begin{figure}[!h]
    \centering  
    \begin{scriptsize} 
\rotatebox{90}{\hskip-.3cm AC Base}
    \begin{minipage}{2.55cm} \centering
    \includegraphics[width=2.555cm,height=2.55cm]{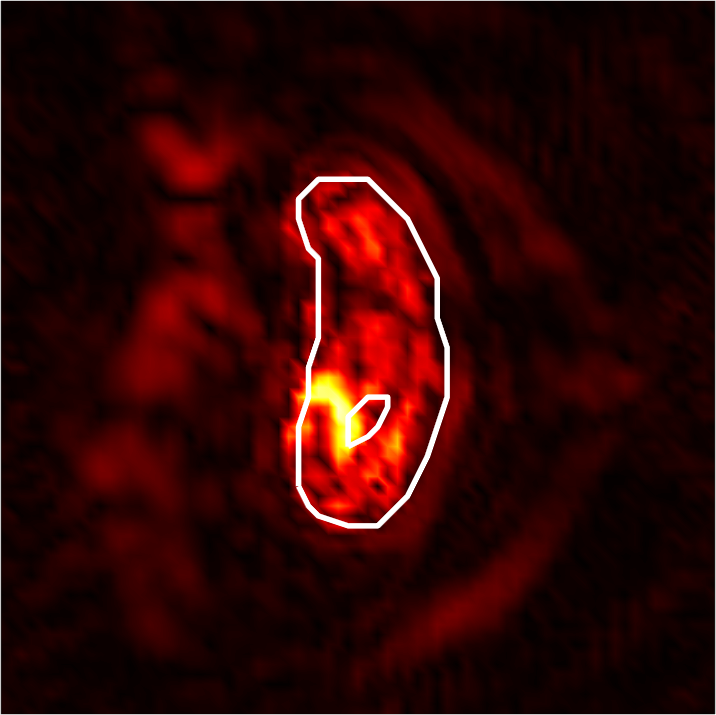} 
\end{minipage}
\begin{minipage}{2.55cm} \centering
    \includegraphics[width=2.555cm,height=2.55cm]{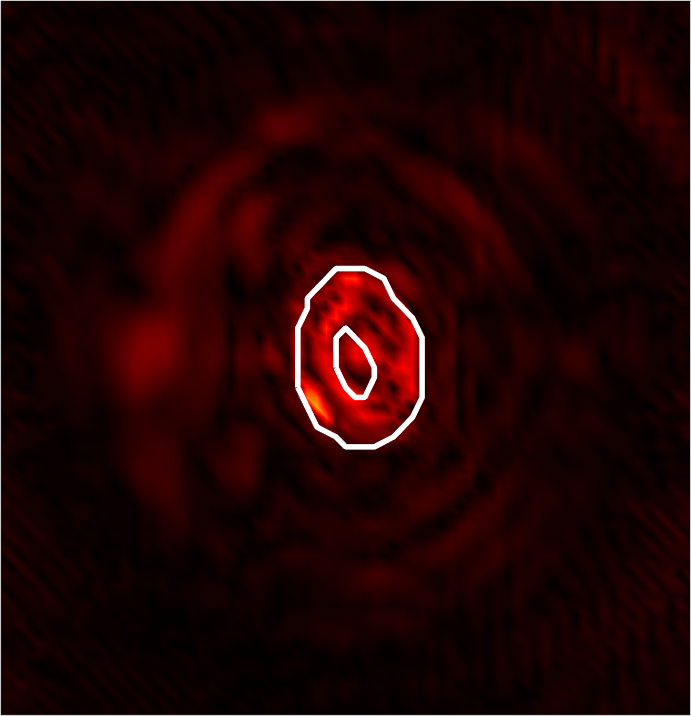} 
\end{minipage}
\begin{minipage}{2.55cm} \centering \vskip-0.05cm
    \includegraphics[width=2.55cm,height=2.55cm]{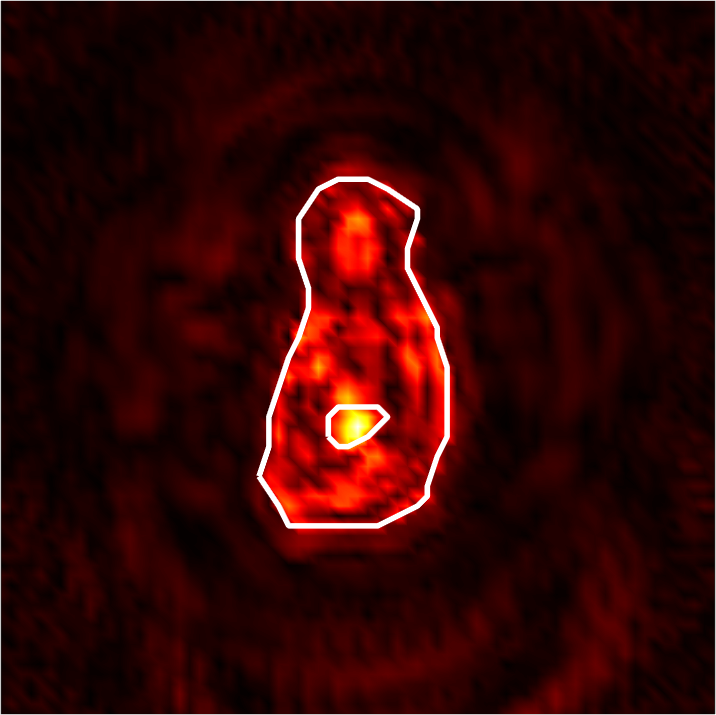} 
\end{minipage}
\\ \vskip0.05cm
\rotatebox{90}{\hskip-.3cm AC Data Max}
    \begin{minipage}{2.55cm} \centering
    \includegraphics[width=2.55cm,height=2.55cm]{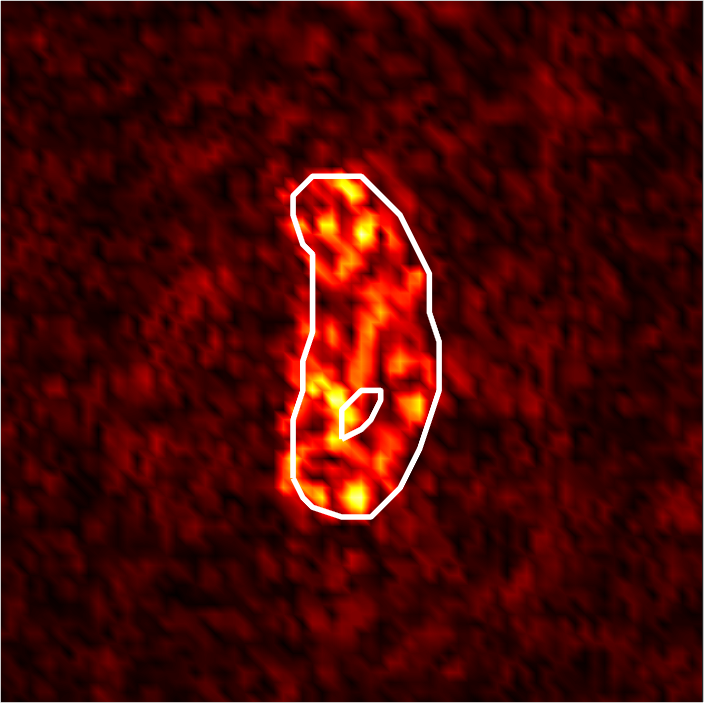} 
\end{minipage}
\begin{minipage}{2.55cm} \centering
    \includegraphics[width=2.55cm]{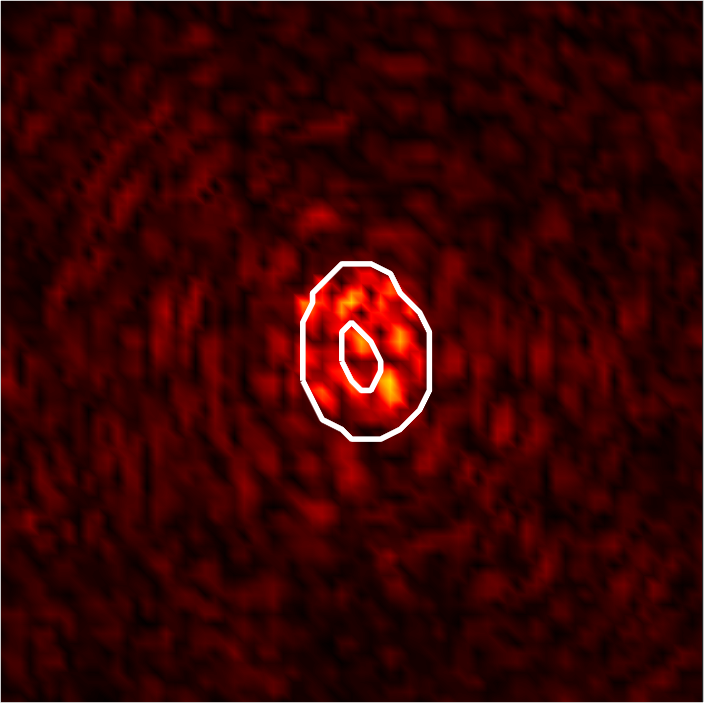} 
\end{minipage}
\begin{minipage}{2.55cm} \centering \vskip-0.04cm
    \includegraphics[width=2.55cm,height=2.55cm]{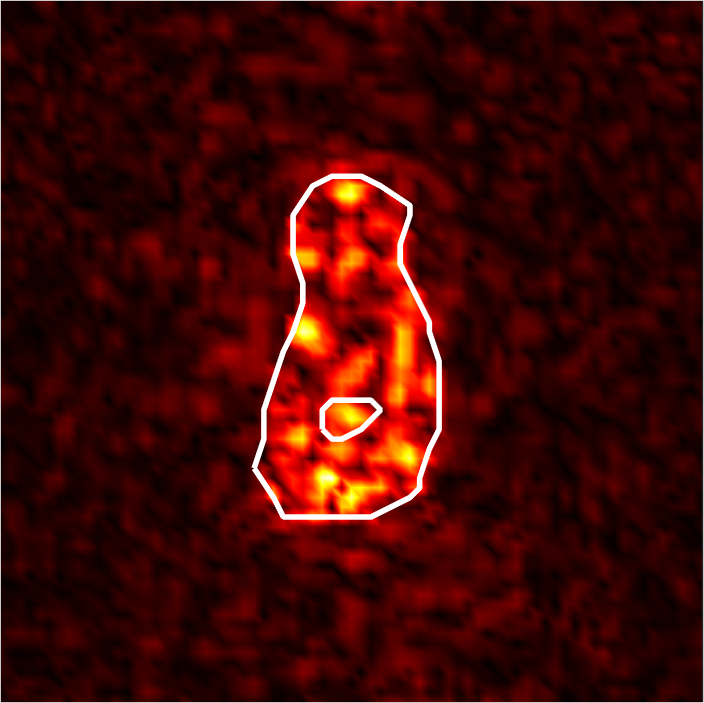} 
\end{minipage}
\\ \vskip0.05cm
\rotatebox{90}{\hskip-.3cm AC TF Max}
    \begin{minipage}{2.55cm} \centering
    \includegraphics[width=2.55cm,height=2.55cm]{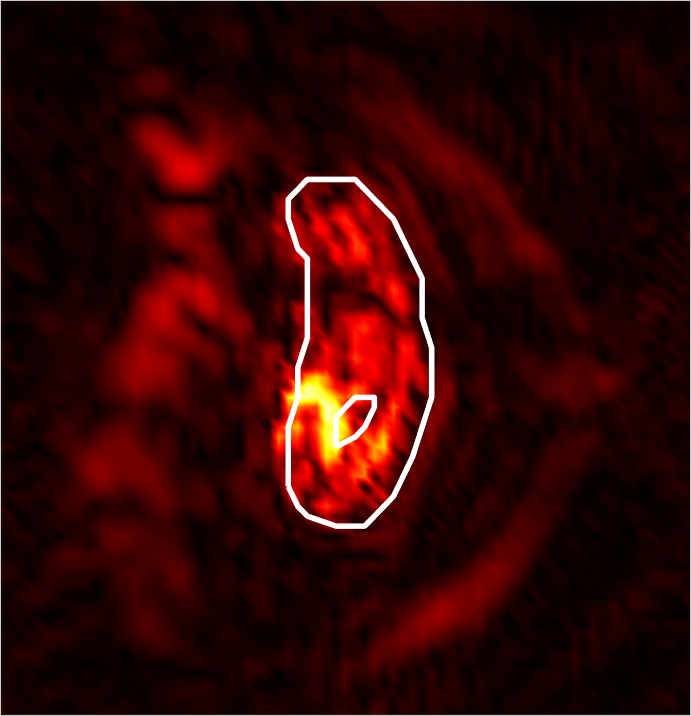} 
\end{minipage}
\begin{minipage}{2.55cm} \centering
    \includegraphics[width=2.55cm,height=2.55cm]{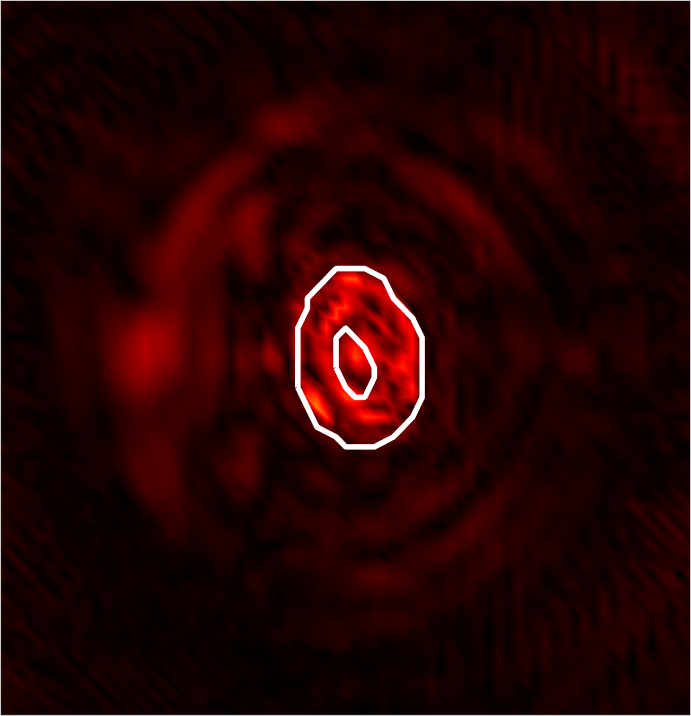} 
\end{minipage}
\begin{minipage}{2.55cm} \centering \vskip-0.05cm
    \includegraphics[width=2.55cm,height=2.55cm]{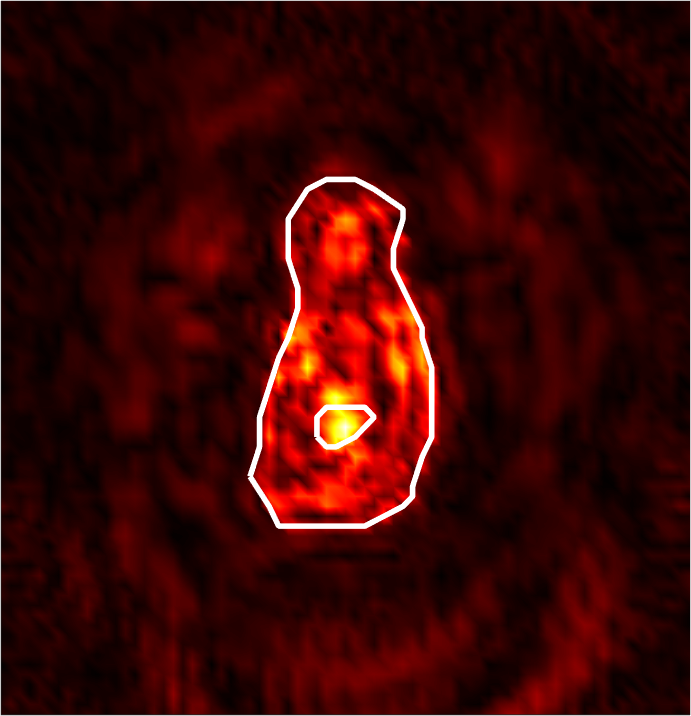} 
\end{minipage}
\\  \vskip0.05cm
\rotatebox{90}{\hskip-.3cm AC PCA}
    \begin{minipage}{2.55cm} \centering
    \includegraphics[width=2.55cm,height=2.55cm]{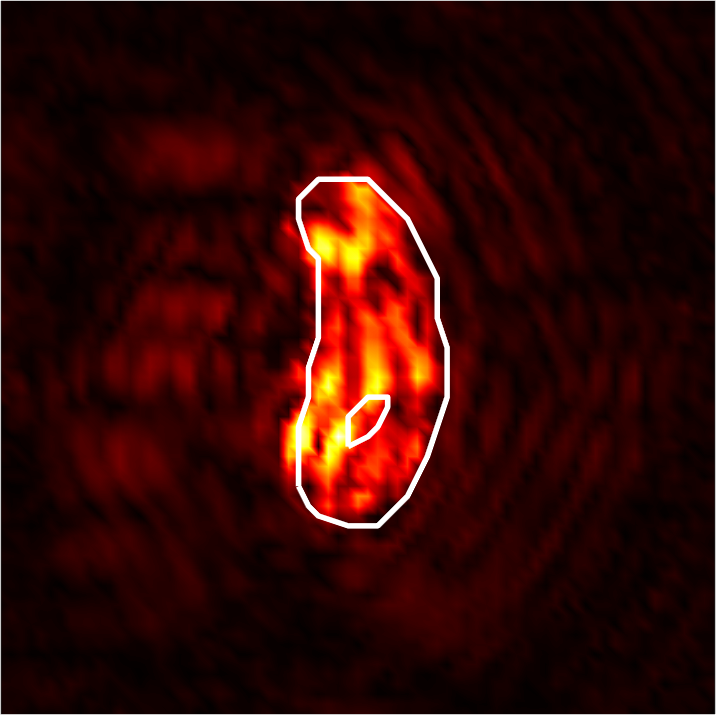} 
\end{minipage}
\begin{minipage}{2.55cm} \centering
    \includegraphics[width=2.55cm,height=2.55cm]{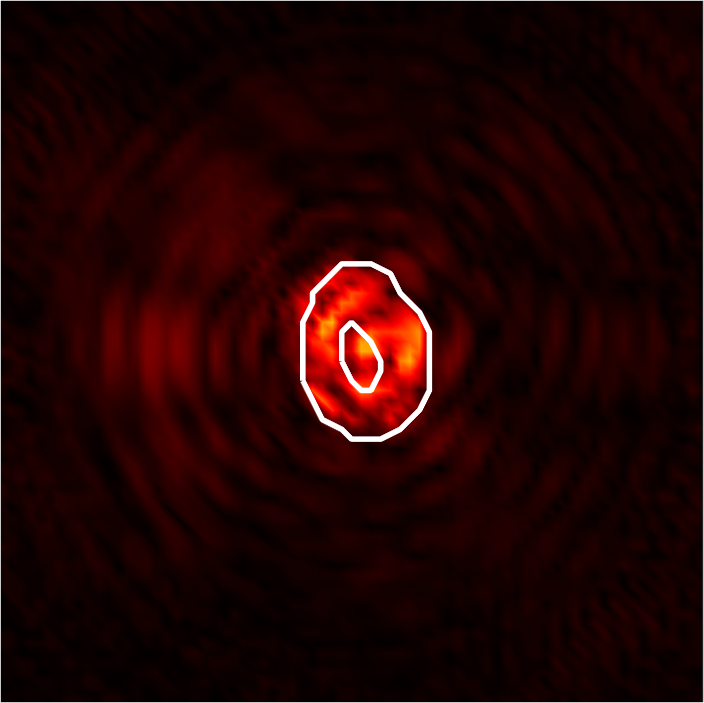} 
\end{minipage}
\begin{minipage}{2.55cm} \centering \vskip-0.04cm
    \includegraphics[width=2.55cm,height=2.55cm]{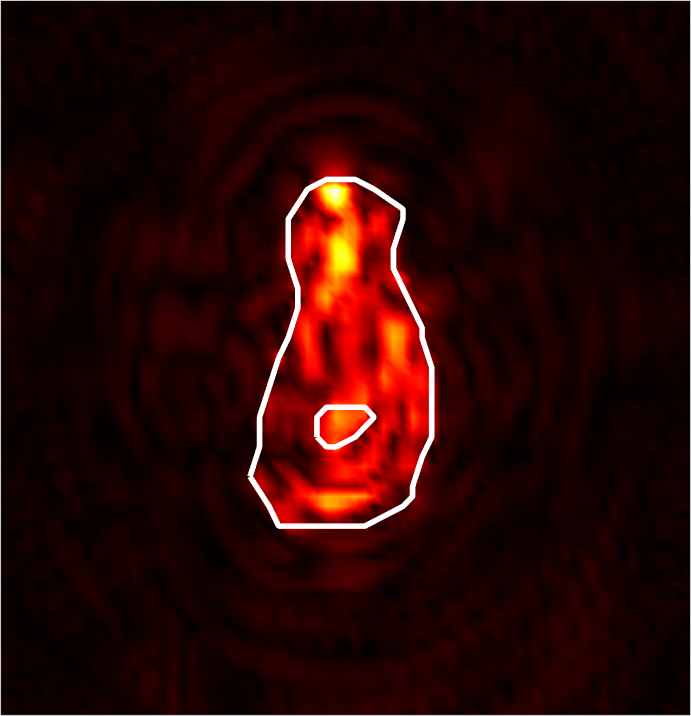} 
\end{minipage}
\\ 
\vskip0.2cm 
          \begin{minipage}{2.55cm} \centering        
        ({\bf I}): ZY
             \end{minipage}
    \begin{minipage}{2.55cm} \centering 
   ({\bf II}): ZX 
            \end{minipage}
    \begin{minipage}{2.55cm} \centering 
   ({\bf III}): XY
               \end{minipage}
\end{scriptsize}
\caption{Backpropagated reconstructions of the detailed analogue  with attenuation correction (AC) over a limited frequency band (3.5--12 GHz). Of the applied filters, Base refers to a plain bandpass filter, while data max evaluates, for the given frequency range, the maximum energy of the signal, TF Max evaluates the maximum energy of the continuous wavelet transform, and PCA is based on the $\ell$ largest PCs. Cut in the ZY, ZX, and XY directions (from left to right). }
\label{result5b}
\end{figure}

\begin{figure}[!ht]
    \centering   
    \begin{scriptsize}  \rotatebox{90}{\hskip-.3cm Base}
    \begin{minipage}{2.55cm} \centering
    \includegraphics[width=2.55cm,height=2.55cm]{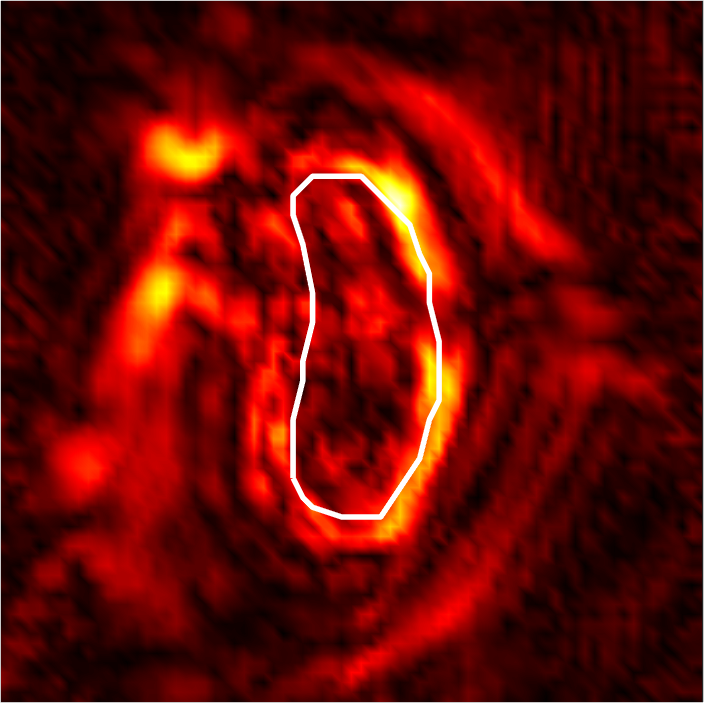} 
\end{minipage}
\begin{minipage}{2.55cm} \centering
    \includegraphics[width=2.55cm,height=2.55cm]{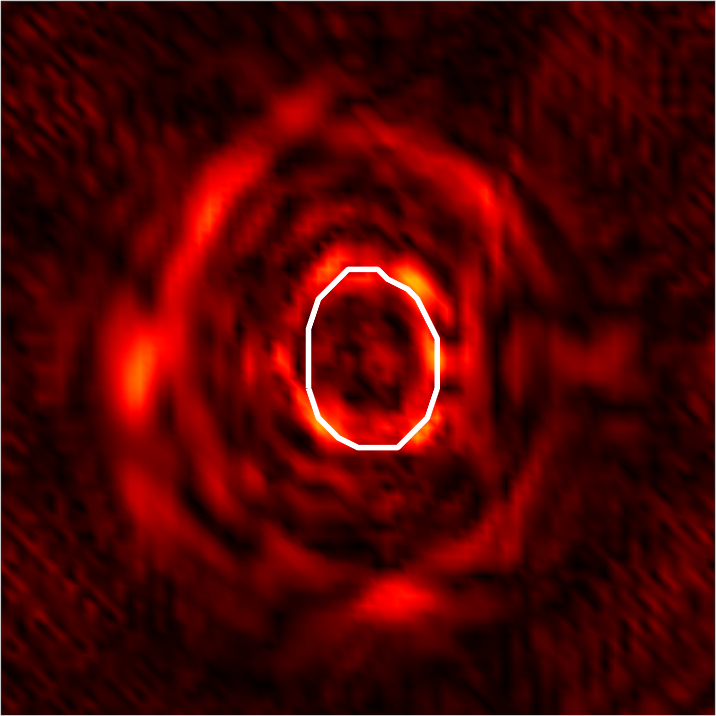} 
\end{minipage}
\begin{minipage}{2.55cm} \centering
    \includegraphics[width=2.55cm,height=2.55cm]{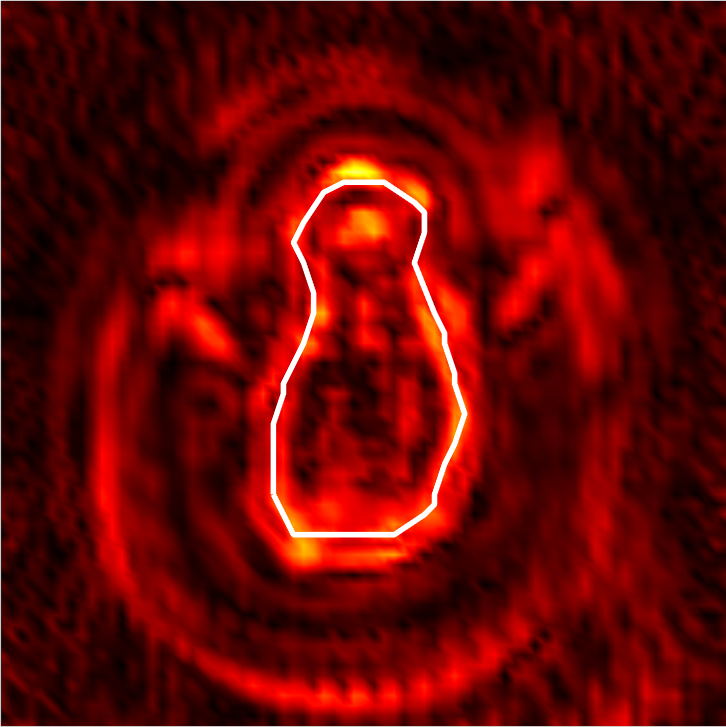} 
\end{minipage}
\\ \vskip0.05cm 
\rotatebox{90}{\hskip-.3cm Data Max}  
    \begin{minipage}{2.55cm} \centering
    \includegraphics[width=2.55cm,height=2.55cm]{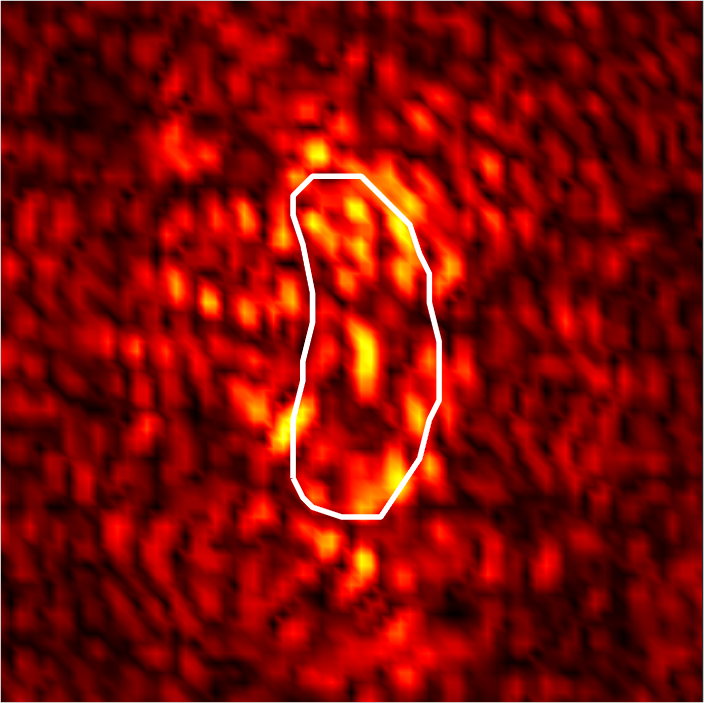} 
\end{minipage}
\begin{minipage}{2.55cm} \centering
    \includegraphics[width=2.55cm,height=2.55cm]{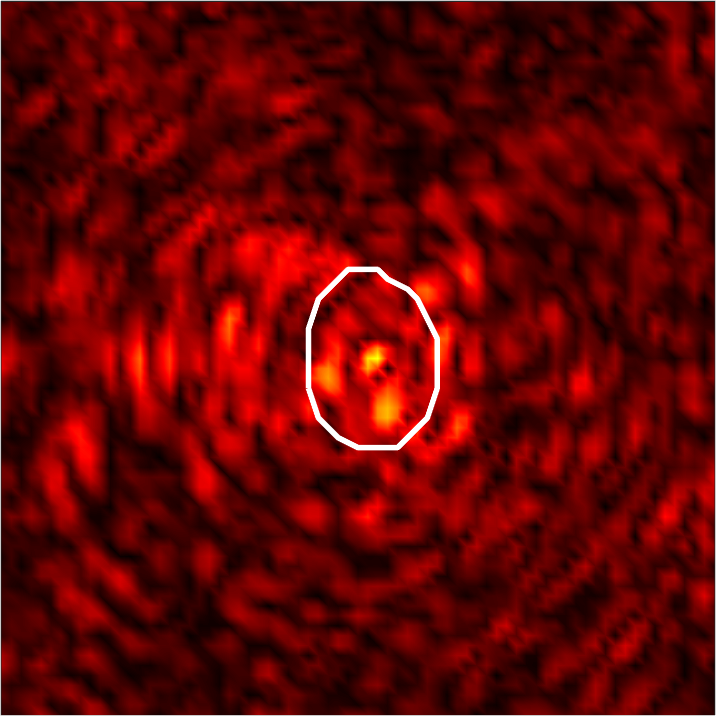} 
\end{minipage}
\begin{minipage}{2.55cm} \centering
    \includegraphics[width=2.55cm,height=2.55cm]{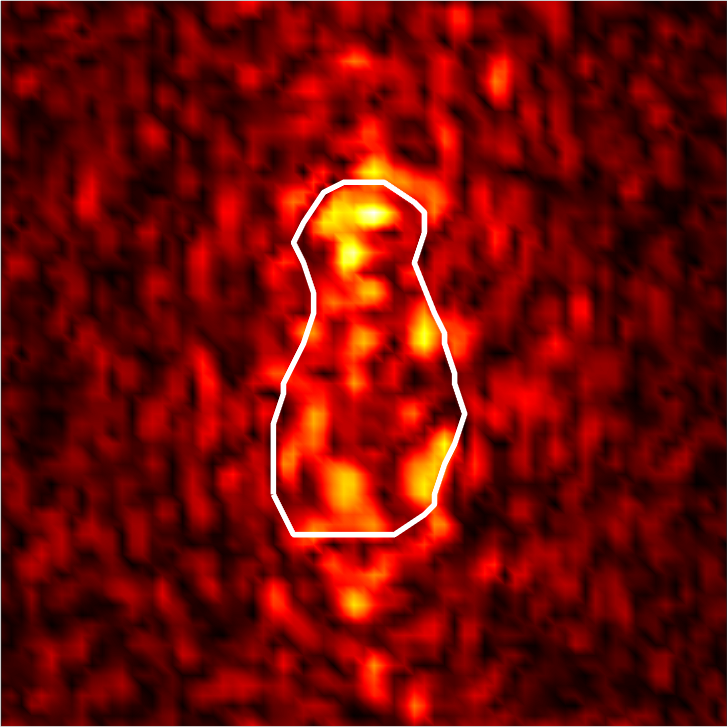} 
\end{minipage}
\\ \vskip0.05cm
\rotatebox{90}{\hskip-.3cm TF Max} 
    \begin{minipage}{2.55cm} \centering
    \includegraphics[width=2.55cm,height=2.55cm]{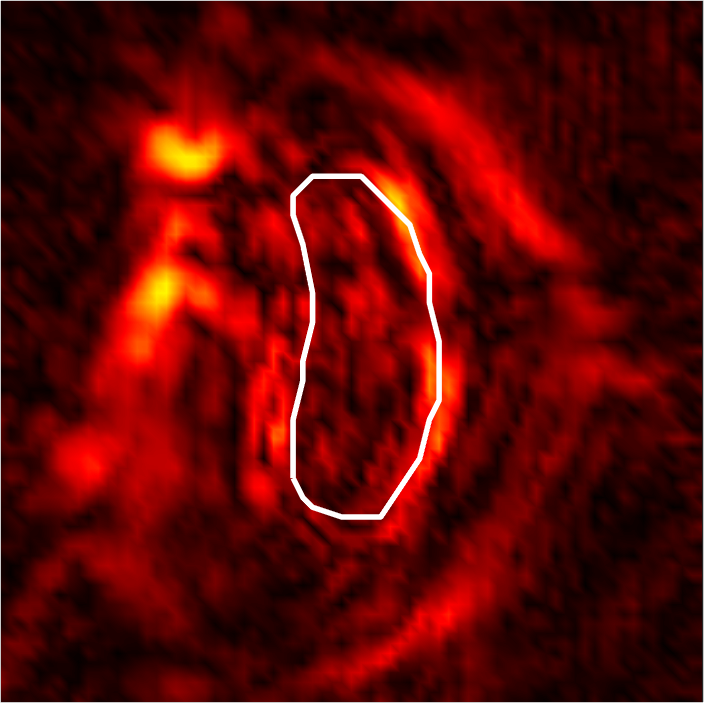} 
\end{minipage}
\begin{minipage}{2.55cm} \centering
    \includegraphics[width=2.55cm,height=2.55cm]{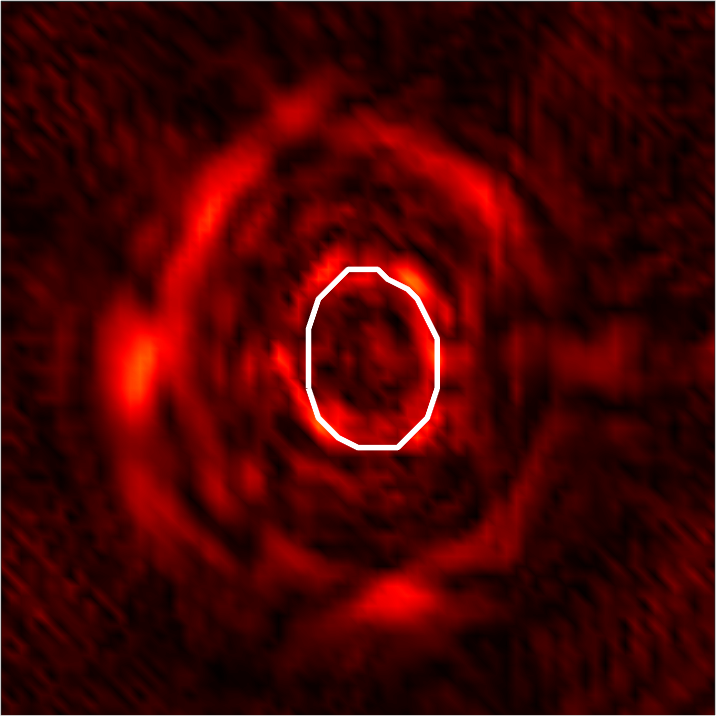} 
\end{minipage}
\begin{minipage}{2.55cm} \centering
    \includegraphics[width=2.55cm,height=2.55cm]{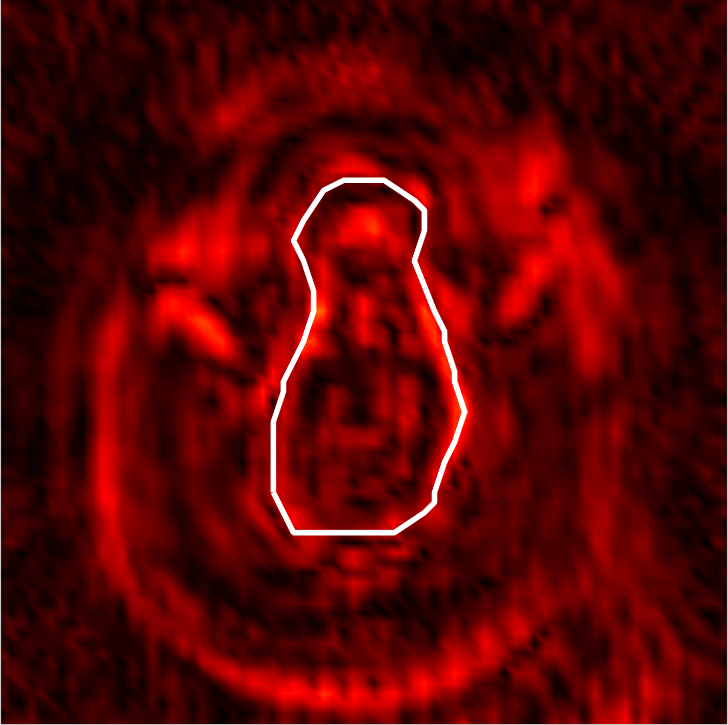} 
\end{minipage}
\\  \vskip0.05cm
\rotatebox{90}{\hskip-.3cm PCA} 
   \begin{minipage}{2.55cm} \centering
    \includegraphics[width=2.55cm,height=2.55cm]{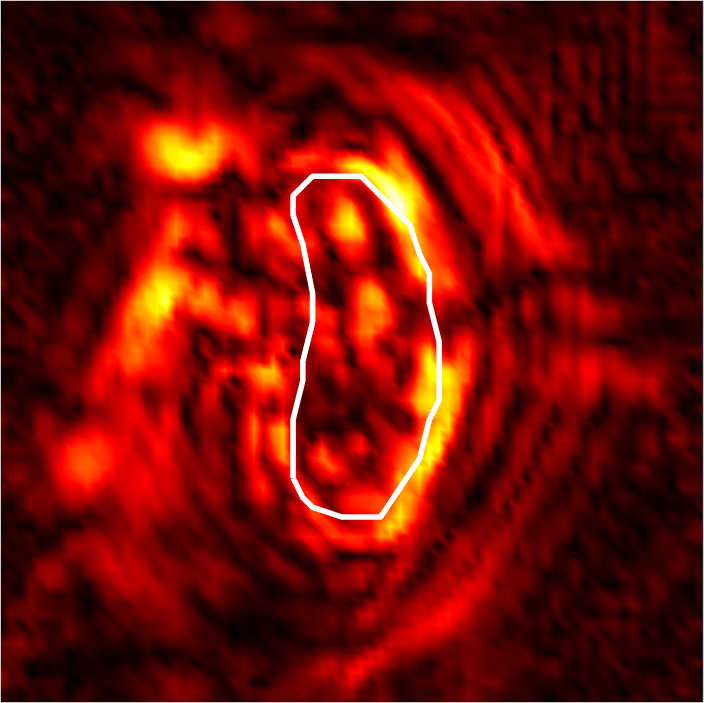} 
\end{minipage}
\begin{minipage}{2.55cm} \centering
    \includegraphics[width=2.55cm,height=2.55cm]{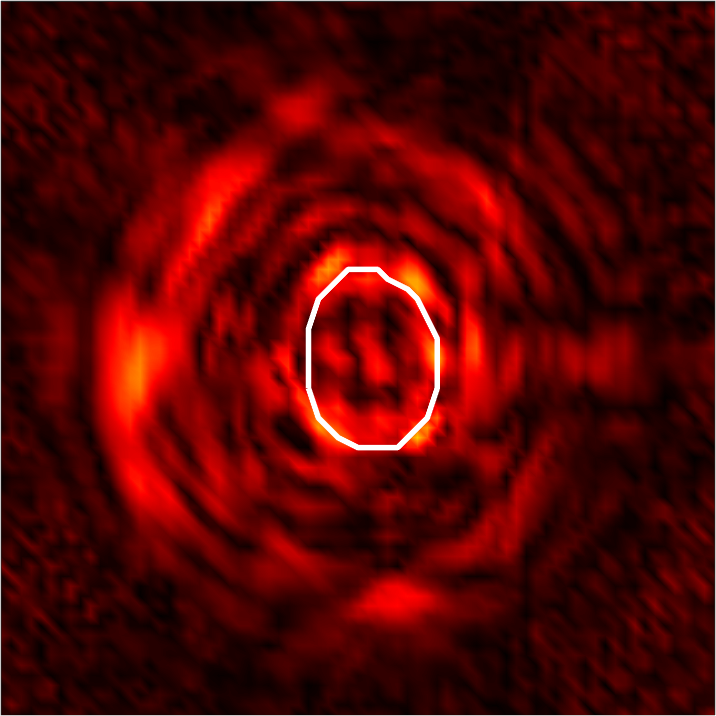} 
\end{minipage}
\begin{minipage}{2.55cm} \centering
    \includegraphics[width=2.55cm,height=2.55cm]{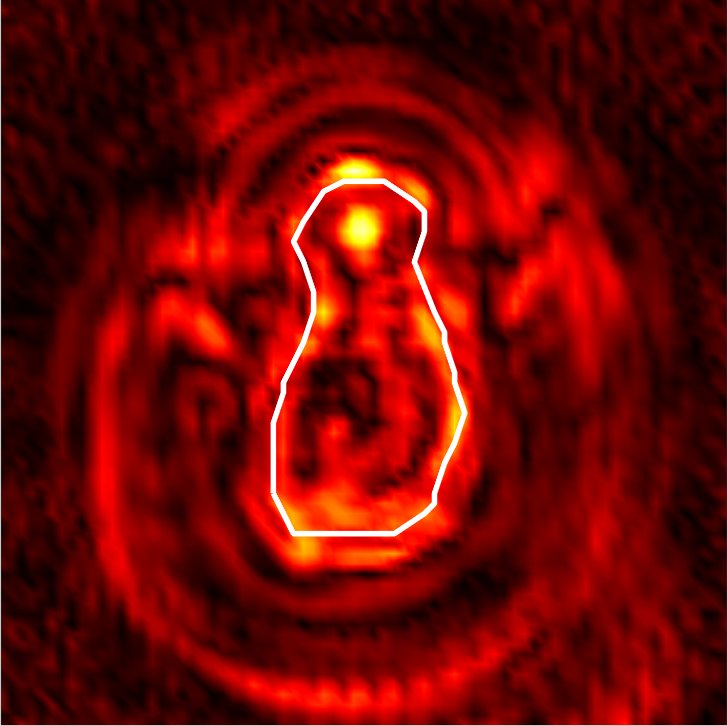} 
\end{minipage}
\\ \vskip0.25cm

          \begin{minipage}{2.55cm} \centering        
        ({\bf I}): ZY
             \end{minipage}
    \begin{minipage}{2.55cm} \centering 
   ({\bf II}): ZX 
            \end{minipage}
    \begin{minipage}{2.55cm} \centering 
   ({\bf III}): XY
               \end{minipage}
\end{scriptsize}
\caption{Backpropagated reconstructions of the homogeneous analogue without attenuation correction (AC) over a limited frequency band (3.5--12 GHz). }
\label{result6}
\end{figure}

\begin{figure}[!ht]
    \centering   
    \begin{scriptsize}
\rotatebox{90}{\hskip-.3cm AC Base}
    \begin{minipage}{2.55cm} \centering
    \includegraphics[width=2.55cm,height=2.55cm]{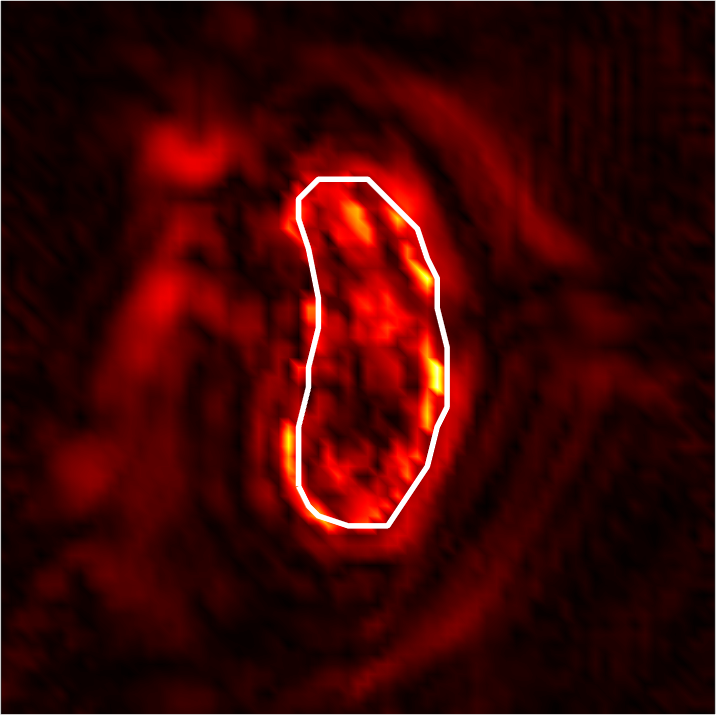} 
\end{minipage}
\begin{minipage}{2.55cm} \centering
    \includegraphics[width=2.55cm,height=2.55cm]{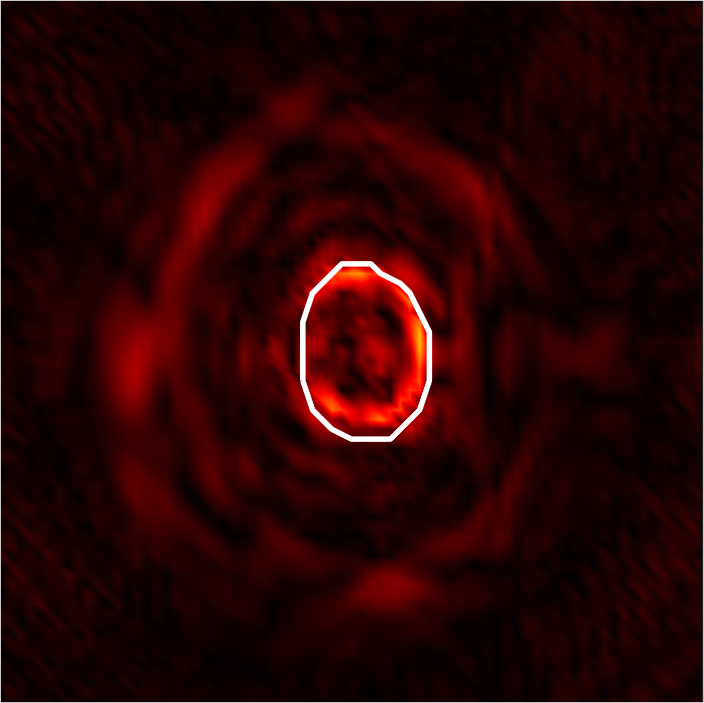} 
\end{minipage}
\begin{minipage}{2.55cm} \centering
    \includegraphics[width=2.55cm,height=2.55cm]{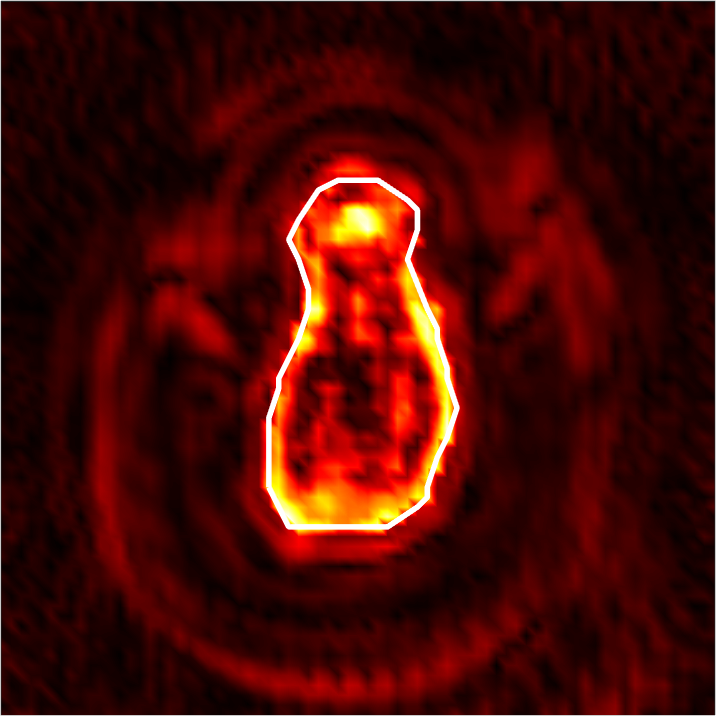} 
\end{minipage}
\\ \vskip0.05cm
\rotatebox{90}{\hskip-.3cm AC Data Max}
    \begin{minipage}{2.55cm} \centering
    \includegraphics[width=2.55cm,height=2.55cm]{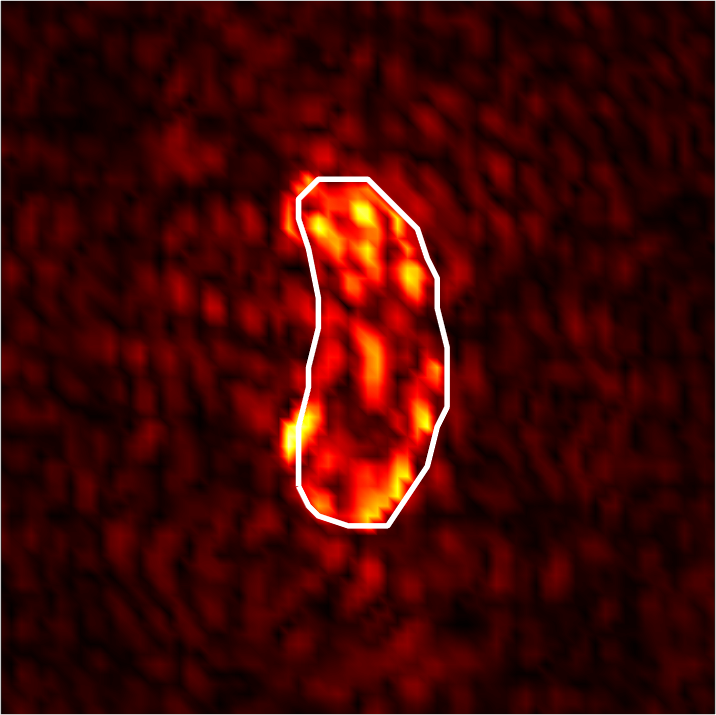} 
\end{minipage}
\begin{minipage}{2.55cm} \centering
    \includegraphics[width=2.55cm,height=2.55cm]{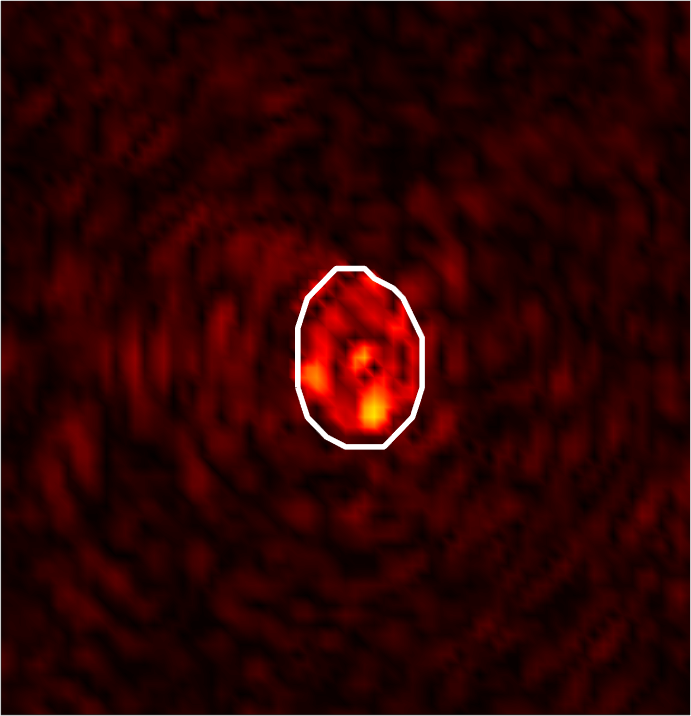} 
\end{minipage}
\begin{minipage}{2.55cm} \centering
    \includegraphics[width=2.55cm,height=2.55cm]{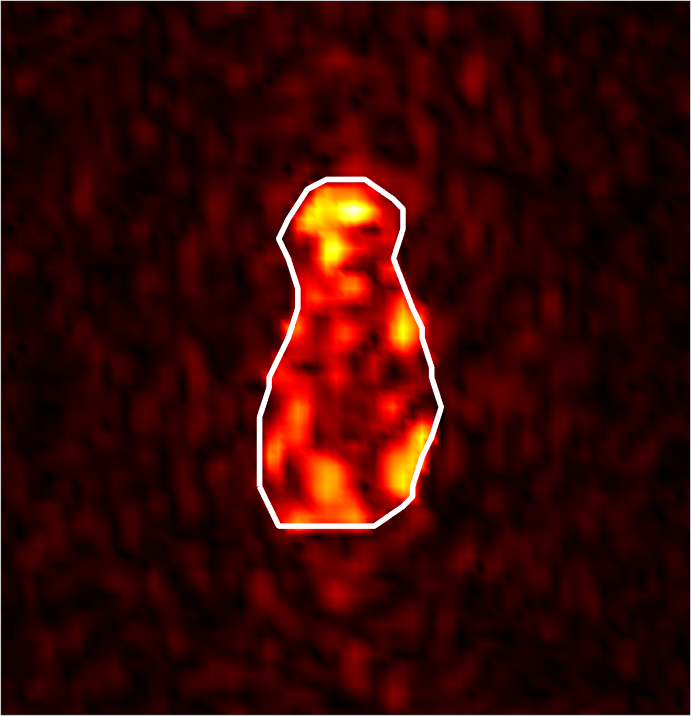} 
\end{minipage}
\\ \vskip0.05cm
\rotatebox{90}{\hskip-.3cm AC TF Max}
    \begin{minipage}{2.55cm} \centering
    \includegraphics[width=2.55cm,height=2.55cm]{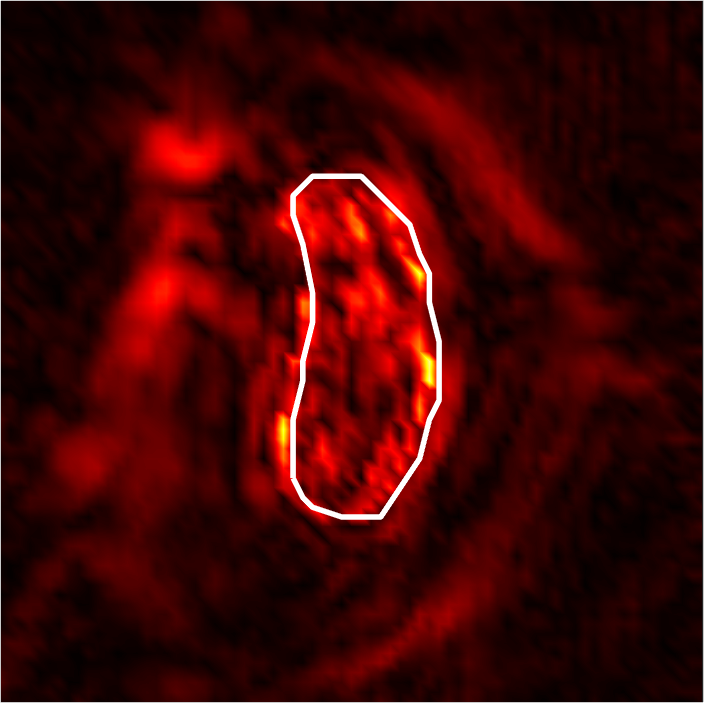} 
\end{minipage}
\begin{minipage}{2.55cm} \centering
    \includegraphics[width=2.55cm,height=2.55cm]{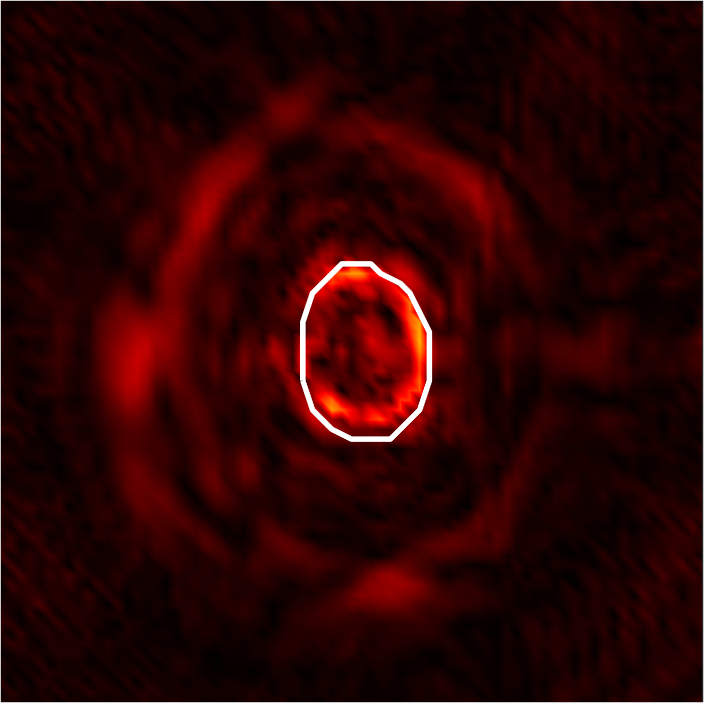} 
\end{minipage}
\begin{minipage}{2.55cm} \centering
    \includegraphics[width=2.55cm,height=2.55cm]{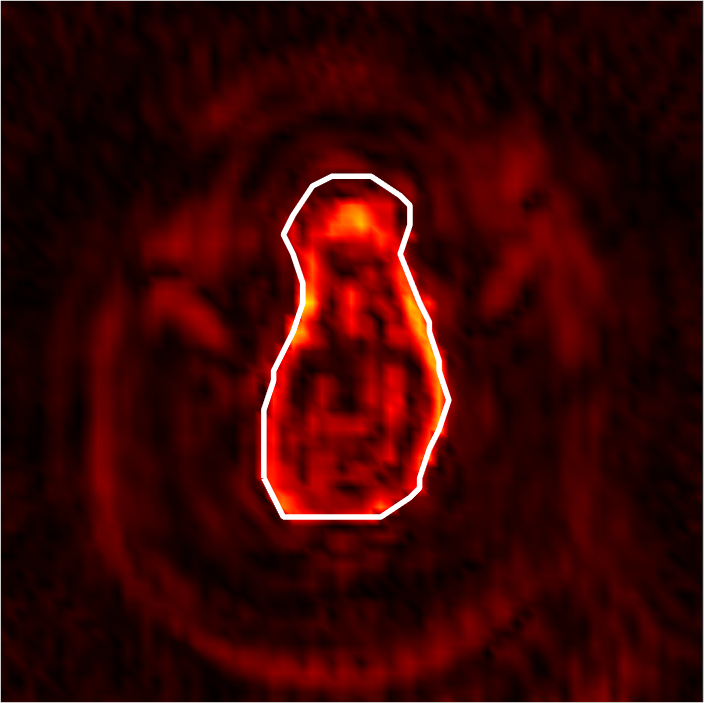} 
\end{minipage}
\\ \vskip0.05cm
\rotatebox{90}{\hskip-.3cm AC PCA}
    \begin{minipage}{2.55cm} \centering
    \includegraphics[width=2.55cm,height=2.55cm]{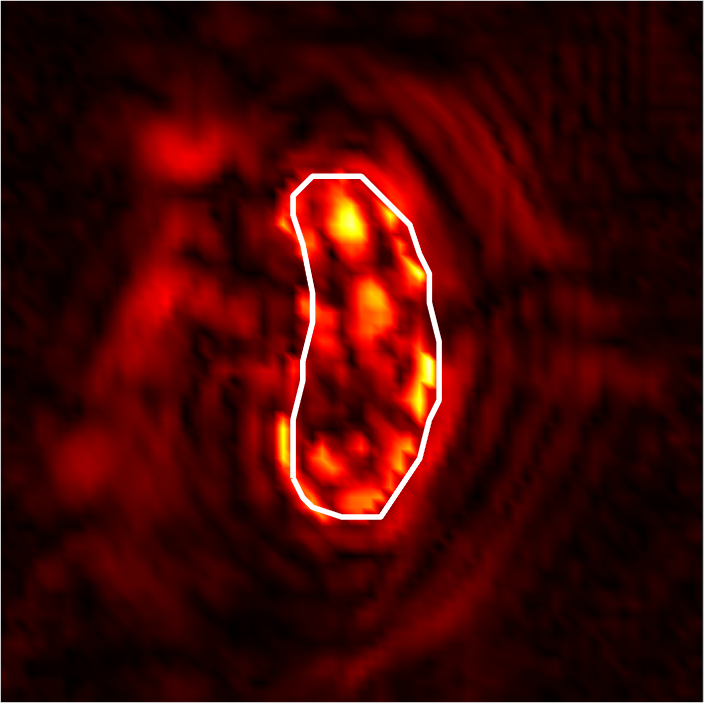} 
\end{minipage}
\begin{minipage}{2.55cm} \centering
    \includegraphics[width=2.55cm,height=2.55cm]{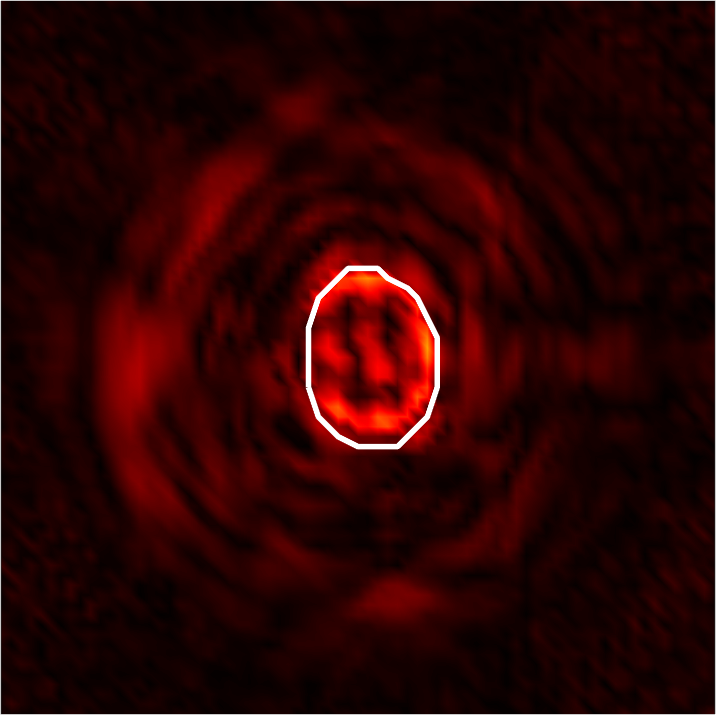} 
\end{minipage}
\begin{minipage}{2.55cm} \centering
    \includegraphics[width=2.55cm,height=2.55cm]{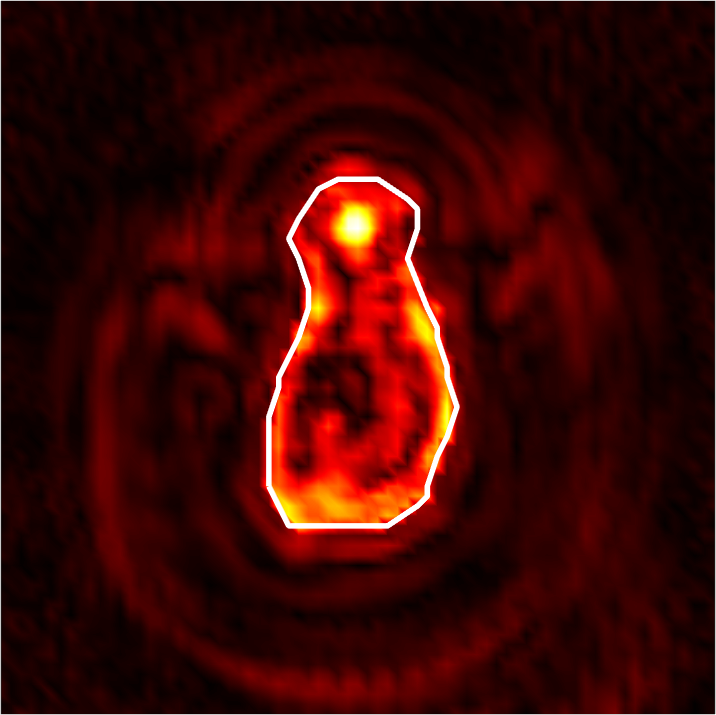} 
\end{minipage}
\\ 
\vskip0.2cm 
          \begin{minipage}{2.55cm} \centering        
        ({\bf I}): ZY
             \end{minipage}
    \begin{minipage}{2.55cm} \centering 
   ({\bf II}): ZX 
            \end{minipage}
    \begin{minipage}{2.55cm} \centering 
   ({\bf III}): XY
               \end{minipage}
\end{scriptsize}
\caption{Backpropagated reconstructions of the homogeneous analogue with attenuation correction (AC) over a limited frequency band (3.5--12 GHz).}
\label{result6b}
\end{figure}

\subsection{Topographic Imaging}
\label{quali}

A 2D TFR map of the measurements in section \ref{analogue} is presented in Fig. \ref{result1}. Fig. \ref{result3} and \ref{result4}  show the topographic reconstruction of the travel time for the signal peak with the maximum energy. The 2D TFR maps presented in Fig. \ref{result1} were obtained from the first three principal components of the full measurements using the spectrogram and wavelet approaches. Since the PCA can be implemented on the measurements along the frequency (set of frequencies) and observation (set of measurement positions), we used the latter to obtain the principal components with the highest variation in the data. In this case, the first three PCs amount to 37\% of the variance in the detailed measurement and 47\% of the variance in the homogeneous measurement. Here, each principal component contains all the frequency points and is ordered according to their variance, hence, the result is presented in Fig. \ref{result1}. 

The sampling frequency $f_{s}$ was selected according to the Nyquist condition $f_{s} \geq 2f_{\max}$, hence, the time and frequency axes are resolved in nanoseconds (ns) and gigahertz (GHz) respectively, as shown in Fig. \ref{result1}. The spectrogram and wavelet TFRs have 146 and 161 frequency bands corresponding to a width of $\approx 0.11$ and $\approx 0.1$ GHz, respectively. The difference in TFR maps for the homogeneous and detailed analogue is pronounced in the cases of 1st and 3rd PC, where the homogeneous signatures are more elongated along the frequency axis while the detailed signatures present discontinuities and, consequently, a greater number of ridges (see Fig. \ref{result1}, 1st \& 3rd PCs, IV). The similarity of the TFR patterns is also obvious, however, they differ in the time-frequency localisation. The pattern in the wavelet cases is elongated along the time axis, while that of the spectrogram presents more features in the decibel range considered (20 dB to -20 dB, see Fig. \ref{result1}, 1st PC).  The TFR maps of Fig. \ref{result1} highlight the presence of weak signatures in some frequencies of the maps corresponding to the detailed analogue which suggests the presence of a distinguishing feature between the two analogues (see Fig. \ref{result1}, 1st \& 3rd PCs).  We observe that the higher-frequency features become evident in the 2nd \& 3rd PCs suggesting that PCA can be used to filter out higher frequency contribution from the data (see Fig. \ref{result1}, 1st PC). Moreover, at higher frequencies, the mesh of the analogue is too large compared to the wavelength, hence, introducing artefacts into the data. This is supported by the findings in \cite{dufaure2023imaging} where the optimal frequency band is 3.5--12 GHz, hence, utilised in the tomographic imaging part in Section \ref{quant}. 


The topographic reconstruction of the maximum energy travel time for each measurement point is presented in Fig. \ref{result3}. The reconstructions for detailed and homogeneous analogue and their difference reflect the presence of an interior structure in the detailed measurement; the area very close to the void location has the lowest travel time indicating that the void structure has a lower permittivity value. Fig. \ref{result4} which is an average reconstruction (mean of reconstructions) of 25 realisations of noisy measurements (additive and uncorrelated Gaussian random noise) at 10 dB peak SNR presents a similar interior structure. The wavelet realisation in Fig. \ref{result4} presents the best localisation of the void structure in the detailed and the difference reconstructions as compared to the spectrogram.

\subsection{Tomographic Imaging}
\label{quant}

The backpropagation approach provides the opportunity to quantitatively measure the reconstruction's goodness of fit based on the analogue's structure while also presenting the interior cross-section of the target; this is advantageous when compared to the topographic imaging approach. We compare the results from \cite{dufaure2023imaging} with results obtained through FBP discussed in Section \ref{inverse}. The reconstructions of detailed and homogeneous analogues without and with attenuation correction (AC) for a limited frequency band (3.5--12 GHz) are presented in Fig. \ref{result5} -- \ref{result6b}, respectively. The figures correspond to a cut along the XY, YZ, and XZ planes for the 3D tomographic reconstruction of the detailed and homogeneous analogues. The cuts are selected such that they present the centre of the void along all three directions. The theoretical shape of the target and void structure is indicated by the white lines while the reconstruction is a contrast between 0 and 1, where the bright red spots are closer to 1, indicating a higher visibility. 

As filtering techniques, we examined a plain bandpass (Base), as well as evaluated the maximum energy of the signal for each measurement point in the frequency domain (data max), the maximum energy of the continuous wavelet transform in time for each frequency in the measurement point (TF Max), and PCA implemented on the measurements along the frequency (set of frequencies as variables) based on the $\ell = 33$ largest PCs for the detailed and homogeneous measurements which correspond to 90\% variance in the data. The comparison of an observation point after different processing techniques (data max, TF max, and PCA) is shown in Fig \ref{result8}.


 The maximum energy of the signal (data max) gives a reconstruction that supports the results obtained from the topographic imaging in section \ref{quali} where the travel time of the maximum energy ${\mathtt{t}}$ in equation \eqref{project} is used for topographic reconstruction. This is expected since ${\mathtt{t}}$ and data max are related, hence explaining a similar phenomenon. The maximum energy of the signal (data max) gives a reconstruction that localises the void structure in the case of the detailed analogue while also presenting a lot of artefacts and discontinuities.  The time-frequency filtering technique (TF Max) was observed to reduce the artefacts around curvatures in the reconstructions to varying degrees (see Fig. \ref{result5} and \ref{result6}, the top part of the homogeneous and detailed analogue). The PCA filtering presents a well-localised void structure in the detailed case but has pronounced artefacts at the top of the detailed model. These artefacts are also present at the top and curvatures of the homogeneous model indicating that the selected principal components are influenced by curved-surface reflections, and possibly reflections from the target support which is evident in the top of the analogues (see Figs. \ref{result6} and \ref{result6b}, Base XY and AC Base XY).

\begin{table}[h]
	\centering
	\small
    \resizebox{0.99\columnwidth}{!}{%
	\begin{tabular}{@{}l |c c c c c |c c c r@{}}
		\toprule
   &\multicolumn{4}{c}{\textbf{Detailed}} & & & \multicolumn{2}{c}{\textbf{Homogeneous}}   \\ 
   \midrule 
		&  \textbf{RMSE} & \textbf{RMSE}  &\textbf{RMSE} &\textbf{RMSE}  &\textbf{RO} &  &\textbf{RMSE} & \textbf{RMSE} & \textbf{RMSE} \\ 
        &  \textbf{all} & \textbf{ext} & \textbf{int} &\textbf{void}  &  & & \textbf{all} & \textbf{ext} & \textbf{int} \\  \hline
		 &  &  & \\ 
		\textbf{Base} &  0.116 &      0.086  & 0.648  &    0.720  &     0.254 & &  0.150     &  0.122 & 0.733\\ 
   &  &  & \\ 
		 \textbf{Data Max}   & 0.181   &    0.169  & 0.566 &    0.668 &      0.106 &  & 0.171   &     0.150 & 0.690\\ 
    &  &  & \\ 
		  \textbf{TF Max}  &  0.129   &    0.103  &0.645  &    0.709   &    {0.259} & & {0.146} &      {0.112} & 0.788\\ 
		    &  &  & \\ 
		\textbf{PCA}    &   {0.109} &      {0.081} & 0.604 &     { 0.657 }   & {0.201} &  &  {0.158} &      {0.134}  & 0.704\\  &  &  & \\ 
  \midrule 
		\textbf{AC Base}    & 0.087   &    0.041   & 0.630 &     0.714   &  0.217 & & 0.104      & 0.058 & 0.719\\ 
     &  &  & \\ 
		\textbf{AC Data Max}    & 0.100     &  0.073 & 0.558 &      0.679   &  0.064 & & 0.104      & 0.062 & 0.701\\ 
  &  &  & \\
  \textbf{AC TF Max}    & 0.090   &    0.049  & 0.628  &      0.702   &  0.191 & & 0.110      & 0.065 & 0.738\\ 
     &  &  & \\ 
		\textbf{AC PCA}    & 0.081   &    0.036 & 0.593 &      0.664   &  0.143 & & 0.103      & 0.061 & 0.693\\ 
             \bottomrule
	\end{tabular} }
	\caption{Relative overlap (RO) and root mean squared error (RMSE) of the reconstructions obtained for the detailed and homogeneous analogue over a limited frequency band (3.5--12 GHz).}
	\label{tab:result7}
\end{table} 

Table \ref{tab:result7} gives the quantitative performance of these methods in terms of the RMSE and relative overlap with respect to the theoretical map of each analogue. The RMSE values are given for the different parts of the analogue i.e., exterior, interior, void, and whole structure (all) while RO is given for the void structure only. Expectedly, the homogeneous analogue does not present measures with regard to a void structure, hence, the RMSE values are given for the exterior, interior, and the whole structure. The measures in table \ref{tab:result7} show that the PCA outperforms the other methods regarding the RMSE of the whole structure, exterior and void for the detailed measurement while the TF Max performs best in terms of the relative overlap. The Data Max outperforms the other methods in terms of the RMSE interior for both the homogeneous and detailed measurements. In the homogeneous case, the TF Max also outperforms the other methods with regard to the RMSE of the whole and exterior structure.

The introduction of the attenuation correction was seen to reduce the artefacts in the exterior while also enhancing the overall reconstruction quality. The effect on the interior and void part of the analogues was minimal suggesting a low variation in the attenuation values in the Green's function correction model. The RO of Base was seen to be the largest in table \ref{tab:result7} while the RMSE void of the PCA was seen to be the lowest indicating the best approaches in terms of void distinguishability when the attenuation correction is or is not considered. The PCA approach also wins in the other measures of the detailed model such as RMSE of the whole structure, RMSE exterior and in two of the considered measures in the homogeneous case. The Data Max again wins in the RMSE interior when the attenuation correction is applied. 

It is important to note that the reconstructions would have however been justified without the introduction of the filters since the Base approach gives complete information on the target structures with minimal concerns such as deviation of the void structure and artefact at the target surface which we sort to remove through filters and model correction. The improvements made with the filtering methods and model correction are seen to be tangible when compared to the Base. This emphasises the need for filtering the measurement to obtain better reconstruction quality.

\begin{figure}[!ht]
    \centering  
    \begin{scriptsize}   
    \begin{minipage}{7.4cm}  \centering
  \begin{minipage}{5.55cm} \centering
    \includegraphics[width=4.85cm]{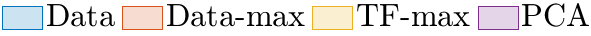} 
\end{minipage}\vskip0.2cm
\rotatebox{90}{\hskip.9cm RMSE all} 
    \includegraphics[width=7.0cm]{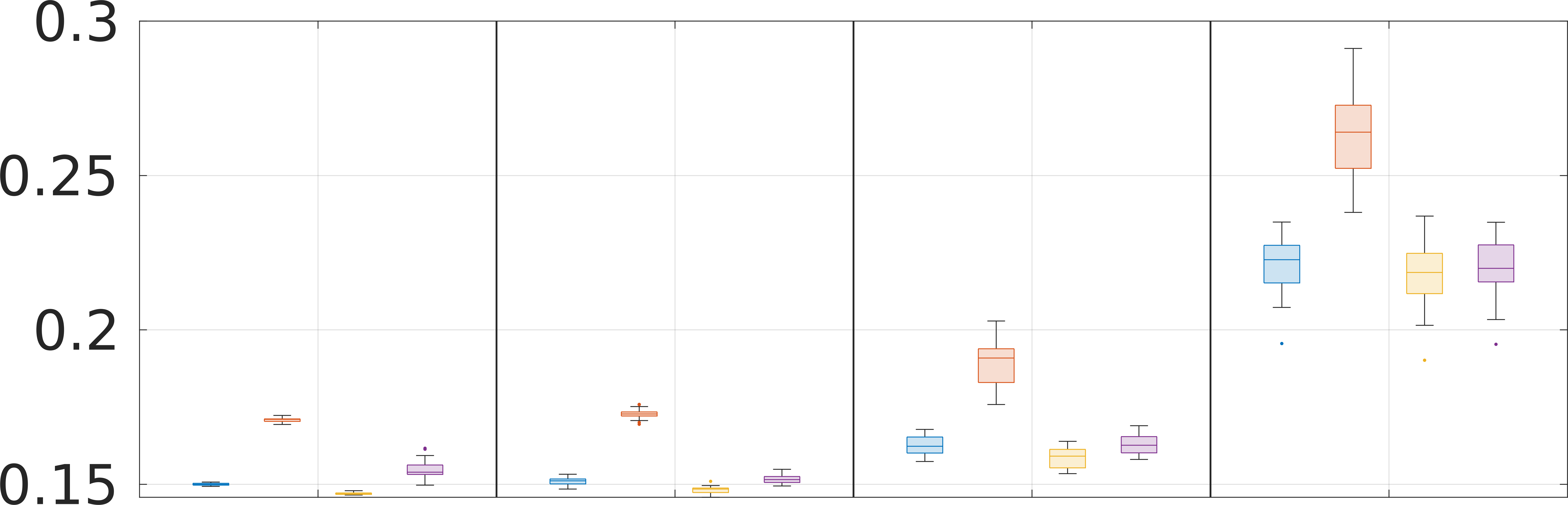} \end{minipage}  
    \\ \vskip0.2cm 
    \begin{minipage}{7.4cm} \centering 
       \rotatebox{90}{\hskip.9cm RMSE int} 
           \includegraphics[width=7.0cm]{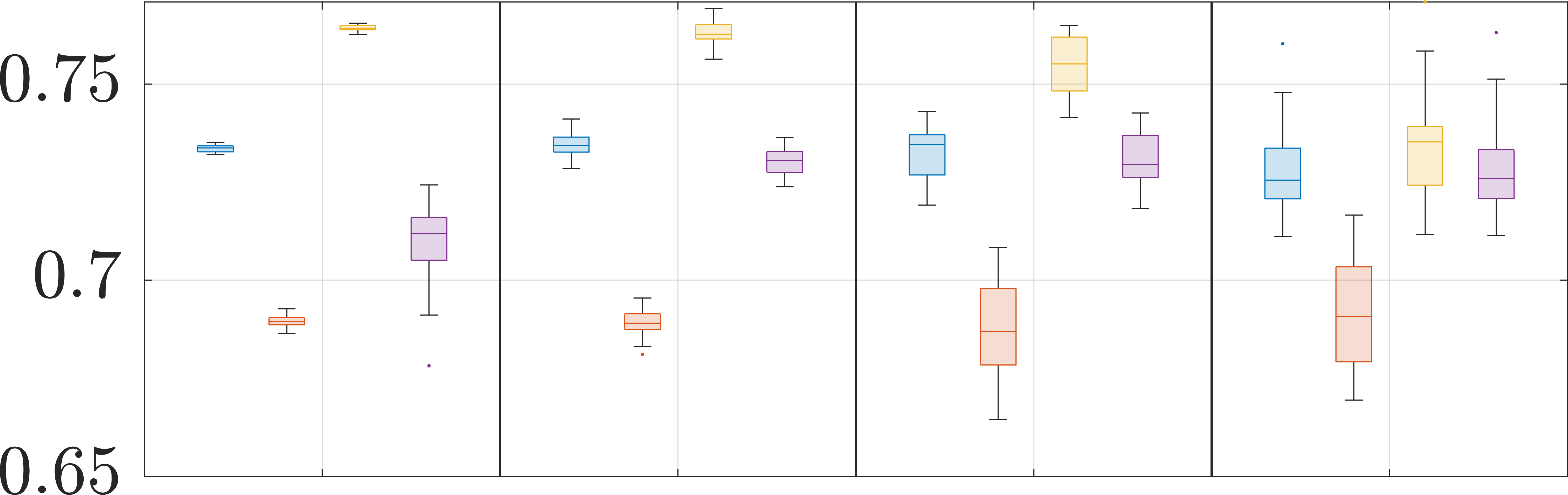} \end{minipage}  \\ \vskip0.2cm 
       \begin{minipage}{7.4cm} \centering 
       \rotatebox{90}{\hskip.9cm RMSE ext} 
           \includegraphics[width=7.0cm]{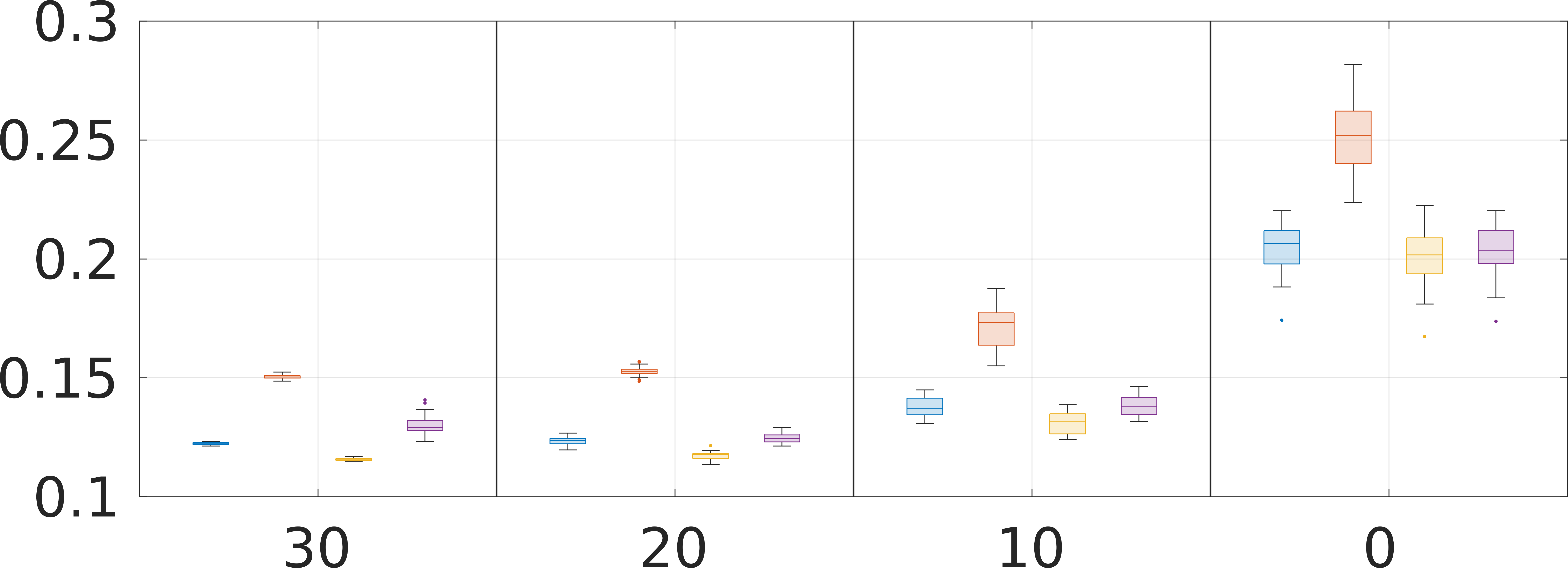} \end{minipage} 
         \end{scriptsize}
    \caption{ The performance of the filtering methods with the RMSE and RO measures for the bandlimited homogeneous measurement of 3.5--12 GHZ.} 
    \label{result10}
\end{figure}

 \begin{figure}[!h]
    \centering  
    \begin{scriptsize}  
    \begin{minipage}{7.4cm} \centering 
    \begin{minipage}{5.55cm} \centering
    \includegraphics[width=4.85cm]{figs/measure/legend_box.png} 
\end{minipage}\vskip0.2cm
    \rotatebox{90}{\hskip.9cm RMSE all} 
    \includegraphics[width=7.0cm]{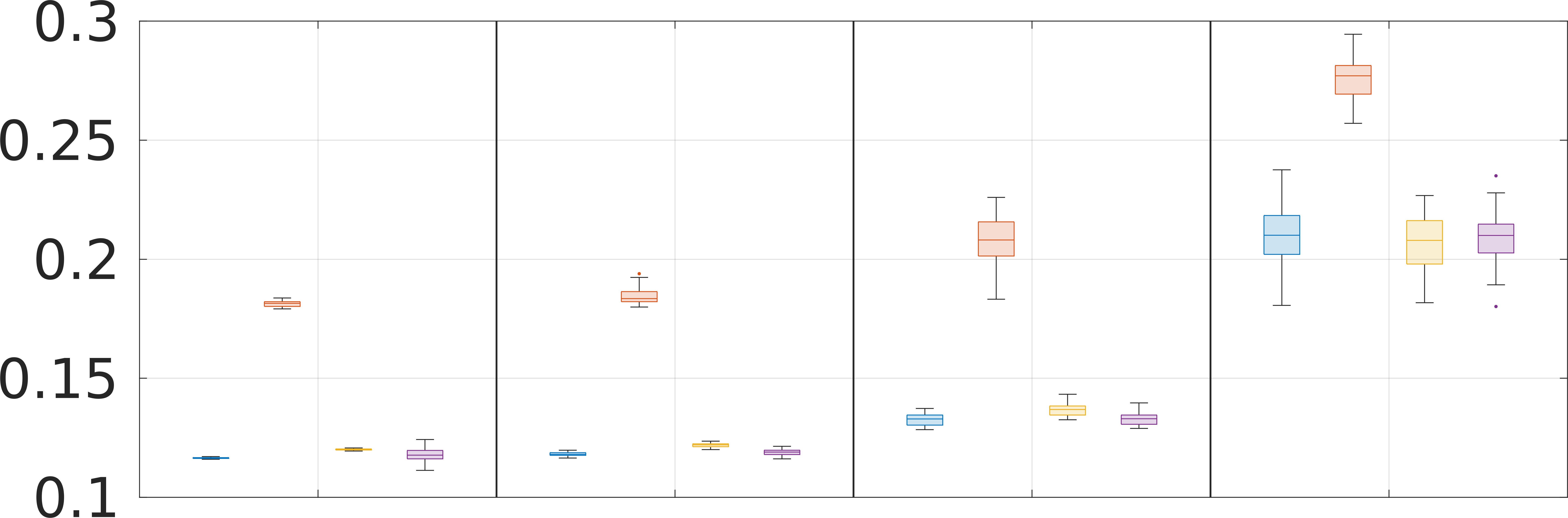} \end{minipage} 
    \\ \vskip0.2cm  
    \begin{minipage}{7.4cm} \centering 
       \rotatebox{90}{\hskip.9cm RMSE int} 
           \includegraphics[width=7.0cm]{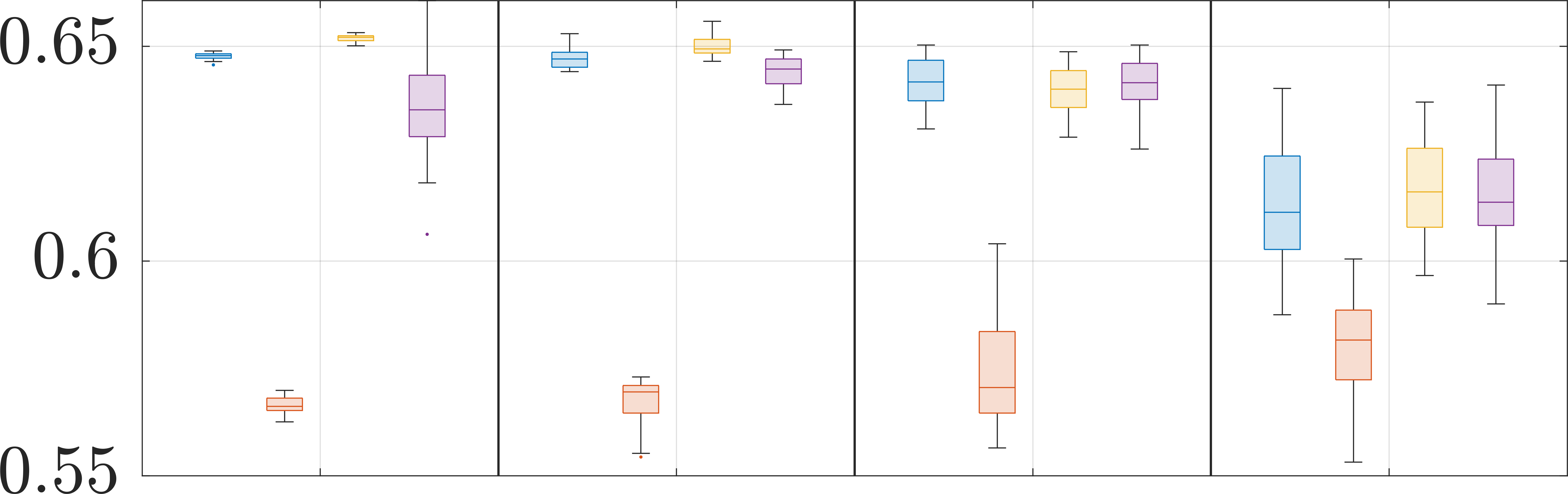} \end{minipage}  \\ \vskip0.2cm 
       \begin{minipage}{7.4cm} \centering
       \rotatebox{90}{\hskip.9cm RMSE ext} 
           \includegraphics[width=7.0cm]{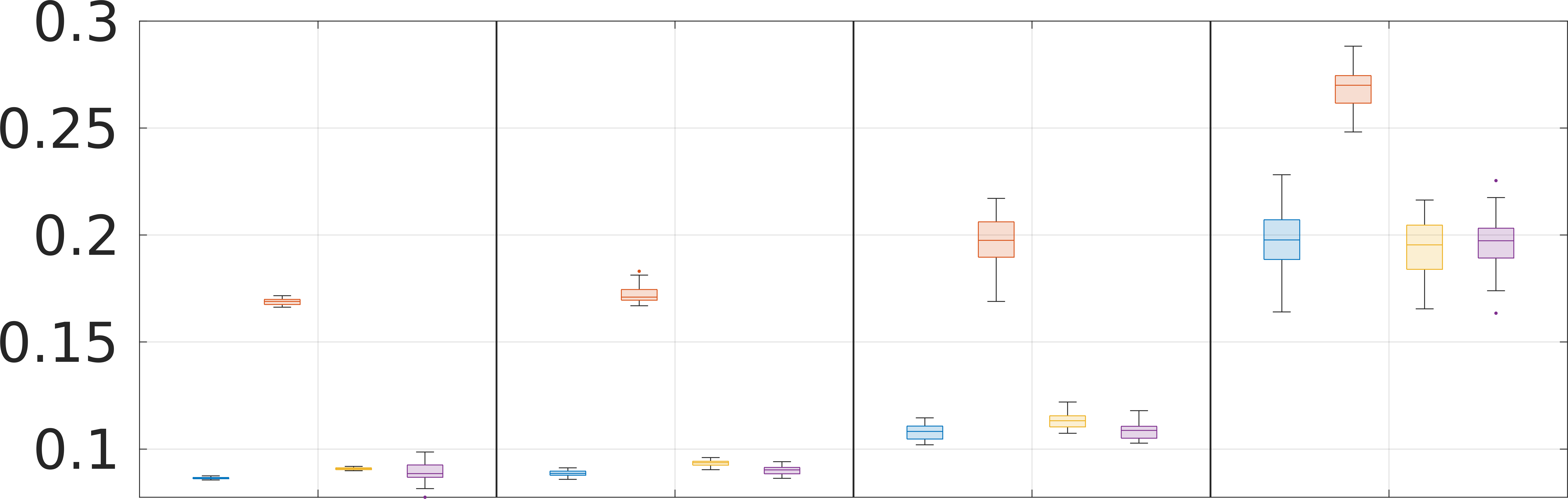} \end{minipage}  \\ \vskip0.2cm 
       \begin{minipage}{7.4cm} \centering 
       \rotatebox{90}{\hskip.9cm RMSE void} 
           \includegraphics[width=7.0cm]{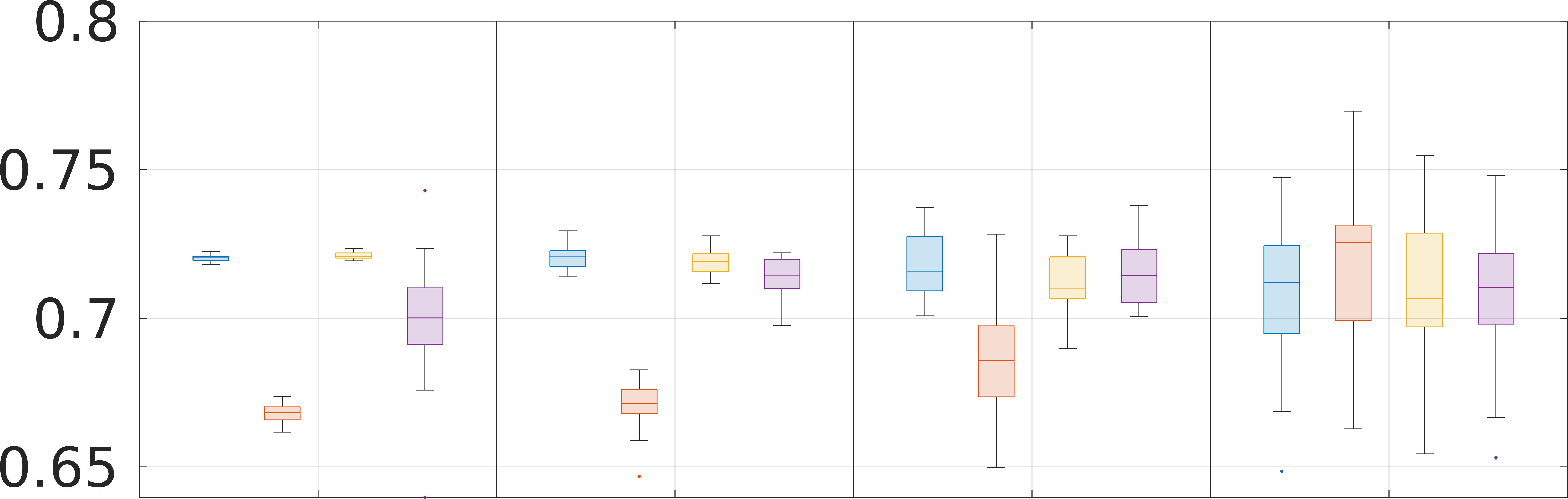} \end{minipage}  \\ \vskip0.2cm 
       \begin{minipage}{7.4cm} \centering 
       \rotatebox{90}{\hskip.9cm Overlap} 
           \includegraphics[width=7.0cm]{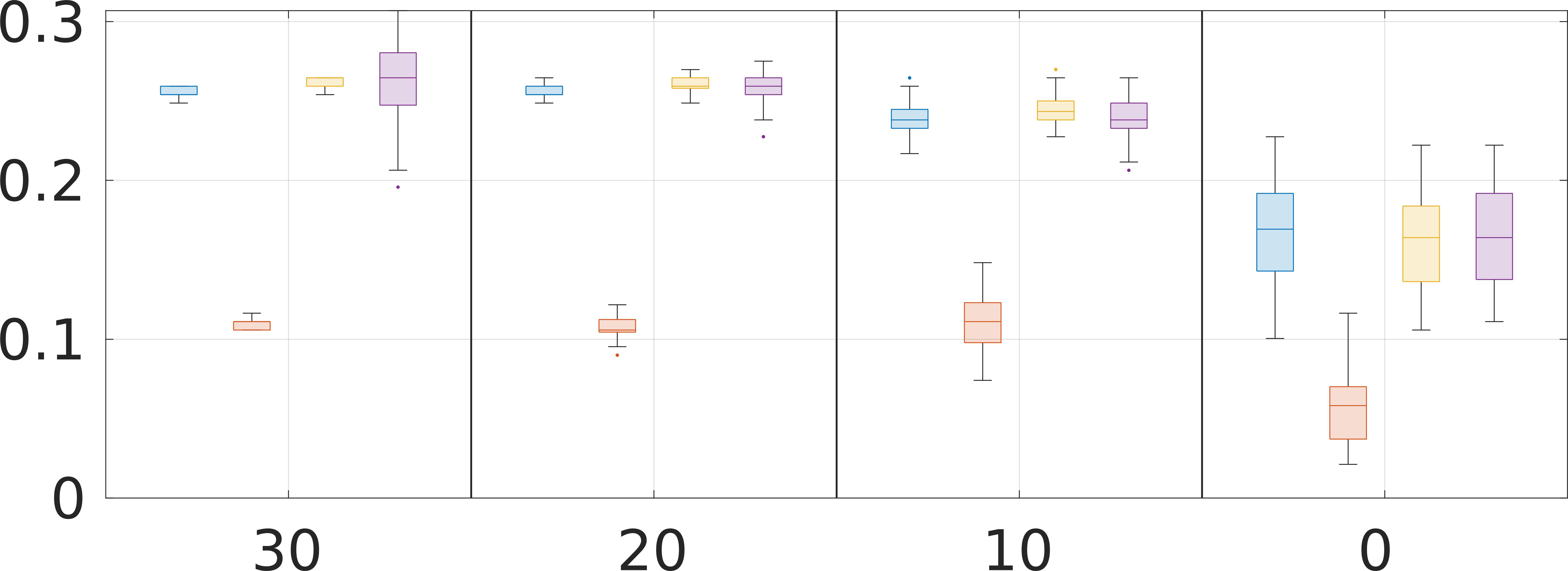} \end{minipage}  
           
         \end{scriptsize}
    \caption{The performance of the filtering methods with the RMSE and RO measures for the bandlimited detailed measurement of 3.5--12 GHZ.} 
    \label{result9}
\end{figure}

\subsection{Noise characterisation}
The robustness of the tomographic imaging methods was validated by adding different levels of noise to the detailed and homogeneous measurements.  The noisy data corresponding to additive and uncorrelated Gaussian random errors relative to the amplitude of the maximum signal (peak SNR) for 30, 20, 10 and 0-dB SNR, averaged over 25 sample realisation. The performance of the methods is evaluated with the  RMSE and RO measures for the bandlimited measurement of 3.5--12 GHZ as presented in Fig. \ref{result10} \& \ref{result9}. 

The RMSE of the whole structure and RMSE exterior were seen to have an increasing trend as the noise level increased for all methods under consideration. The spread (uncertainty) of the measures was also seen to increase for all the imaging methods as the noise level increased. This is expected since increased noise levels corrupt the reconstruction hence, creating more variance in the estimates. The mean of the measures was seen to be consistent with the result in Table \ref{tab:result7}. The data max technique outperforms the other methods with respect to the RMSE interior and RMSE void while showing reduced performance with respect to the other measures (RMSE all, RMSE exterior, and RO). The PCA filtering outperforms the Base and TF max with respect to all the measures under consideration, although the difference in performance with the Base is relatively marginal in the RMSE all and RMSE exterior cases. A closer inspection of the RMSE interior and RMSE void shows a convergence of the upward trend of the data max and the downward trend of the other methods as the noise level increases. The downward trend is explained by the discrepancies caused by the filtering method (such as mislocalisation of the void) rather than the added noise, hence the RMSE values above the level obtained with 0 dB noise.

\section{Discussion}
\label{sec:discussion}

This study is focused on fast imaging approaches for anomaly distinguishability in an asteroid analogue using experimental measurements.  The aim was to find a topographic and backpropagated tomographic reconstruction via the time-frequency techniques and principal component analysis while also investigating the robustness of the imaging methods in the presence of different noise levels. FBP is well-motivated in this study: namely, it has been used successfully in many laboratory studies \cite{chang1978method,koljonen2019mathematical}. Here it can be harnessed via the means of (1) data filtering to find the peak with maximum energy and, thereby, suppress the effect of higher-order scattering, and (2) correcting the signal propagation estimates provided by Green's function for vacuum. It can be also shown to have a connection to travel-time tomography, which is obvious based on the topographic analysis.

The time-frequency maps are excellent in representing signal signatures which are different with regards to the scattering target. The TF maps are reassuring since we see the effect of higher frequencies occurring in the later principal components. This confirms the hypothesis of PCA being able to separate the data into components based on their ordered contribution to the variance. Hence, one can distinguish the targets by their respective scattering features while also observing the frequencies at which these scattering features occur. In particular, the signal maximum energy depicts the detection of the target, thus, projecting this on the surface of the target gives a quick insight into the interior structure. The results of the averaged topographic projection suggest the robustness of the method to the noise level of 10 dB SNR by quantitatively distinguishing a well-localised projection of the void.

Our approach for correcting the Green's function operator in the filtered backpropagation accounts for the wave attenuation in the target. This approach significantly suppresses the external artefacts with an enhancement of details in the interior and void structures. While results from the filtered backpropagation are consistent with that of \cite{dufaure2023imaging}, some of the filtering techniques provide additional enhancement of the reconstruction by reducing the artefacts present in the reconstruction.

Regarding the differences between the filters, the results suggest that,  on one hand, the simplest filter for anomaly distinguishability is data max, suggesting that it approximates the first-order direct-path scattering from the voids. However, on the other hand, compared to TF max or PCA, the data max filter seems also more vulnerable to outliers due to higher-order scattering effects, e.g., multi-path surface scattering, which is why it results in a comparably spotty reconstruction lowering the relative overlap measure. The data max gives in overall a lower RMSE in the interior and void with a noise level below 0 dB (total noise corruption) due to method-related discrepancies (e.g., mislocalisation) of the other filtering techniques. Filtering out those outliers might be possible via an improved forward simulation, e.g.,  finite element time-domain simulations, volume integral method or complementary ray-tracing estimates for the travel time, which were successfully applied in \cite{sorsa2023imaging}.

The findings presented in this study are relevant to the analysis of tomographic planetary mission measurements of small bodies such as the HERA mission, which is planned for probing the binary asteroid didymos using the Juventas radar \cite{herique2019juventas}. Our result shows that the measurements from such a mission can be quantitatively analysed to infer information about the interior structure of such an asteroid since this has been achieved with laboratory measurements. We have also accounted for noisy measurements up to 10 dB SNR which is consistent with earlier findings \cite{takala2018far} on the usability of noisy measurements with noise levels higher than 10 dB being corrupt and unusable. 
\label{sec:conclusion}




\bibliographystyle{IEEEtran}

\bibliography{cas-refs}

\begin{thebibliography}{10}
\providecommand{\url}[1]{#1}
\csname url@samestyle\endcsname
\providecommand{\newblock}{\relax}
\providecommand{\bibinfo}[2]{#2}
\providecommand{\BIBentrySTDinterwordspacing}{\spaceskip=0pt\relax}
\providecommand{\BIBentryALTinterwordstretchfactor}{4}
\providecommand{\BIBentryALTinterwordspacing}{\spaceskip=\fontdimen2\font plus
\BIBentryALTinterwordstretchfactor\fontdimen3\font minus \fontdimen4\font\relax}
\providecommand{\BIBforeignlanguage}[2]{{%
\expandafter\ifx\csname l@#1\endcsname\relax
\typeout{** WARNING: IEEEtran.bst: No hyphenation pattern has been}%
\typeout{** loaded for the language `#1'. Using the pattern for}%
\typeout{** the default language instead.}%
\else
\language=\csname l@#1\endcsname
\fi
#2}}
\providecommand{\BIBdecl}{\relax}
\BIBdecl

\bibitem{michel2015asteroids}
P.~Michel, F.~E. DeMeo, and W.~F. Bottke, \emph{Asteroids iv}.\hskip 1em plus 0.5em minus 0.4em\relax University of Arizona Press, 2015.

\bibitem{herique2018direct}
A.~H{\'e}rique, B.~Agnus, E.~Asphaug, A.~Barucci, P.~Beck, J.~Bellerose, J.~Biele, L.~Bonal, P.~Bousquet, L.~Bruzzone \emph{et~al.}, ``Direct observations of asteroid interior and regolith structure: Science measurement requirements,'' \emph{Advances in Space Research}, vol.~62, no.~8, pp. 2141--2162, 2018.

\bibitem{michel2022esa}
P.~Michel, M.~K{\"u}ppers, A.~C. Bagatin, B.~Carry, S.~Charnoz, J.~De~Leon, A.~Fitzsimmons, P.~Gordo, S.~F. Green, A.~H{\'e}rique \emph{et~al.}, ``The esa hera mission: detailed characterization of the dart impact outcome and of the binary asteroid (65803) didymos,'' \emph{The planetary science journal}, vol.~3, no.~7, p. 160, 2022.

\bibitem{knaell1995radar}
K.~Knaell and G.~Cardillo, ``Radar tomography for the generation of three-dimensional images,'' \emph{IEE Proceedings-Radar, Sonar and Navigation}, vol. 142, no.~2, pp. 54--60, 1995.

\bibitem{Kofman2007}
W.~Kofman, A.~Herique, J.-P. Goutail, T.~Hagfors, I.~Williams, E.~Nielsen, J.-P. Barriot, Y.~Barbin, C.~Elachi, P.~Edenhofer, A.-C. Levasseur-Regourd, D.~Plettemeier, G.~Picardi, R.~Seu, and V.~Svedhem, ``The comet nucleus sounding experiment by radiowave transmission (consert): A short description of the instrument and of the commissioning stages.'' \emph{Space Science REviews}, vol. 128, no. 413-432, 2007.

\bibitem{Kofman2015}
W.~Kofman, A.~Herique, Y.~Barbin, J.-P. Barriot, V.~Ciarletti, S.~Clifford, P.~Edenhofer, C.~Elachi, C.~Eyraud, J.-P. Goutail, E.~Heggy, L.~Jorda, J.~Lasue, A.-C. Levasseur-Regourd, E.~Nielsen, P.~Pasquero, F.~Preusker, P.~Puget, D.~Plettemeier, Y.~Rogez, H.~Sierks, C.~Statz, H.~Svedhem, I.~Williams, S.~Zine, and J.~Van~Zyl, ``{Properties of the 67P/Churyumov-Gerasimenko interior revealed by CONSERT radar},'' \emph{Science}, vol. 349, no. 6247, pp. aab0639--1 -- aab0639--6, 07 2015.

\bibitem{Kofman2020}
W.~Kofman, S.~Zine, A.~Herique, Y.~Rogez, L.~Jorda, and A.-C. Levasseur-Regourd, ``The interior of comet 67p/c–g; revisiting consert results with the exact position of the philae lander.'' \emph{Monthly Notices of the Royal Astronomical Society}, vol. 497, p. 2616–2622, 2020.

\bibitem{Herique2016}
A.~Herique, W.~Kofman, P.~Beck, L.~Bonal, I.~Buttarazzi, E.~Heggy, J.~Lasue, A.~Levasseur-Regourd, E.~Quirico, and S.~Zine, ``Cosmochemical implications of consert permittivity characterization of 67p/cg,'' \emph{Monthly Notices of the Royal Astronomical Society}, vol. 462, p. S516–S532, 2016.

\bibitem{Herique2019}
A.~Herique, W.~Kofman, S.~Zine, J.~Blum, J.-B. Vincent, and V.~Ciarletti, ``Homogeneity of 67p/churyumov-gerasimenko as seen by consert: implication on composition and formation,'' \emph{A\&A}, vol. 630, 2016.

\bibitem{kofman1998}
W.~Kofman, Y.~Barbin, J.~Klinger, A.-C. Levasseur-Regourd, J.-P. Barriot, A.~Herique, T.~Hagfors, E.~Nielsen, E.~Grün, P.~Edenhofer, H.~Kochan, G.~Picardi, R.~Seu, J.~van Zyll, C.~Elachi, J.~Melosh, J.~Veverka, P.~Weissman, L.~Svedhem, S.~Hamran, and I.~Williams, ``Comet nucleus sounding experiment by radiowave transmission,'' \emph{Advances in Space Research}, vol.~21, pp. 1589--1598, 1998.

\bibitem{barriot1999two}
J.-P. Barriot, W.~Kofman, A.~Herique, S.~Leblanc, and A.~Portal, ``A two dimensional simulation of the consert experiment (radio tomography of comet wirtanen),'' \emph{Advances in Space Research}, vol.~24, no.~9, pp. 1127--1138, 1999.

\bibitem{sava2015}
P.~Sava, D.~Ittharat, R.~Grimm, and D.~Stillman, ``Radio reflection imaging of asteroid and comet interiors i: Acquisition and imaging theory,'' \emph{Advances in Space Research}, vol.~55, no.~9, pp. 2149 -- 2165, 2015.

\bibitem{sava2018tomography}
P.~Sava and E.~Asphaug, ``3d radar wavefield tomography of comet interiors,'' \emph{Advances in Space Research}, vol.~61, no.~8, pp. 2198 -- 2213, 2018.

\bibitem{eyraud2018imaging}
C.~Eyraud, A.~H{\'e}rique, J.-M. Geffrin, and W.~Kofman, ``Imaging the interior of a comet from bistatic microwave measurements: case of a scale comet model,'' \emph{Advances in Space Research}, vol.~62, no.~8, pp. 1977--1986, 2018.

\bibitem{Sorsa2019}
L.-I. Sorsa, M.~Takala, P.~Bambach, J.~Deller, E.~Vilenius, and S.~Pursiainen, ``Bistatic full-wave radar tomography detects deep interior voids, cracks and boulders in a rubble-pile asteroid model,'' \emph{The Astrophysical Journal}, vol. 872(1):44, 2019.

\bibitem{deng2021ei}
J.~Deng, W.~Kofman, P.~Zhu, A.~H{\'e}rique, R.~Liu, and S.~Zheng, ``{EI+ FWI Method for Reconstructing Interior Structure of Asteroid Using Lander-to-Orbiter Bistatic Radar System},'' \emph{IEEE Transactions on Geoscience and Remote Sensing}, 2021.

\bibitem{gassot2021ultra}
O.~Gassot, A.~Herique, W.~Fa, J.~Du, and W.~Kofman, ``Ultra-wideband sar tomography on asteroids,'' \emph{Radio Science}, vol.~56, no.~8, pp. 1--17, 2021.

\bibitem{herique2019radar}
A.~Herique, D.~Plettemeier, C.~Lange, J.~T. Grundmann, V.~Ciarletti, T.-M. Ho, W.~Kofman, B.~Agnus, J.~Du, W.~Fa \emph{et~al.}, ``A radar package for asteroid subsurface investigations: Implications of implementing and integration into the mascot nanoscale landing platform from science requirements to baseline design,'' \emph{Acta Astronautica}, vol. 156, pp. 317--329, 2019.

\bibitem{herique2019juventas}
A.~Herique, D.~Plettemeier, W.~Kofman, Y.~Rogez, C.~Buck, and H.~Goldberg, ``A low frequency radar to fathom asteroids from juventas cubesat on hera,'' in \emph{Proc. EPSC-DPS Joint Meeting}, vol.~13.\hskip 1em plus 0.5em minus 0.4em\relax EPSC-DPS Joint Meeting 2019, Geneva, Switzerland, 2019, p. 807.

\bibitem{rivkin2021double}
A.~S. Rivkin, N.~L. Chabot, A.~M. Stickle, C.~A. Thomas, D.~C. Richardson, O.~Barnouin, E.~G. Fahnestock, C.~M. Ernst, A.~F. Cheng, S.~Chesley \emph{et~al.}, ``The double asteroid redirection test (dart): planetary defense investigations and requirements,'' \emph{The Planetary Science Journal}, vol.~2, no.~5, p. 173, 2021.

\bibitem{hirabayashi2022double}
M.~Hirabayashi, F.~Ferrari, M.~Jutzi, R.~Nakano, S.~D. Raducan, P.~S{\'a}nchez, S.~Soldini, Y.~Zhang, O.~S. Barnouin, D.~C. Richardson \emph{et~al.}, ``Double asteroid redirection test (dart): Structural and dynamic interactions between asteroidal elements of binary asteroid (65803) didymos,'' \emph{The planetary science journal}, vol.~3, no.~6, p. 140, 2022.

\bibitem{daly2023successful}
R.~T. Daly, C.~M. Ernst, O.~S. Barnouin, N.~L. Chabot, A.~S. Rivkin, A.~F. Cheng, E.~Y. Adams, H.~F. Agrusa, E.~D. Abel, A.~L. Alford \emph{et~al.}, ``Successful kinetic impact into an asteroid for planetary defence,'' \emph{Nature}, vol. 616, no. 7957, pp. 443--447, 2023.

\bibitem{pursiainen2016orbiter}
S.~Pursiainen and M.~Kaasalainen, ``Orbiter-to-orbiter tomography: a potential approach for small solar system bodies,'' \emph{IEEE Transactions on Aerospace and Electronic Systems}, vol.~52, no.~6, pp. 2747--2759, 2016.

\bibitem{sobolev1999robust}
S.~V. Sobolev, A.~Gr{\'e}sillaud, and M.~Cara, ``How robust is isotropic delay time tomography for anisotropic mantle?'' \emph{Geophysical Research Letters}, vol.~26, no.~4, pp. 509--512, 1999.

\bibitem{daubechies1990wavelet}
I.~Daubechies, ``The wavelet transform, time-frequency localization and signal analysis,'' \emph{IEEE transactions on information theory}, vol.~36, no.~5, pp. 961--1005, 1990.

\bibitem{cohen1995time}
L.~Cohen, \emph{Time-frequency analysis}.\hskip 1em plus 0.5em minus 0.4em\relax Prentice hall New Jersey, 1995, vol. 778.

\bibitem{grochenig2001foundations}
K.~Gr{\"o}chenig, \emph{Foundations of time-frequency analysis}.\hskip 1em plus 0.5em minus 0.4em\relax Springer Science \& Business Media, 2001.

\bibitem{hlawatsch1992linear}
F.~Hlawatsch and G.~F. Boudreaux-Bartels, ``Linear and quadratic time-frequency signal representations,'' \emph{IEEE signal processing magazine}, vol.~9, no.~2, pp. 21--67, 1992.

\bibitem{qian1999joint}
S.~Qian and D.~Chen, ``Joint time-frequency analysis,'' \emph{IEEE Signal Processing Magazine}, vol.~16, no.~2, pp. 52--67, 1999.

\bibitem{boashash2015time}
B.~Boashash, \emph{Time-frequency signal analysis and processing: a comprehensive reference}.\hskip 1em plus 0.5em minus 0.4em\relax Academic press, 2015.

\bibitem{esmersoy1989backprojection}
C.~Esmersoy and D.~Miller, ``Backprojection versus backpropagation in multidimensional linearized inversion,'' \emph{Geophysics}, vol.~54, no.~7, pp. 921--926, 1989.

\bibitem{l2012filtered}
G.~L.~Zeng, ``A filtered backprojection algorithm with characteristics of the iterative landweber algorithm,'' \emph{Medical physics}, vol.~39, no.~2, pp. 603--607, 2012.

\bibitem{koljonen2019mathematical}
V.~Koljonen, O.~Koskela, T.~Montonen, A.~Rezaei, B.~Belay, E.~Figueiras, J.~Hyttinen, and S.~Pursiainen, ``A mathematical model and iterative inversion for fluorescent optical projection tomography,'' \emph{Physics in Medicine \& Biology}, vol.~64, no.~4, p. 045017, 2019.

\bibitem{dufaure2023imaging}
A.~Dufaure, C.~Eyraud, L.-I. Sorsa, Y.~Yusuf, S.~Pursiainen, and J.-M. Geffrin, ``Imaging of the internal structure of an asteroid analogue from quasi-monostatic microwave measurement data. i. the frequency domain approach,'' \emph{Astronomy and Astrophysics}, vol. 674, p. A72, 2023.

\bibitem{sorsa2023imaging}
L.-I. Sorsa, Y.~O. Yusuf, A.~Dufaure, J.-M. Geffrin, C.~Eyraud, and S.~Pursiainen, ``Imaging of the internal structure of an asteroid analogue from quasi-monostatic microwave measurement data-ii. the time domain approach,'' \emph{Astronomy \& Astrophysics}, vol. 674, p. A73, 2023.

\bibitem{kawaguchi2008hayabusa}
J.~Kawaguchi, A.~Fujiwara, and T.~Uesugi, ``Hayabusa—its technology and science accomplishment summary and hayabusa-2,'' \emph{Acta Astronautica}, vol.~62, no. 10-11, pp. 639--647, 2008.

\bibitem{sorsa2021analogue}
L.-I. Sorsa, C.~Eyraud, A.~H\'{e}rique, M.~Takala, S.~Pursiainen, and J.-M. Geffrin, ``{Complex-structured 3D-printed wireframes as asteroid analogues for tomographic microwave radar measurements},'' \emph{Materials \& Design}, vol. 198, p. 109364, 2021.

\bibitem{Eyraud2020analog}
C.~Eyraud, L.-I. Sorsa, J.-M. Geffrin, M.~Takala, G.~Henry, and S.~Pursiainen, ``Full wavefield simulation versus measurement of microwave scattering by a complex 3d-printed asteroid analogue,'' \emph{Astronomy \& Astrophysics}, vol. 643, p. A68, 2020.

\bibitem{sorsa2021analysis}
L.-I. Sorsa, S.~Pursiainen, and C.~Eyraud, ``Analysis of full microwave propagation and backpropagation for a complex asteroid analogue via single-point quasi-monostatic data,'' \emph{Astronomy \& Astrophysics}, vol. 645, p. A73, 2021.

\bibitem{daubechies1992ten}
I.~Daubechies, \emph{Ten lectures on wavelets}.\hskip 1em plus 0.5em minus 0.4em\relax SIAM, 1992.

\bibitem{jaffard2001wavelets}
S.~Jaffard, Y.~Meyer, and R.~D. Ryan, \emph{Wavelets: tools for science and technology}.\hskip 1em plus 0.5em minus 0.4em\relax SIAM, 2001.

\bibitem{stankovic1994method}
L.~Stankovic, ``A method for time-frequency analysis,'' \emph{IEEE Transactions on Signal Processing}, vol.~42, no.~1, pp. 225--229, 1994.

\bibitem{debnath2003wavelets}
L.~Debnath, \emph{Wavelets and signal processing}.\hskip 1em plus 0.5em minus 0.4em\relax Springer Science \& Business Media, 2003.

\bibitem{jolliffe2002principal}
I.~T. Jolliffe, \emph{Principal component analysis for special types of data}.\hskip 1em plus 0.5em minus 0.4em\relax Springer, 2002.

\bibitem{jolliffe2016principal}
I.~T. Jolliffe and J.~Cadima, ``Principal component analysis: a review and recent developments,'' \emph{Philosophical transactions of the royal society A: Mathematical, Physical and Engineering Sciences}, vol. 374, no. 2065, p. 20150202, 2016.

\bibitem{bording1987applications}
R.~P. Bording, A.~Gersztenkorn, L.~R. Lines, J.~A. Scales, and S.~Treitel, ``Applications of seismic travel-time tomography,'' \emph{Geophysical Journal International}, vol.~90, no.~2, pp. 285--303, 1987.

\bibitem{tarantola1984seismic}
A.~Tarantola and F.~Santosa, ``The seismic reflection inverse problem,'' \emph{Inverse problems of acoustic and elastic waves}, pp. 104--181, 1984.

\bibitem{schuster2017seismic}
G.~T. Schuster, \emph{Seismic inversion}.\hskip 1em plus 0.5em minus 0.4em\relax Society of Exploration Geophysicists, 2017.

\bibitem{bertero2021introduction}
M.~Bertero, P.~Boccacci, and C.~De~Mol, \emph{Introduction to inverse problems in imaging}.\hskip 1em plus 0.5em minus 0.4em\relax CRC press, 2021.

\bibitem{chang1978method}
L.-T. Chang, ``A method for attenuation correction in radionuclide computed tomography,'' \emph{IEEE Transactions on nuclear Science}, vol.~25, no.~1, pp. 638--643, 1978.

\bibitem{yusuf2022investigation}
Y.~O. Yusuf, A.~Dufaure, L.-I. Sorsa, C.~Eyraud, and S.~Pursiainen, ``Investigation of wavelength-induced uncertainties in full-wave radar tomography of high contrast domain: An application to small solar system bodies,'' \emph{Icarus}, vol. 387, p. 115173, 2022.

\bibitem{takala2018far}
M.~Takala, P.~Bambach, J.~Deller, E.~Vilenius, M.~Wittig, H.~Lentz, H.~Braun, M.~Kaasalainen, and S.~Pursiainen, ``Far-field inversion for the deep interior scanning cubesat,'' \emph{IEEE Transactions on Aerospace and Electronic Systems}, 2018.

\end{thebibliography}

\end{document}